%% file: main.tex
\documentclass[12pt]{article}

\usepackage{scicite}

\usepackage{times}
\usepackage[utf8]{inputenc} 
\usepackage[T1]{fontenc}    
\usepackage{hyperref}       
\usepackage{url}            
\usepackage{booktabs}       
\usepackage{amsfonts}       
\usepackage{nicefrac}       
\usepackage{microtype}      
\usepackage{multicol}
\usepackage{color}
\usepackage{pgfplots}
\usepackage{graphicx}
\usepackage{makecell} 
\usepackage{multirow}
\usepackage{booktabs, graphicx, amssymb}
\usepackage{longtable}
\usepackage{array}
\usepackage{stfloats}
\usepackage{balance}
\usepackage{multicol}
\usepackage{color}
\usepackage{epstopdf}
\usepackage{bm}
\usepackage{amsmath}
\usepackage{amssymb}
\usepackage{booktabs} 
\usepackage{epsfig}
\usepackage{enumitem}
\usepackage{cleveref}
\usepackage{arydshln}
\usepackage{balance}
\usepackage{xspace}
\usepackage{wrapfig}
\usepackage{enumitem}
\pgfplotsset{compat=newest}
\usepackage{subfigure}
\usepackage{multibib}
\usepackage{minitoc}
\usepackage[subfigure]{tocloft}
\newcommand{\tabincell}[2]{\begin{tabular}{@{}#1@{}}#2\end{tabular}}
\newcommand{\hide}[1]{} 
\newcommand{\vpara}[1]{\vspace{0.05in}\noindent \textbf{#1 }}
\newcommand{\ipara}[1]{\vspace{0.03in}\noindent \textit{#1 }}

\newcommand{\beq}[1]{\vspace{-0.03in}\begin{equation}#1\end{equation}\vspace{-0.03in}}

\cftsetindents{figure}{0em}{4.5em}
\cftsetindents{table}{0em}{4em}

\newcommand{\model}{xTrimoPGLM}
\newcommand{\smodel}{xTrimoPGLM\space}

\newenvironment{sciabstract}{%
\begin{quote} \bf}
{\end{quote}}

\title{\model: Unified 100B-Scale Pre-trained Transformer for Deciphering the Language of Protein}

\author{
Bo Chen$^{2}$\thanks{Equal Contribution. Emails: \href{mailto:cb21@mails.tsinghua.edu.cn}{cb21@mails.tsinghua.edu.cn}, \href{mailto:xingyi@biomap.com}{derrickzy@gmail.com}}~~\thanks{Work done during interning at BioMap Research, California, USA.},
Xingyi Cheng$^{1}$\footnotemark[1]~~\thanks{The project leader at BioMap Research, California, USA.} \\
Pan Li$^{1}$, 
Yangli-ao Geng$^{2}$\footnotemark[2], 
Jing Gong$^{1}$, 
Shen Li$^{1}$, 
Zhilei Bei$^{2}$\footnotemark[2], 
Xu Tan$^{1}$,  \\
Boyan Wang$^{2}$\footnotemark[2], 
Xin Zeng$^{1}$,
Chiming Liu$^{1}$, 
Aohan Zeng$^{2}$, Yuxiao Dong$^{2}$  \\
Jie Tang$^{2}$\footnotemark[4], Le Song$^{1,3}$\thanks{The corresponding authors. Emails: \href{mailto:songle@biomap.com}{dasongle@gmail.com}, \href{mailto:jietang@tsinghua.edu.cn}{jietang@tsinghua.edu.cn}} \\
\normalsize{$^{1}$BioMap Research}
\normalsize{$^{2}$Tsinghua University}
\normalsize{$^{3}$MBZUAI} \\
}

\begin{document} 

\baselineskip24pt


\maketitle

\input{abs}
\input{intro}
\input{exp}

\clearpage
\input{method}
\bibliography{ref_all} 
\bibliographystyle{Science}
\clearpage
\input{extend_fig}
\clearpage
\input{app} 

\end{document}

%% file: abs.tex
\begin{sciabstract}

Protein language models have shown remarkable success in learning biological information from protein sequences. However, most existing models are limited by either autoencoding or autoregressive pre-training objectives, which makes them struggle to handle protein understanding and generation tasks concurrently. We propose a unified protein language model, \model, to address these two types of tasks simultaneously through an innovative pre-training framework. Our key technical contribution is an exploration of the compatibility and the potential for joint optimization of the two types of objectives, which has led to a strategy for training \smodel at an unprecedented scale of 100 billion parameters and 1 trillion training tokens. Our extensive experiments reveal that 
1) \smodel significantly outperforms other advanced baselines in 18 protein understanding benchmarks across four categories. The model also facilitates an atomic-resolution view of protein structures, leading to an advanced 3D structural prediction model that surpasses existing language model-based tools.
2) \smodel not only can generate de novo protein sequences following the principles of natural ones, but also can perform programmable generation after supervised fine-tuning (SFT) on curated sequences. 
These results highlight the substantial capability and versatility of \smodel in understanding and generating protein sequences, contributing to the evolving landscape of foundation models in protein science. Trained weight for the \smodel model, and downstream datasets are available at \url{https://huggingface.co/proteinglm}.
\end{sciabstract}

%% file: intro.tex
\section*{Introduction}

Proteins play vital roles in the sustenance, growth, and defense mechanisms of living organisms. They provide structural support for many essential biological processes such as synthesizing enzymes, facilitating transportation, regulating gene expression, and contributing to immune function. 
Therefore, understanding the biological information encoded within proteins is crucial for unraveling the intricate workings of life and advancing fields such as medicine and biotechnology~\cite{jumper2021highly, abramson2024accurate, baek2021accurate}.
As protein sequences serve as the blueprint for protein structure and function~\cite{anfinsen1959molecular}, 
pre-trained techniques on sequences, known as \textbf{P}rotein \textbf{L}anguage \textbf{M}odels (PLMs), e.g., the family of ESM models~\cite{rives2021biological, lin2023evolutionary}, ProtTrans~\cite{elnaggarrost}, PROGEN~\cite{madani2023large}, etc., offer a powerful tool for characterizing the properties and distributions of general protein sequences. These models are trained on large-scale protein datasets~\cite{apweiler2004uniprot, finn2014pfam, steinegger2019protein} that encompass billions of sequences, allowing them to capture evolutionary patterns and sequence features that are inherent in protein structures. 
As a result, these models achieve state-of-the-art results in predicting protein functions and structures~\cite{jumper2021highly, abramson2024accurate, lin2023evolutionary} or generating novel sequences with faithful three-dimensional structures~\cite{madani2023large, nijkamp2023progen2}. 

It is worth noting that different categories of protein-related tasks necessitate divergent outputs from PLMs, such as protein understanding tasks call for PLMs to yield accurate residue-level or protein-level representations, while protein design tasks depend heavily on the potent generation capabilities of PLMs.
Despite these varying outputs, all tasks reveal a consistent underlying dependency among protein sequences~\cite{anfinsen1959molecular, verkuil2022language}, which suggests the possibility of characterizing these tasks within one unified framework, potentially mitigating the disparity between task types and further augmenting the modeling power of PLMs. 
Unfortunately, existing PLMs are designed to address specific tasks depending on their pre-training framework. This presents a significant challenge to selecting appropriate PLMs for specific task types.
Consequently, we explore the feasibility of integrating tasks of understanding and generation, dictated by autoencoding and autoregressive pre-training objectives, respectively, into one unified framework. This unified approach aims to encapsulate the intricate dependencies inherent in protein sequences, potentially resulting in more versatile and robust protein foundation models.

Large Language Models (LLMs) have explored the revenue of developing unified pre-training paradigms. 
However, these studies typically adopt analogous training patterns. For instance, all pre-training objectives are commonly optimized using either the BERT-style~\cite{bao2020unilmv2} or GPT-style regime~\cite{tay2023ul2}. 
A balanced approach incorporating both bi-directional auto-encoding and uni-directional auto-regressive objectives could fulfill the requirements of unified PLMs, yet the feasibility of such integration remains an open question.
Practically, the current landscape of natural language processing is dominated by generative models, which afford various types of tasks via mapping task labels into a unified text space for zero/few-shot learning~\cite{brown2020language} or instruction-tuning~\cite{wei2022finetuned, chung2024scaling}. 
However, this capability is currently beyond the reach of PLMs. In practice, applications of protein modeling still rely on the bridging of representations with downstream task-specific labels, such as discrete values of categories or continuous values of 3D coordinates~\cite{lin2023evolutionary, wu2022high}. 
These tasks heavily rely on bi-directional auto-encoding training to tackle protein understanding tasks. Consequently, this highlights the need for a unified model that incorporates both training objectives.

In this work, we develop the first, to our knowledge, the xTrimo Protein General Language Model (\model), a unified pre-training framework and foundation model that scales up to 100 billion parameters, designed for various protein-related tasks, including understanding and generation (or design). The model differs from previous encoder-only (e.g., ESM) or causal decode-only (e.g., PROGEN) protein language models by leveraging the General Language Model (GLM)~\cite{du2022glm} as the backbone for its bidirectional attention and auto-regressive objective. 
To enhance the representation capacity of \model, we further introduce the Masked Language Model (MLM) objective to the bidirectional prefix region, building upon the generation ability encapsulated within the GLM objective.
Additionally, we compiled a large pre-training dataset, comprising approximately 940 million unique protein sequences with roughly 200 billion residues, and trained a model with 100 billion parameters over 1 trillion tokens over a cluster of 96 NVIDIA DGX machines each with 8$\times$A100 GPU cards. 

\model-100B demonstrates the significant enhancement in the realm of protein understanding.
By conducting extensive empirical experiments with linear probing and advanced fine-tuning techniques, we elevated the performance benchmarks in this domain. \model-100B has significantly surpassed previous state-of-the-art (SOTA) methods in 15 out of 18 tasks, covering a comprehensive range of areas including protein structure, interactions, functionality, and developability (Figure~\ref{fig:understand_task}A). We also illustrate that \smodel achieves lower Perplexity (PPL) on two Out-Of-Distribution (OOD) protein sets over other models (Figure~\ref{fig:1_xtglm}B). 
These results empirically validate the scaling behavior, demonstrating that larger models commonly tend to yield better performance (Figure~\ref{fig:1_xtglm}C and Figure~\ref{fig:understand_task}B). 


\smodel can serve as the base for developing  a high-performance 3D structural prediction tool. Inspired by methodologies similar to those in ESMFold, merge folding modules with a protein language model, thereby refining protein structure training. Our version named xTrimoPGLM-Fold~(xT-Fold for short) has shown promising results with impressive TM-scores in both CAMEO~(n=194) and CASP15~(n=56) protein benchmarks. Additionally, we optimized xT-Fold through 4-bit quantization, enhancing its performance and efficiency, which makes xT-Fold a leading option in PLM-based structure prediction tools. As a result, xT-Fold achieves a 5-point TM-score increase over ESMFold in the CASP15 dataset, coupled with a faster inference speed across various scenarios (Figure~\ref{fig:structure_performance_and_speed}).


\smodel also showcases an extraordinary ability to generate de novo protein sequences. These sequences not only exhibit diverse structures closely akin to natural counterparts, as evidenced by a median sequence identity of just 11.7\% (Figure~\ref{fig:show_general_cases}), but can also be tailored towards specific structural and biophysical properties through supervised fine-tuning (Figure~\ref{fig:sft_res} and Figure~\ref{fig:show_sft_cases}). This ``super alignment'' capability of \smodel underscores its potential as a programmable model for exploring and synthesizing the vast protein space.

Lastly, we discuss the key limitations of our protein language model in practical protein applications. Although our study confirms the potential of protein language models, it also highlights that critical enhancements are necessary for their effective deployment in real-world drug design. These enhancements include adapting models to diverse protein tasks, improving prediction accuracy for protein structures, and reducing generative protein hallucinations. Overcoming these challenges is essential to bridge the gap between theoretical capabilities and their practical application in drug discovery and development.

\label{sec:intro} 

%% file: exp.tex
\section*{Results}
\subsection*{The \smodel Framework}
\label{sec: framework}

\begin{figure*}
\centering
\includegraphics[width=1.0\linewidth]{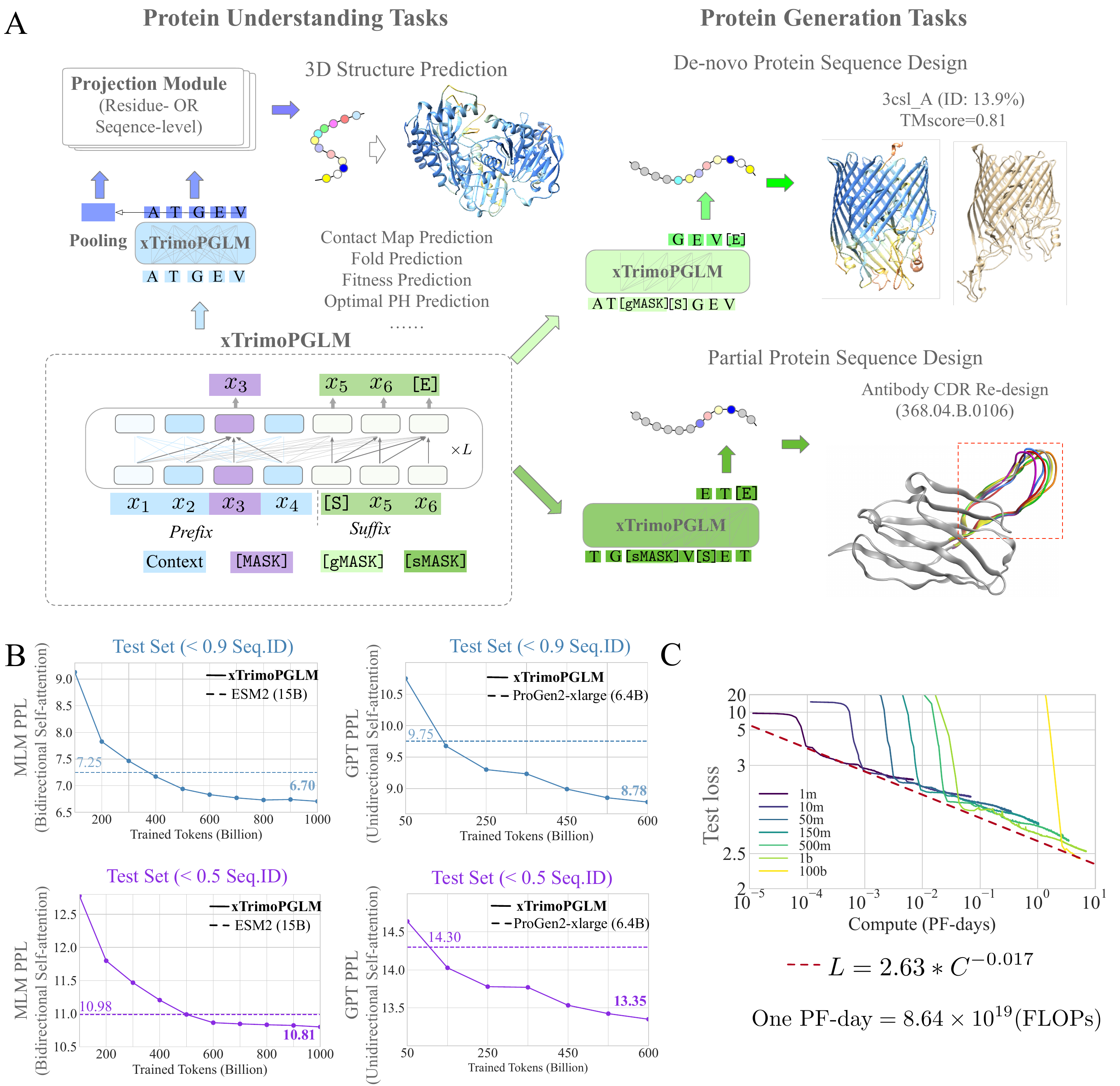} 
\caption[Comprehensive Insights into \model]{
\textbf{Comprehensive Insights into \model.}
\textbf{A.} The pre-training and fine-tuning stages of \model, combining BERT-style (blue and purple, for masking and predicting tokens) and GPT-style (green, from [S] to [E], for autoregressive generation) objectives. The prefix's bidirectional attention facilitates protein understanding tasks like structure prediction, while the suffix supports both de novo and conditional protein design through sequence generation.
\textbf{B.} \smodel shows lower perplexity than other leading PLMs like ESM2 and PROGEN2-xlarge in evaluations on two distinct out-of-distribution datasets, indicating its advanced performance.
\textbf{C.} The scaling behavior of \model-series from 1 million to 1 billion parameters, trained with 100 billion tokens, demonstrating \model-100B's efficiency through a power law fit of training losses against computational resources.
}
\label{fig:1_xtglm}
\end{figure*}

We adopt the GLM as our foundational framework to exploit its strengths in autoregressive blank infilling for training, while simultaneously processing input text bi-directionally. This dual approach is enhanced by the integration of a Masked Language Model (MLM) objective~\cite{kenton2019bert} to enhance its understanding capacity.
The core of our model is the simultaneous optimization of two distinct pre-training objectives, each characterized by unique indicator tokens, ensuring the proficiency in both understanding and generation capacities:
\begin{itemize}
    \item \textbf{Masked Language Model (MLM) Objective}: This task involves the prediction of tokens that have been randomly masked within sequences. These tokens are indicated by the special marker \texttt{[MASK]}. This task aligns with the functionality of BERT \cite{kenton2019bert} and ESM~\cite{lin2023evolutionary}, focusing on bidirectional contextual understanding.
    \item \textbf{General Language Model (GLM) Objective}: This task entails predicting subsequent tokens in a sequence, including short masked spans (indicated by \texttt{[sMASK]}) and longer spans at sequence ends (marked by \texttt{[gMASK]}). While the GLM objective takes into account the unidirectional context for predicting subsequent words, the prefix encoding portion remains bidirectional.
\end{itemize}


The framework of \model, along with its application in downstream tasks, is depicted in the lower panel of Figure~\ref{fig:1_xtglm}A. Motivated by the philosophy of curriculum learning, the pre-training stage is conducted in two distinct phases:
1). Initial pre-training with the MLM objective, focusing on rapid loss minimization across approximately 400 billion tokens. This phase is geared towards enhancing the model's understanding capabilities,
2). Subsequent training employs a unified approach, merging MLM and GLM objectives at a specific ratio (20\% MLM, 80\% GLM). This stage, utilizing an additional 600 billion tokens, is dedicated to refining both the model's representational and generative abilities.

\vpara{Quantification Scaling Law of \smodel Family Models.}
\label{subsec:scaling_law}
The scaling law is a crucial concept for understanding the performance of large language models (LLMs) during pre-training. It suggests a power-law relationship between a model's performance, typically measured by cross-entropy test loss $L(C)$, adheres to a power-law relationship with the computer resources $C$ used in training.
To investigate the scaling law for \smodel in the context of protein data, we employ the formula $L = a \times C^b$ where $C$ as approximated by $6ND$, where $N$ is the model size and  $D$ is the pre-trained dataset size (set to 100 billion tokens in this case for all the \smodel family models)
~\cite{kaplan2020scaling, hoffmann2022training}. 
Thus the scaling behaviour between ($L(C)$) and total compute $C$ can also be viewed as the rule between ($L(C)$) and model scale ($N$). 
We quantify $C$ in terms of training floating point operations (FLOPs), using PF-days (PetaFLOP-days) as the unit of measure, where one PF-day$=$ 8.64 × $10^{19}$ FLOPs. 
We extract a range of $(C, L(C))$ data points from the loss trajectories of models varying from 1 million to 1 billion parameters, each trained with 100 billion tokens (Figure~\ref{fig:1_xtglm}C).
Remarkably, this curve demonstrates close alignment with the actual training loss observed for the 100 billion parameter model of \model, trained with the same volume of tokens. 
This observation substantiates the model's adherence to the anticipated scaling law, 
thereby reinforcing the credibility of this principle as a pivotal guideline in the development of large-scale protein language models.
\vpara{\smodel Achieves Low Perplexity on Out-Of-Distribution Protein Sequences.}
\label{subsec:ppl}
Perplexity is a metric in language modeling that quantifies the uncertainty of a probability model in predicting a text sample, where lower values signify greater predictive accuracy. 
For a comprehensive evaluation, we construct two OOD datasets using a specific process: 
(1) Sampling an extensive collection of protein sequences from UniProt that were uploaded after January 2023, which is also the cutoff date for the training data.
And (2) Filtering these sequences based on sequence identity thresholds of 0.5 and 0.9 with the pre-training datasets. 
Each OOD dataset comprises approximately 10,000 protein sequences (Figure~\ref{fig:1_xtglm}B).
\model-100B achieves perplexity scores of 10.81 and 6.70 on the 0.5 and 0.9 sequence identity datasets, respectively, outperforming 15 billion parameters ESM2-15B (10.98 and 7.25). 
Similarly, against PROGEN2-xlarge with 6.4 billion parameters, \smodel recorders scores of 13.35 and 8.78, compared to PROGEN2-xlarge’s 14.30 and 9.75.
We observe that large-scale models are more sample efficient, reaching these perplexity levels with substantially less training data with respect to the corresponding pre-training objective: less than the actual 480 billion trained tokens for MLM pre-training (400 billion tokens at phases-1 and 80 out of 100 billion at phases-2) to match ESM2 trained on 1 trillion tokens, and less than actual 120 billion tokens for the GLM pre-training (120 out of 150 billion tokens at phase-2) to parallel PROGEN2-xlarge trained with 350 billion tokens, even if the learning rate schedule has not yet ended, resulting in an overestimated loss value at this stage. Noted that, the empirical studies in the Supplementary Section~\ref{sec:ana} confirm that the GLM model, when continued from MLM pre-training, converges faster than GLM model pre-trained from scratch. Thus the initial 400 billion MLM tokens serve as a warm-up stage, enhancing the model’s ability to understand sequence distributions. This foundational understanding accelerates the convergence during the subsequent GLM pre-training phase to enhance the model's generation capacity.


\begin{figure*}[!t]
\centering
\includegraphics[width=1.0\linewidth]{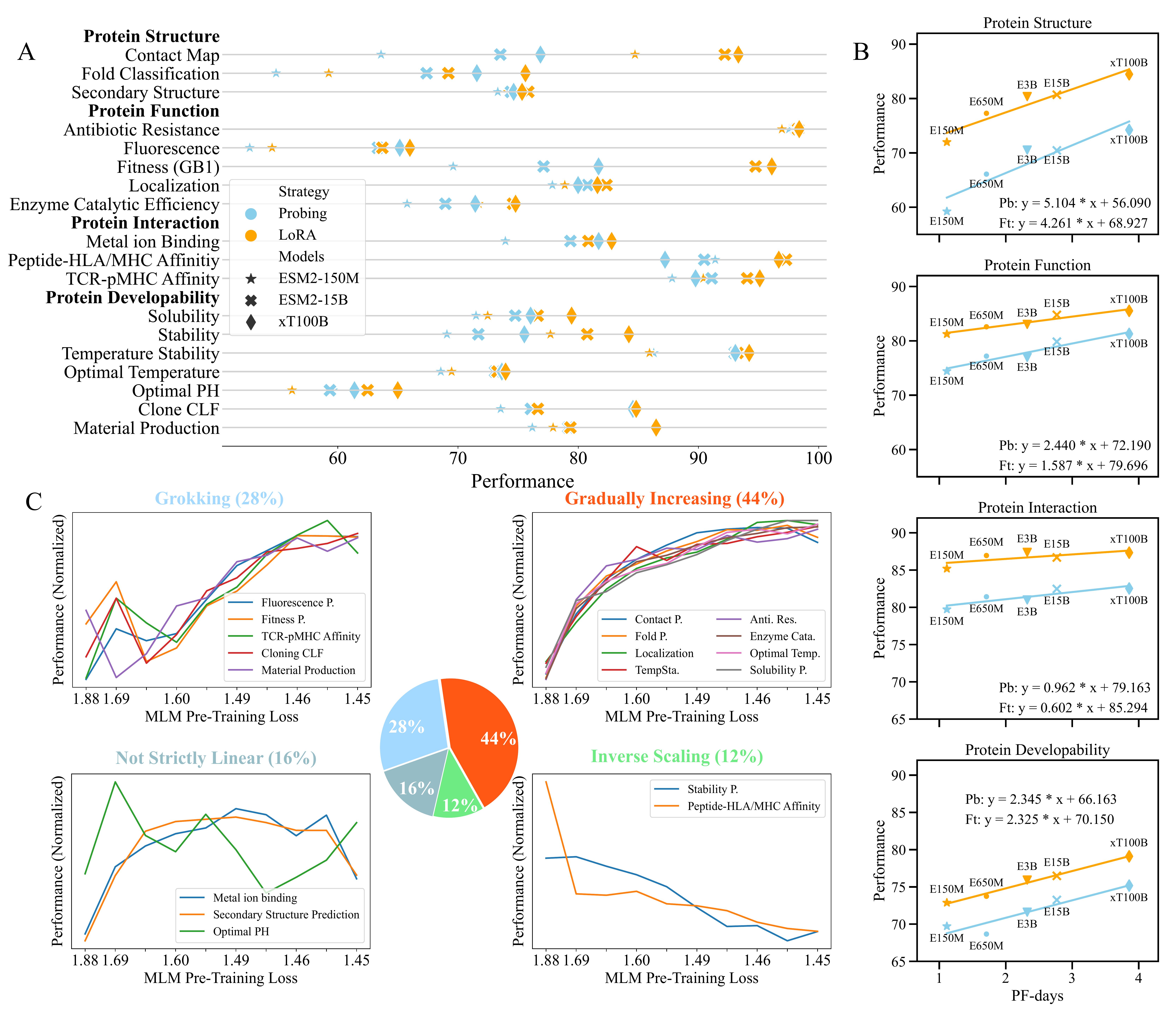}
\caption[The Performance of Protein Understanding Benchmark]{\textbf{The Performance of Protein Understanding Benchmark.}
\textbf{A.} For the classification task, four metrics are employed (Supplementary Section~\ref{sec::downstream_tasks}) including TopL/5 accuracy (Contact map), accuracy (Fold classification, Secondary structure, Antibiotic resistance, Solubility, Localization, Metal ion binding), AUC (Peptide-HLA/MHC affinity, TCR-pMHC affinity, Clone CLF, Material production) and Matthews Correlation Coef. (Temperature stability). For the regression task, two metrics are used including the Spearman Correlation Coef. (Fluorescence, Fitness, Stability, Optimal temperature, Optimal PH) and the Pearson Correlation Coef. (Enzyme catalytic efficiency). 
\textbf{B.} The scaling trend between the computational cost of model training, quantified by PF-days, 
where one PF-day$=$ 8.64 × $10^{19}$ FLOPs,
and the model performance 
Each data point symbolizes the mean performance metric for a specific task category (Pb for Probing and Ft for Fine-tuning with LoRA). E150M/650M/3B/15B, and xT100B represent ESM2-150M/650M/3B/15B, and \model-100B, respectively. 
\textbf{C.} Correlations between the pre-training validation loss measured by MLM objective and the performance of the downstream tasks. 
To facilitate comparison, we normalize this performance by subtracting the mean value and dividing it by the standard deviation. 
}
\label{fig:understand_task}
\end{figure*}

\subsection*{Evaluation on Protein Understanding benchmarks}
\label{sec:protein_und_19tasks}
To comprehensively assess the understanding capabilities of \model-100B, we benchmark it against 18 downstream protein-related tasks. These tasks span four primary categories: protein structure, developability, interactions, and functions (Figure~\ref{fig:understand_task}A and Supplementary Section~\ref{sec::downstream_tasks}).
Protein Structure: This category involves modeling the protein structure from the primary sequence, including secondary and tertiary structures.
Protein Developability: This category focuses on the engineering characteristics of proteins, such as their stability and manufacturability.
Protein Interaction: This category includes scenarios where proteins interact with other molecules, such as short peptides or other proteins.
Protein Function: This category encompasses tasks that predict the intrinsic cellular features and activities of proteins, including enzyme catalytic efficiency.
We compare \model-100B with two other protein language models in the field, ESM2-150M and ESM2-15B~\cite{lin2023evolutionary}, to provide a well-rounded evaluation. Our assessment methodology encompasses two distinct approaches to evaluate the effectiveness of the models' representations:
\begin{itemize}
    \item \textbf{Probing with MLP.} We utilize a trainable multi-layer perceptron (MLP) as a probe to examine the evolutionary information encoded in the pre-trained representations. This method simply and efficiently identifies the types of protein information captured by the models, where pre-trained PLMs' parameters remain fixed, and only the MLP is trained. For comparisons, the embeddings from pre-trained models are projected into 128 dimensions followed by ReLU activation before passing to the next layer of MLP.
    \item \textbf{Fine-tuning with LoRA.} Considering the constraints of GPU memory, full-scale fine-tuning is not feasible for models with parameters on the scale of 100 billion. 
    Hence, we employ Low-Rank Adaptation (LoRA)~\cite{hu2022lora} as a parameter-efficient alternative. This technique involves freezing the pre-trained model's weights and adding trainable low-rank matrices to each transformer layer. LoRA significantly reduces the number of trainable parameters for downstream tasks while preserving the adaptability of the learned representations. 
    The fine-tuning architecture and settings are similar to those in MLP probing, with only the transformer's ${W_q, W_k, W_v, W_o }$ parameters fine-tuned. 
\end{itemize}


Figure~\ref{fig:understand_task}A provides a comprehensive visualization of performance across all benchmarked protein-related tasks. Distinct colors and shapes in the figure represent different evaluation strategies and models, respectively. The ESM2-150M model, being relatively smaller, serves as an indicator to assess the difficulty of these tasks.
The performance distribution highlights the inherent relationships between the complexity of tasks and the advantages brought by the scale of the model. 
The distribution of performance across tasks underscores the relationship between task complexity and the benefits derived from the scale of the model. In more complex tasks, such as Contact Map prediction (Protein Structure category), Fluorescence (Protein Function), Metal Binding (Protein Interaction), and Stability (Protein Development), the larger models (\model-100B and ESM2-15B) significantly outperform the smaller ESM2-150M.
This disparity in performance highlights the necessity for more advanced models to effectively address complex tasks. Conversely, for simpler tasks, like Antibiotic Resistance (Protein Function category), the performance gap between the large and small models is notably smaller.
This observed pattern suggests that larger models are more adept at capturing the intrinsic evolutionary information of protein sequences. Consequently, as models scale up, they exhibit marked improvements in performance, particularly for complex tasks.
The application of LoRA consistently enhances overall task performance compared to the static Probing method, which limits the capacity of pre-trained models (Figure~\ref{fig:understand_task}B). LoRA's efficiency lies in its ability to refine and utilize key features without significantly increasing the number of trainable parameters (less than 1\% increase with LoRA rank set to 8).
The empirical findings also highlight a scaling trend in task performance with supervised fine-tuning, suggesting a strong correlation between model scale and performance. The following sections will delve deeper into the varying scaling behaviors observed across different types of tasks.

\vpara{Scaling Behaviors of Downstream Tasks.}
We extend the investigation of scaling laws to downstream tasks in four distinct categories of protein-related tasks from three perspectives.

\noindent \textit{Comparison Between ESM and \smodel Family models.}
We observe the scaling behavior of downstream tasks among the ESM and \smodel family models using both Linear Probing with MLP (Pb) and Fine-tuning with LoRA (Ft) techniques (Figure~\ref{fig:understand_task}B). The x-axis measures the total computational cost in Petaflop Days (PF-Days), while the y-axis represents the performance of various task categories. The results indicate that, although the intensity of the scaling effect varies among task types, a common trend persists: an exponential increase in computational resources during pre-training correlates with linear improvements in downstream task performance.
While the intensity of the scaling effect differs among task types, a common trend observed from pre-training extends to these downstream tasks: an exponential increase in computational resources during pre-training correlates with linear improvements in task performance. This extension of the scaling law to downstream tasks in protein modeling is a novel observation.

\noindent \textit{Comparison Among Different Training States of \model.}
To eliminate the differences in backbone architectures, pre-train datasets, etc, we further take a deep insight into the correlations between the downstream task performance with the ongoing pre-training process (measured by the MLM validation loss, Figure~\ref{fig:understand_task}C). 
We observe that most tasks demonstrate positive correlations. Specifically, 44\% show a gradual increase in performance, 28\% exhibit a ``Grokking'' phenomenon, and 16\% do not follow a strictly linear pattern.
The 'Grokking' phenomenon mirrors emergent abilities seen in large NLP models~\cite{wei2022emergent}.
It occurs when models initially overfit the training data, especially when task datasets have limited overlap with pre-training data. As the model's knowledge base expands, it begins to apply its understanding to out-of-distribution (OOD) scenarios, resulting in a sudden increase in test data performance. 
Thus the performance of the test data suddenly increased, denoted as the Grokking phenomenon.
However, \textcolor{blue}{12}\% of tasks indicate a potential negative impact from increased computational resources, suggesting an inverse scaling effect. 

\noindent \textit{Comparison Among Different Scales of \smodel Family Models.}
Additionally, we further analyze the scaling behavior by comparing \smodel family models in the  Supplementary Figure~\ref{fig:glm_scales}, which show a similar trend that most tasks exhibit a positive relationship between task performance and the scaling of training FLOPs and model size.

Our findings highlight the potential of scaling up models to enhance performance across a broad spectrum of protein-related tasks. This approach contrasts with other methods that seek data-efficient, cost-reduced, and knowledge-guided PLMs without relying on large-scale language models~\cite{elnaggar2023ankh}. 
These empirical observations provide valuable insights for future research in model development and advancement in the field of protein language modeling.

\begin{figure*}[!t]
\centering
\includegraphics[width=1.0\linewidth]{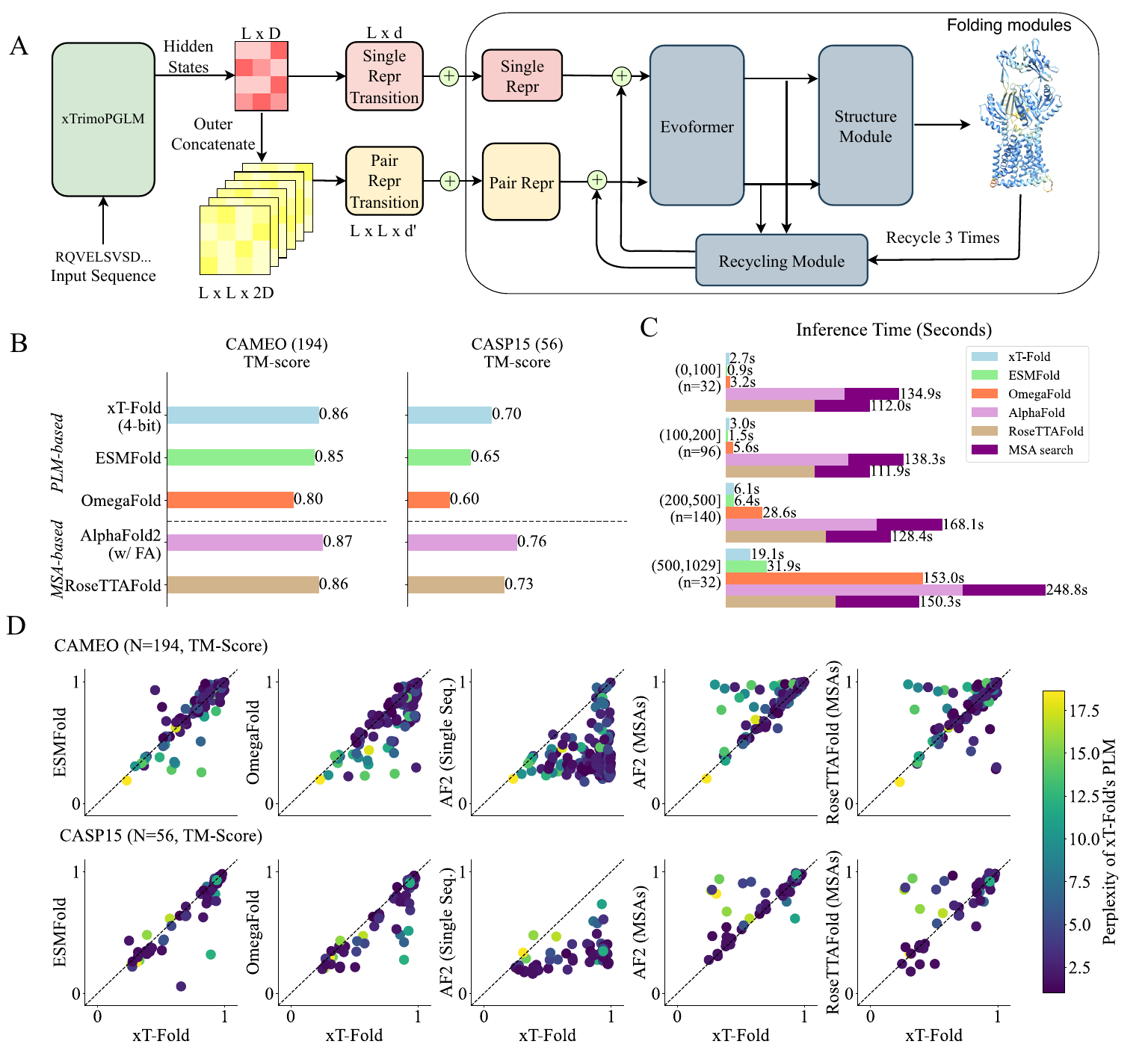} 
\caption[Structure Prediction with xT-Fold]{
\textbf{Structure Prediction with xT-Fold.}
\textbf{A.} xT-Fold architecture leverages a Multi-Layer Perceptron (MLP) to convert PLM representations into inputs for the folding modules, which generate 3D coordinates and pLDDT confidence scores.
\textbf{B.}
TM-score benchmarks for structure prediction models. The bar chart shows the performance of single-sequence PLM-based models and MSA-based models on CAMEO and CASP15 datasets.
\textbf{C.} 
 Inference time comparison across models for varying sequence length intervals, showing xT-Fold, ESMFold, OmegaFold, AlphaFold, RoseTTAFold, and MSA search times in seconds.
\textbf{D.}
Scatter plots compare xT-Fold predictions (x-axis) to other models (y-axis), color-coded by perplexity~(green for high, purple for low). 
}
\label{fig:structure_performance_and_speed}
\end{figure*}

\subsection*{Evaluation on Protein Structure Prediction}
\label{sec:protein_struct}
Protein sequences encode rich information about their 3D structures and function through evolutionary history~\cite{anfinsen1959molecular}. 
Advanced methods using multiple sequence alignments (MSAs) like AlphaFold2~\cite{jumper2021highly} and RoseTTAFold~\cite{baek2023efficient} are highly accurate in predicting protein structures and are key tools in computational biology. 
Meanwhile, PLM-based models such as ESMFold~\cite{lin2023evolutionary} and OmegaFold~\cite{wu2022high}, while not as precise as MSA-based models, provide faster predictions. 
This rapid prediction is crucial for high-throughput applications, accelerating our understanding of biological mechanisms and hastening drug discovery efforts.

In this section, we propose xT-Fold, where building on the xTrimoPLGM-100B framework, marks a significant advancement by achieving SOTA results for the PLM-based structure prediction model on benchmarks such as CAMEO and the latest CASP15.
Notably, the model offers both high accuracy and computational efficiency, utilizing 4-bit quantization and FlashAttention~\cite{dao2022flashattention} techniques to run effectively on a single A100 GPU card~(Supplementary Section~\ref{xtfold_config})
This balance of speed and precision makes xT-Fold a tool choice for fast-paced research and drug discovery.

The overall architecture of xT-Fold closely resembles that of ESMFold~(Figure~\ref{fig:structure_performance_and_speed}A). It involves training Multi-Layer Perceptron (MLP) layers to map the respective single representation and pair representation to $d$ and $d'$ from $D$ dimension, which are fed into the structure module for 3D coordinates prediction. We used a 48-layer evoformer with approximately 88 million parameters. The structure module accounts for about 2 million parameters, plus additional heads (81K), bringing the total to around 90 million parameters. The maximum recycle count is 3. During training, the recycle count was randomly selected between 0 and 3, consistent with the original AF2 and ESMFold approaches.

We evaluated two individual test sets, CAMEO and CASP15, both of which are out-of-distribution in terms of their timelines, ensuring differences from our training set~(Supplementary Section~\ref{xtfold_data}). CAMEO includes 194 samples (released date between April and June 2022), while CASP15 consists of 56 publicly available proteins~(released in May 2022). 
Initially, we observed the perplexity~(the lower are better) of these two data sets in PLM, which served as the backbone of the folding models~(Supplementary Figure~\ref{fig:ppl_comp}). 
On CAMEO and CASP15, xTrimoPGLM achieved a perplexity of 4.01 and 4.45, respectively, in contrast to ESMFold's PLM~(ESM2-3B), which scored 5.21 and 6.18. 
The perplexity~(PPL) demonstrates that language models generally have a better understanding of the CAMEO dataset compared to CASP15. This suggests that CAMEO might be less challenging than CASP15, indicating that CASP15, relative to CAMEO, is closer to OOD data for language models.

We conducted a comparative analysis against other MSA-free PLM-based models including ESMFold, OmegaFold, and 4-bit xT-Fold (Figure~\ref{fig:structure_performance_and_speed}B and Supplementary Table~\ref{tb:gdt}). 
In performance evaluation, xT-Fold achieved a TM-score of 0.86 on the CAMEO dataset and 0.70 on CASP15. The scores for ESMFold were 0.85 and 0.65, respectively, while OmegaFold scored 0.80 and 0.60 on these datasets. 
When compared to methods that integrate MSAs and template retrievals, such as AlphaFold2 and RosettaFold, PLM-based models like xT-Fold and ESMFold showed performance closely matching that of RosettaFold on the CAMEO dataset. However, on the more OOD dataset, i.e. CASP15, the overall efficacy of the PLM-based methods still lagged behind MSA-augmented approaches. 

On the other hand, when benchmarked on the CASP15 and CAMEO test sets across various sequence length intervals, the inference speed of MSAs/template-based models is notably slower than that of PLM-based models (Figure~\ref{fig:structure_performance_and_speed}C), lagging by approximately 10x to 50x. 
This disparity remains even for accelerated AlphaFold2, which is optimized with FlashAttention (FA), achieving an acceleration of 2x to 8x compared to the open-source variant (available at \url{https://github.com/google-deepmind/alphafold}) across sequence lengths ranging from 200 to 1000 and kept the output consistent with the canonical self-attention. 
For the PLM-based methods, the open-source versions of ESMFold (\url{https://github.com/facebookresearch/esm}) and OmegaFold (\url{https://github.com/HeliXonProtein/OmegaFold}) were utilized. 
As result, xT-Fold overall exhibits a marginal speed advantage over ESMFold and OmegaFold, due to the optimization with FA. Particularly on longer sequence intervals, xT-Fold demonstrates its superiority. The relative slowness of OmegaFold is attributed to its default setting of using 10 recycling passes to achieve better results.


Upon further examination of the scatter plots~(Figure~\ref{fig:structure_performance_and_speed}D), a perceptible correlation between the perplexity from xTrimoPGLM and the structural metric TM-score is observed. Notably, for samples with intermediate PPL values, xT-Fold demonstrates enhanced performance compared to ESMFold and OmegaFold. 
xT-Fold is also juxtaposed with AlphaFold2 with single-sequence input. Given that AlphaFold2 is not trained on single sequences, a significant drop in overall effectiveness is evident. The last two columns compare xT-Fold with AlphaFold2 and RosettaFold. It is apparent that where PPL is moderately high, corresponding to less accurate predictions, xT-Fold's performance is less effective than the two methods, For samples with lower PPL predictions, xT-Fold occasionally surpasses RosettaFold. 
To provide a more comprehensive and quantified understanding of the relationship between sequence perplexities (PPL) and the corresponding TM-scores predicted by xT-Fold, we have calculated the Pearson correlation coefficient and P-value for the CAMEO, CASP14, and CASP15 datasets. The Pearson coefficients and p-value pair are (-0.415, 1.7e-9), (-0.579, 1e-5), and (-0.239, 8e-2), respectively. These results clearly indicate negative correlations between PPLs and TM-scores on CAMEO and CASP14, but not strictly correlations in CASP15.
Overall, these quantified metrics demonstrate that lower perplexities are associated with more accurate predicted structures.
In summary, while scaling single-sequence models enhances the performance, it still struggles with OOD data, which is effectively addressed by MSAs augmentation.






\begin{figure*}[!t]
\centering
\includegraphics[width=1.0\linewidth]{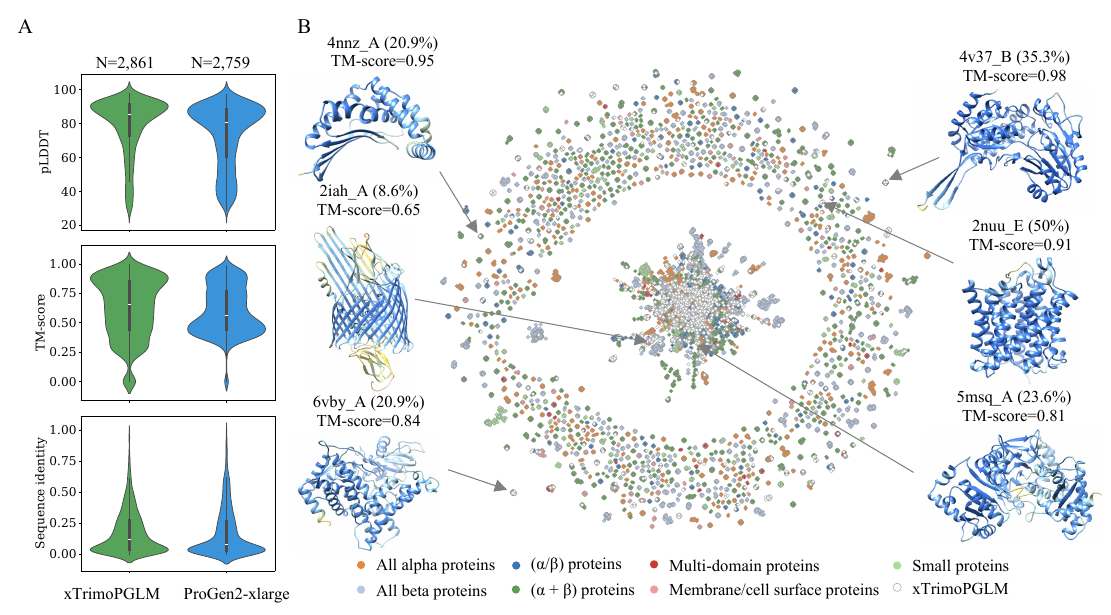} 
\caption[Diversification of Protein Structures by \model]{
\label{fig:show_general_cases}
\textbf{Diversification of Generated Proteins by \model.} 
\textbf{A.} Violin plots comparing ESMFold-predicted confidence (pLDDT scores) and similarity to Protein Data Bank (PDB) entries—measured by TM-score and sequence identity—for sequences generated by \smodel (green, $N$ = 14,626) and PROGEN2-xlarge (blue, $N$ = 8,466). The plots display the median, upper and lower quartiles, and whiskers representing 1.5$\times$ the interquartile range.
\textbf{B.} The comprehensive mapping of protein structural space as informed by sequences generated by \model. Each node represents a sequence generated by \smodel or a sequence from SCOPe70\_2.08. 
Two nodes are linked when one of them can be searched from SCOPe70\_2.08 with an alignment of at least 20 amino acids and 70\% hhsearch probability. The color coding corresponds to distinct SCOP structural classes, with \model-generated sequences highlighted in white. For illustrations (Supplementary Figure~\ref{fig:app_gen}), we showcase 6 examples from generated sequences. The PDB chain ID with the highest structural similarity to the generated sequence, their sequence identity, and TM-score are displayed above each example. The color of the structure matches the xT-Fold pLDDT values. The blue color represents high confidence (pLDDT>90). 4nnz\_A: Probable zinc protease. 2iah\_A: Ferripyoverdine receptor. 6vby\_A: Cinnamic acid 4-hydroxylase. 4v37\_B: Betaine aldehyde dehydrogenase. 2nuu\_E: Ammonia channel. 5msq\_A: Carboxylic acid reductase.
}
\end{figure*}

\subsection*{Evaluation on Protein Sequence Generation}
\label{sec:protein_gen}
Autoregressive models have emerged as powerful tools for representing the diverse array of evolutionary sequences found in nature. 
This capability facilitates the generation of novel protein sequences, exhibiting diverse folding patterns that significantly diverge from naturally occurring proteins~\cite{nijkamp2023progen2, ferruz2022protgpt2}. To validate the generative ability of \model-100B, we conduct an extensive analysis of the properties of protein sequences synthesized by \model-100B under various generative scenarios.
Our investigation spans several generative contexts: 
universal protein synthesis utilizing pre-training data and
task-specific sequence generation via Supervised Fine-Tuning (SFT), sequence creation following Reinforcement Self-Training (ReST).
This multifaceted approach provides deep insights into the potential and versatility of \model-100B in advancing protein sequence generation.

\vpara{\smodel Generate Sequences with Diverse Structures.}
To evaluate the generative capacity of \smodel, we generate 14,626 sequences with the \model-100B model utilizing \texttt{[gMASK]} indicators as the inserting prompt. This process generates new sequences by continuously predicting the next token in an auto-regressive manner until the \texttt{<eos>} token is predicted or the pre-set maximum length is reached. We use nucleus sampling by combining different top P (0.5, 0.7, 0.9, 1.0) and sampling temperature (0.2, 0.4, 0.6, 0.8, 1.0) parameters. For each parameter combination, we generated 2,000 protein sequences and limited the maximum length to 800 tokens. Then we performed a simple filtering on the generated sequences: (1) removing the generated sequences with perplexity>10; (2) removing the sequences containing repeated fragments; (3) MMseqs2 clustering (--min-seq-id 0.9 -c 0.5) and only leave the centroid sequence. We also used same strategy to generate 8,466 sequences for PROGEN2.


We employed ESMFold to predict the structures of all generated sequences. Our model generated proteins with higher confidence scores than PROGEN2-xlarge (mean pLDDT scores 84.0 vs 74.3, Figure~\ref{fig:show_general_cases}A upper). We then used Foldseek to search for the most structurally similar natural proteins from the PDB database. We measured the structural and sequence similarity using TM-score and sequence identity, respectively. Our model exhibited much higher structural resemblance to PDB entries than PROGEN2-xlarge (mean TM-score 0.695 vs 0.522) with very low sequence identity (mean sequence identity 0.224 vs 0.165) and high diversity (Supplementary Figure~\ref{fig:gen_cluster}A). We have stratified the generated sequences into four groups based on perplexity values (<2, 2-5, 5-8, >8). Across all perplexity ranges, our model consistently produces proteins with higher confidence scores and greater structural and sequence resemblance to PDB entries compared to PROGEN2 (Supplementary Figure \ref{fig:generated_vs_ppl}).

\begin{figure}[!h]
\centering
\includegraphics[width=1.0\linewidth]{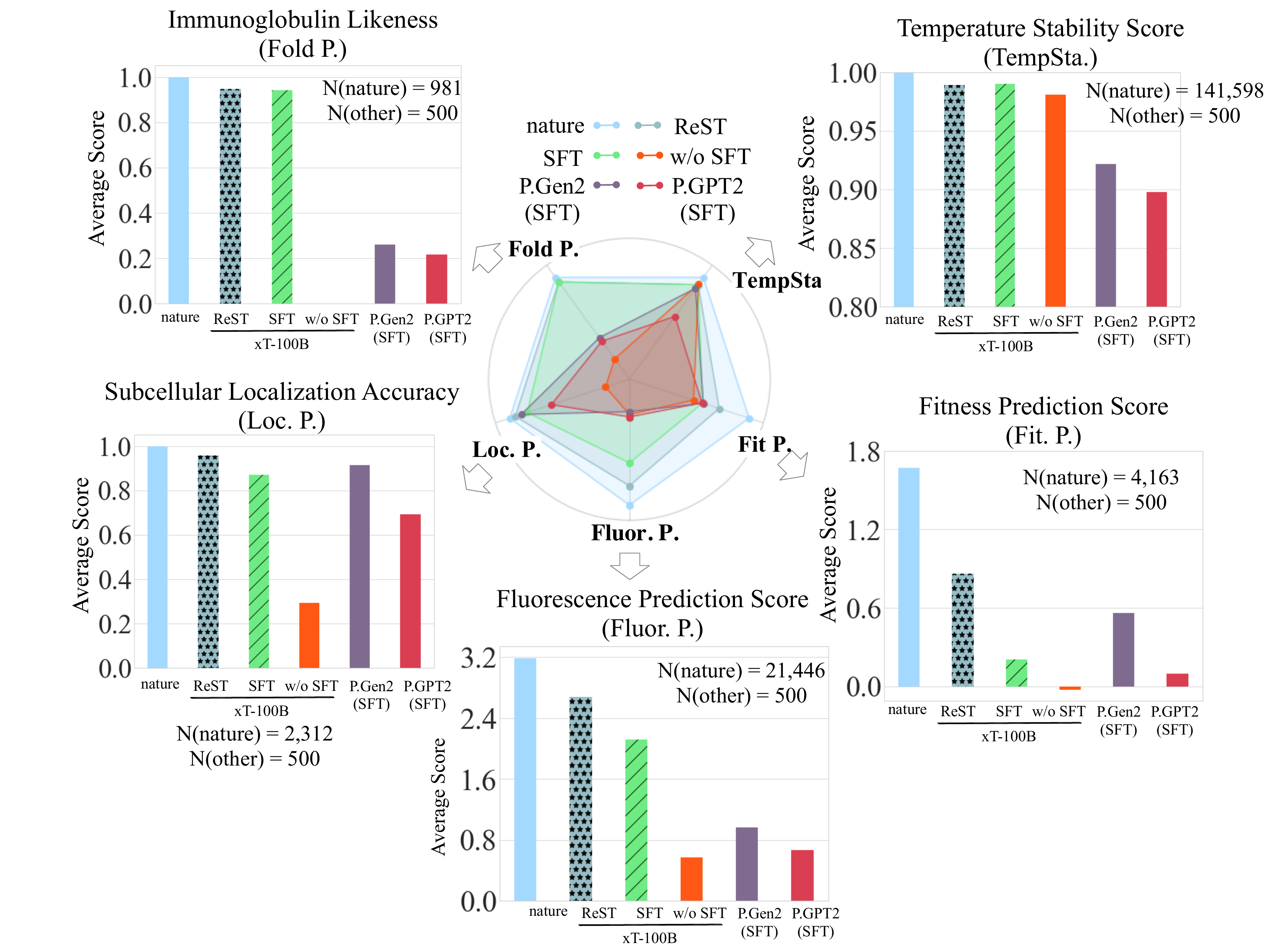} 
\caption[Alignment Capabilities of \smodel in Protein Sequence Generation]{
\textbf{Robust Alignment Capabilities of \smodel in Protein Sequence Generation towards Desired Properties.}
Quantitative analysis of \model, enhanced with Supervised Fine-Tuning (SFT) and Reinforcement Self-Training (ReST), across five selected tasks. The number of sampled generated sequences, $N$(other), and natural sequences, $N$(nature), used in the analysis are illustrated in the figure. The results demonstrate \model’s effectiveness in aligning with specific task objectives, as shown by the average scores (higher scores indicate better alignment). P.Gen2 refers to the PROGEN2-xlarge model~\cite{nijkamp2023progen2} with 6.4 billion parameters, and P.GPT2 denotes the ProtGPT2 model~\cite{ferruz2022protgpt2} with 740 million parameters.
 }
\label{fig:sft_res}
\end{figure}

To measure the generated protein distribution in the protein space, we used Foldseek to align the structures predicted by ESMFold to the AlphaFold/UniProt50-minimal structure dataset to obtain the maximum TM-score for each generated sequence. This dataset is constructed by MMseqs2 with 50\% sequence similarity from the UniProt subset in the AlphaFold Database. To obtain the maximum sequence similarity to natural protein sequences, we also retrieved each generated sequence from UniProt, UniClust30, and BFD database with MMseqs2 and HHblits (Supplementary Figure~\ref{fig:gen_cluster}B). The UniProt and BFD databases contain more than 2.5 billion protein sequences, representing the space of currently known protein sequences in nature. For sequences with ESMFold pLDDT>80 (N=11,048), we show the maximum sequence similarity and maximum TM-score of all query sequences in the scatter plot (Supplementary Figure~\ref{fig:gen_cluster}C), with a mean sequence similarity value of 0.699 and a mean TM-score value of 0.864. The vast majority of generated structures have similar 3D structures to those in UniProt50, which confirms that our pre-trained model comprehensively understands and represents the protein universe.


This potentially allows us to access a larger sequence in the protein manifold while we try to design certain protein structures.
Figure~\ref{fig:show_general_cases}B showcases a comprehensive network of the protein structural space, informed by sequences synthesized by \model. Each node corresponds to a \model-generated sequence or a sequence from SCOPe70\_2.08. Sequences originating from \smodel are distinctly marked in white. This network vividly illustrates that \smodel successfully generates novel protein sequences encompassing a broad spectrum of protein folds, while maintaining low sequence identity.
More analysis refers to Supplementary Section~\ref{subsec: more_gen}.

\vpara{Enhanced Protein Sequence Generation through Supervised Fine-Tuning and Reinforcement Self-Training.}
The \model-100B excels in generating diverse protein sequences but faces challenges in aligning to produce sequences with specific properties or families, such as lysozymes or immunoglobulins. This limitation is a critical bottleneck for applications in various industries, including pharmaceuticals and agriculture.
Adopting strategies from OpenAI's GPT models~\cite{brown2020language}, \model-100B serves as a protein foundational model, equipped with vast knowledge from trillions of residue tokens. We apply established alignment methods like Supervised Fine-Tuning (SFT)~\cite{ouyang2022training} on select protein families and enhance this with Reinforcement Self-Training (ReST)~\cite{gulcehre2023reinforced}, based on the SFT model.
We fine-tune \model-100B on datasets representing common protein structures or chemical properties. We choose five tasks from 18 benchmark protein understanding tasks (Fold Prediction, Temperature Stability, Localization Prediction, Fluorescence Prediction, and Fitness Prediction) for fine-tuning. 
Specifically, we adopted two filtering strategies to obtain the SFT datasets: Regression Tasks (Fluorescence and Fitness Prediction): We filtered samples whose label scores exceed a certain threshold, then fine-tuned the models with these samples. Classification Tasks (Fold Prediction, Temperature Stability, Localization): We used samples from one category to fine-tune the model and generate new samples. 
Moreover, we further filter the protein sequences generated by the SFT models as the ReST datasets. We applied the same filtering and fine-tuning settings during the Reinforcement Self-Training (ReST) stages.
We finetune our model and the baseline models, such as PROGEN2 and ProtGPT2, using the same causal language modeling regime (Supplementary Section~\ref{sec:sft}). Comparative analysis is conducted against ProtGPT2 and PROGEN2 using identical SFT protocols.
To circumvent trivial results, we employ the non-fine-tuned \model-100B as a control. 
Task-specific predictors evaluate the quality of the generated sequences, acting as biased evaluators. Due to the impracticality of in vitro validation, we used in silico simulators, a common approach in prior research~\cite{nijkamp2023progen2, ferruz2022protgpt2}.
More specifically, we use the corresponding task predictor to predict the scores of the desired class (for classification tasks) or the regression scores (for regression tasks) to validate whether the SFT models could generate sequences with the desired properties. 

Our findings reveal that:
1). Sequences from \smodel with SFT consistently score higher on targeted properties than the non-fine-tuned baseline,
2). \smodel surpasses PROGEN2 and ProtGPT2 under the same SFT conditions for most tasks, underscoring its efficacy as a foundational model capable of superior alignment with minimal data or fewer tuning steps, in line with observed scaling behavior (Figure~\ref{fig:sft_res}).
Furthermore, we implement a one-step Reinforcement Self-Training process, which utilizes task predictors as reward models, guiding the self-training of the SFT-enhanced \model-100B as follows:
1). Task predictors evaluate the quality of sequences from the SFT model,
2). Sequences of higher quality are then selected to form a new dataset, which is used for further fine-tuning. This iterative process results in the development of the Reinforcement Self-Training (ReST) model.
Remarkably, the ReST model effectively synthesizes sequences closely resembling natural datasets, highlighting \model's potential as a robust protein synthesizer for industrial applications.

\begin{figure}[!h]
\centering
\includegraphics[width=1.0\linewidth]{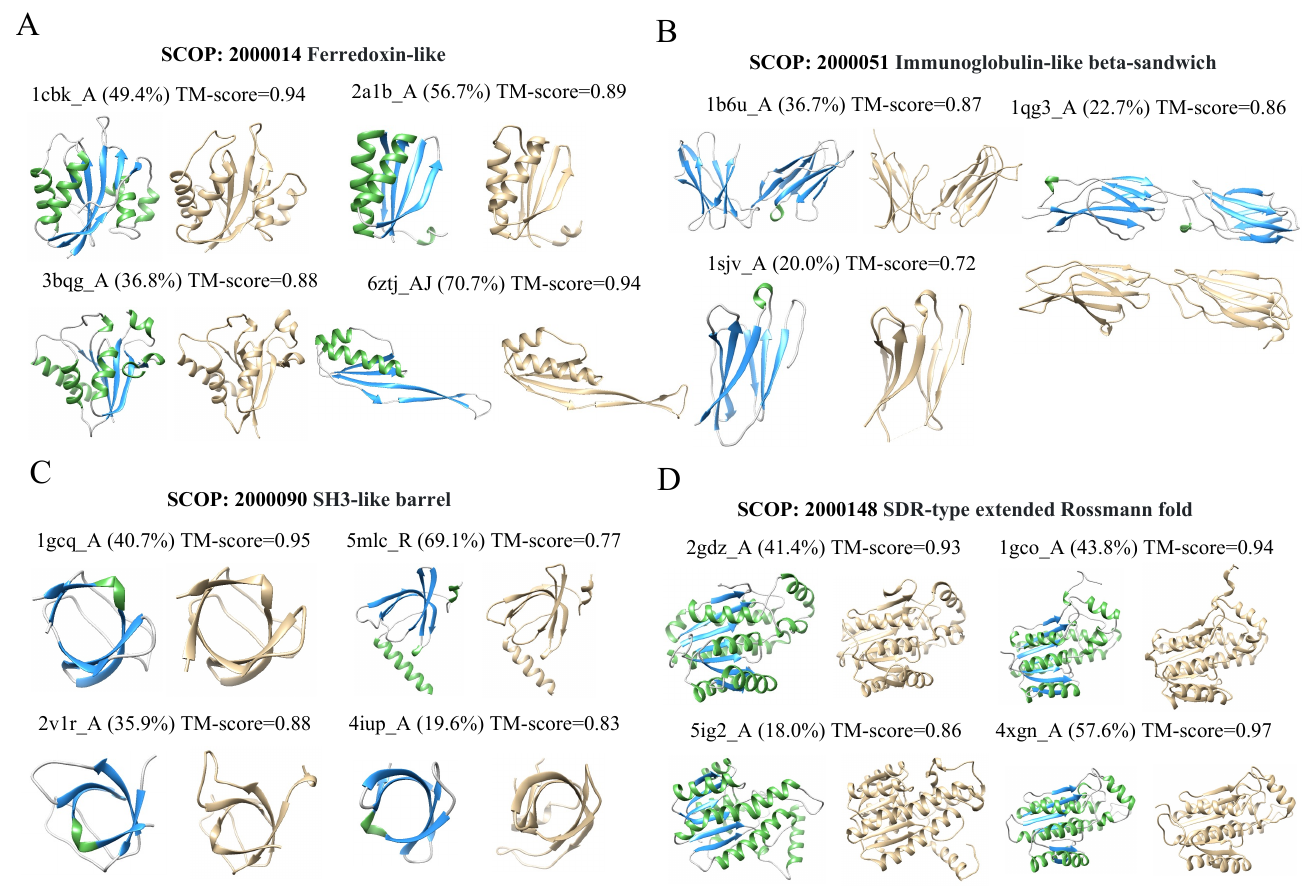} 
\caption[Cases Study of \smodel in Controllable Generation]{
\textbf{Cases Study of \smodel in Controllable Generation.}
Generation of four SCOP fold types by \smodel (SFT): Ferredoxin-like (\textbf{A}), Immunoglobulin-like beta-sandwich (\textbf{B}), SH3-like barrel (\textbf{C}), and SDR-type extended Rossmann fold (\textbf{D}). Generated protein structures are depicted in interleaved green, blue, and gray colors, whereas gold-colored structures represent the most structurally similar proteins from the PDB database. Percentages in parentheses indicate sequence identity.
 }
\label{fig:show_sft_cases}
\end{figure}

To further showcase that \smodel can generate proteins that mimic the natural sequences with the SFT alignment pipeline. We fine-tune the model on four SCOP fold-type sequences. Quantitative analyses exploring the relationship between structural prediction confidence, as indicated by xT-Fold's pLDDT scores, sequence identities, and their structural counterparts in the Protein Data Bank (PDB), as measured by TM-score (Figure~\ref{fig:show_sft_cases}). 
These findings underscore \model's exceptional capability in generating protein sequences that not only embody specific structural characteristics but also align closely with established PDB entries, thereby reinforcing its potential as a tool for synthesizing proteins with targeted structural attributes.

\label{sec:exp}
\section*{Discussion}
\label{limitations}
Although xTrimoPGLM demonstrated impressive performance, it has limitations. 
A significant limitation of xTrimoPGLM-100B is the high computational cost associated with the model, which presents a considerable barrier to their practical application.  Fine-tuning without quantization requires at least four A100 80G GPUs, which may not be feasible for all users. To mitigate this, more advanced efficient compression technologies in terms of parameters or memory, such as quantization~\cite{dettmers2024qlora}, kernel fusion~\cite{dao2022flashattention} and other accelerate technologies~\cite{kwon2023efficient, ainslie2023gqa, chen2023accelerating, leviathan2023fast} could be applied. These methods might enable the training and deployment of larger models with reduced computational resources. In this area, there is a wealth of research exploring various methods. We have only validated a few of these methods so far. However, more research is needed to assess their effectiveness in real-world scenarios.


Moreover, we observed diminishing returns in performance with increasing model parameters. Specifically, while xT-Fold has shown impressive both TM-score and inference speed in PLM-based models, a notable challenge persists in its OOD test performance~(such as CASP proteins). 
This performance gap remains consistent, even as the pre-training model becomes more powerful.  Though the model is gradually improving in compressing OOD sequences, it already diminishes the return (Figure~\ref{fig:1_xtglm}D). We speculate that protein sequences, unlike natural language, are less smooth in semantic space and more akin to a factual nature.
Currently, PLM-based methods still struggle to outperform MSA-based or retrieval-augmented approaches, especially when dealing with significant out-of-distribution (OOD) test data. To bridge this gap, leveraging more abundant data sources—such as sequences, structures, and functional descriptions~\cite{hayes2024simulating}—and expanding data modalities to include proteins, DNA, and RNA~\cite{abramson2024accurate}, along with implementing compute-optimal pre-training strategies~\cite{cheng2024training}, could be crucial.
Therefore, we advocate for proportional scaling of both data and models and emphasize the importance of exploring their efficient frontier further.
Another option is that integrating MSA modules with neural network retrievers~\cite{lewis2020retrieval, chen2024msagpt, borgeaud2022improving}, could lead to better end-to-end optimization and faster inference speeds than traditional MSA methods. Such developments could enhance the model's ability to generalize on OOD data while accelerating its processing capabilities.

Similar to most large language models, \smodel also experiences the issue of generating hallucinations. During the generation process, when the sampling temperature is set low, the model tends to produce fragments with a high repetition of amino acids. Although certain types of repetitive fragments (such as repeated Alanine) might be predicted with high-confidence structures, these fragments do not exist in nature. On the other hand, out of the sequences we generated, approximately 20\% of the sequences could not be confidently predicted by xT-Fold (pLDDT < 70), leaving us uncertain whether these sequences can exist and fold stably. Even after SFT, only 17.8\% to 88\% of the generated sequences could find similar structures in the Protein Data Bank (PDB). To avoid or reduce the hallucination problem of LLMs, augmenting constraints~\cite{lewis2020retrieval} during the training and sampling process can improve the efficiency of the model generation.

To summary, we explored unified understanding and generation pre-training with an extremely large-scale protein language model. Our experiments suggest that such scaling can extend to downstream tasks, including the key 3D structure prediction. We have further unlocked new possibilities in protein sequence design through supervised fine-tuning, paving the way for groundbreaking advancements in this field. Our work serves as a stepping stone for future research in the protein foundation model, and we hope it can facilitate further progress in a broader spectrum of protein-related applications.

\section*{Acknowledgments}
We extend our gratitude to Zhipu.AI for providing the computing resources.
This work has been supported by the NSFC for Distinguished Young Scholar 62425601 and 62276148, New Cornerstone Science Foundation through the XPLORER PRIZE  and research found from Zhipu.

\section*{Author contributions}
B.C conceived the method, implemented and trained the model, investigated the scaling law, and drafted the manuscript. 
X.C. led the project, managed workflows, prepared pre-trained datasets, trained scaling law and xT-Fold, and refined the manuscript.
P.L and Z.B analyzed the generated protein sequences. 
Y.G, J.G, S.L, and B.W evaluated the model on the protein understanding benchmark.
X.T and X.Z helped to develop the xT-Fold.
A.Z and C.L helped implement the model training framework.
L.S and J.T served as the corresponding authors and played a crucial role in integrating all resources, with L.S serving as an intermittent advisor throughout.
All the authors read and approved the manuscript.

\section*{Competing interests}
The authors declare no competing interests.

%% file: method.tex
\section*{Methods}
\label{sec: method}
\subsection*{Backbone Framework: General Language Model (GLM)}
Current PLM frameworks are commonly categorized as either encoder-only, like ESM~\cite{rives2021biological, lin2023evolutionary}, or decoder-only, such as PROGEN~\cite{madani2023large, nijkamp2023progen2}, exhibit limitations in addressing both task categories effectively due to their inherent inductive biases. 
Conceptually, the two types of tasks mirror the broader spectrum of protein sequence distributions~\cite{verkuil2022language}, suggesting the need for a unified model capable of encapsulating this diversity. 
To achieve this, encoder-decoder architectures like T5~\cite{raffel2020exploring} and non-causal decoder-only models like the General Language Model (GLM)~\cite{du2022glm,zeng2022glm} are optimized through an auto-regressive generating and bidirectional input-processing objective, which emerge as promising candidates for this dual capability. However, the GLM, with its parameter efficiency, stands out as a more viable option compared to the T5, which requires significantly 2x larger parameters for similar efficacy.
GLM is a transformer-based language model characterized by its unique training methodology. It employs autoregressive blank infilling while processing input text bi-directionally. This approach involves randomly blanking out continuous spans of tokens from the input and training the model to sequentially reconstruct these spans. This dual focus on autoencoding and autoregressive pre-training differentiates GLM from causal decoder-only language models, which rely solely on unidirectional attention. 

\subsection*{Pre-Training Objectives}
\label{sec:pretrains}
GLM incorporates two distinct pre-training objectives to ensure its generative capabilities: 
1) \textit{Span prediction}. This objective focuses on recovering short blanks within sentences, with the blank lengths cumulatively forming a significant portion of the input.  
and 
2) \textit{Long-text generation}. Aimed at generating extended blanks at sentence ends, this objective works with variable-length blanks, utilizing prefix contexts for guidance.
Additionally, to enhance \model’s comprehension abilities, we have integrated the Masked Language Model (MLM) strategy~\cite{kenton2019bert}. 
This inclusion ensures that \smodel not only excels in accurate residue-level representation but also effectively captures sequence-level representations, providing a comprehensive understanding of protein sequences.

\vpara{Masked Language Models (MLM) for Understanding.}
\label{sec:mlm}
The MLM objective aims at in-place masked token predictions. Formally, for an input protein sequence $\mathbf{x} = [x_1, \cdots, x_n]$ and the positions of masks $M = \{m_1, \cdots, m_{|M|}\}$, then the MLM pre-training loss is defined as

\beq{\mathcal{L}_{\text{MLM}} = \mathbb{E}_{M}\left[\sum_{m\in M}-\log p(x_m|\mathbf{x}_{/M})\right],}

\noindent where $x_{/M}$ denotes all input tokens except the ones that are in $M$.

\vpara{General Language Models (GLM) for Generation.}
\label{sec:gpt}
The GLM objective aims at recovering the masked consecutive tokens, i.e., spans, in an autoregressive manner. Concretely, for an input sequence $\mathbf{x}$, sequence spans $\{\mathbf{s}_1, \cdots, \mathbf{s}_m\}$ are sampled from it. Each span $\mathbf{s}_i$, consisting of a consecutive section of tokens $[s_{i,1}, \cdots, s_{i, l_i} ]$ in $\mathbf{x}$, is replaced with a single mask token \texttt{[sMASK]} or \texttt{[gMASK]} to form $\mathbf{x}_{\text{corrupt}}$. To make sure the interactions among corrupted spans, \smodel randomly permutes the order of spans like GLM, and defines the pre-training objective as
\beq{
\mathcal{L_{\text{GLM}}}=\mathbb{E}_{\mathbf{z}\sim Z_m}\left[\sum_{i=1}^{m}\sum_{j=1}^{l_i} -\log p\left(s_{z_i,j}|\mathbf{x}_{\text{corrupt}}, \mathbf{s}_{z_{<i}}, s_{z_i,<j}\right)\right],
}

\noindent where $Z_m$ denotes the set of the sequence’s all permutations and $\mathbf{s}_{z_{<i}}$ represents $\{\mathbf{s}_{z_1}, \cdots, \mathbf{s}_{z_{i-1}}\}$.

\vpara{Unified Pre-Training.}
\label{sec:uni}
The two types of pre-training objectives are jointly optimized to pre-train the \smodel model.  The unified pre-training objective, which aims to maximize the likelihood of the oracle tokens, is defined as:
\beq{\mathcal{L} = \mathcal{L}_{\text{MLM}} + \alpha \cdot \mathcal{L_{\text{GLM}}},}

\noindent where $\alpha$ is a weighting factor used to balance the different pre-training objectives. As a result, the proposed unified framework effectively takes advantage of the GLM architecture to characterize both the understanding ability via $\mathcal{L}_{\text{MLM}}$ and the generation capacity via $\mathcal{L_{\text{GLM}}}$.

\subsection*{The Pre-Training Strategy}
Motivated by the philosophy of curriculum learning, we begin the pre-training of \smodel-100B with a rather simpler MLM objective, then followed by the GLM objective, which unfolds in two methodically structured stages:

\vpara{Masked Language Model Stage~(400 Billion Pre-Trained Tokens).} 
In this initial stage, the \texttt{[MASK]} token is employed for masking random tokens in the sequence, with these masked tokens comprising 15\% of the total input. This stage, consuming about 400 billion tokens, is dedicated to advancing the model's representation abilities. At this stage, the $\alpha$ for scaling the GLM loss is set to 0, focusing solely on minimizing the MLM loss to enhance the model’s understanding capabilities.

\vpara{Unified Training Stage~(600 Billion Pre-Trained Tokens).} 
Subsequently, the model undergoes training with a combined approach of MLM and GLM objectives, in a ratio of 20\% MLM to 80\% GLM. In this stage, the model processes an additional 600 billion tokens, aiming to further refine both its representational and generative capabilities. Here, the $\alpha$ is set to 1, allowing both objectives to equally contribute to the overall loss function on each training instance, with the GLM objective appearing four times more frequently than the MLM objective.
\begin{itemize}
    \item \textbf{Masked Language Model Component.} Leveraging the \texttt{[MASK]} token, this component focuses on enhancing the model’s understanding of protein sequences. 
    \item  \textbf{General Language Model Component.} This component employs two types of masking: \texttt{[sMASK]} for consecutive span masking, following a Poisson distribution ($\lambda$ = 6), and \texttt{[gMASK]} for masking larger sequence segments based on a uniform distribution (minimally 40\% of the tokens masked). The \texttt{[sMASK]} token aids in blank infilling tasks, while \texttt{[gMASK]} facilitates the model in generating extended masked segments using the unmasked prefix.
\end{itemize}
This dual-stage training strategy meticulously integrates the MLM and GLM objectives, thereby optimizing \model-100B for a comprehensive understanding and generation of protein sequences.

\subsection*{Empirical Analysis of Unified Training}
\label{sec:ana}
This section presents an in-depth analysis of the feasibility of concurrently optimizing two distinct pre-training objectives in \model. 
Unlike prior unified pre-training frameworks~\cite{tay2023ul2,bao2020unilmv2}, which typically adopt similar formulations for diverse objectives, 
our approach extends these methodologies to a broader context. 
We critically examine whether models benefit from joint optimization of in-place token predictions (Masked Language Model) and next-token predictions (General Language Model).
Central to our investigation are two pivotal questions:
\textit{Objective Compatibility.} 
Does the in-place token prediction objective be optimized with the next-token prediction approach simultaneously? This inquiry is essential to understand whether these objectives can be effectively integrated within a single training framework;
\textit{Mutual Contribution.}
Can the in-place token prediction strategy enhance the performance of next-token prediction tasks, and does the reverse also hold true? This question addresses the potential synergistic effects of combining these two objectives in the training regime of \model.
Our exploration into these questions aims to shed light on the intricate dynamics of unified training models, particularly in the context of large-scale language models specialized for protein sequence analysis.

\vpara{Pre-training Settings.}
Our experiments utilize \model-150m, featuring 30 layers, 20 attention heads, 640 embedding dimensions, and FP16 precision. This configuration aligns with \model-100B's hyperparameters. Pre-training is conducted on the Uniref50 database~~\cite{suzek2015uniref}. 
We employ batches of 2,048 sequences, each 1,024 tokens in length. 
To operate within a fixed compute budget, we focus on the number of tokens observed during pre-training (corresponding to the total computational cost), rather than those actually trained (i.e., those on which a loss is calculated). 
These differences are considered intrinsic efficiency trade-offs between training objectives.

\begin{itemize}
    \item  \textbf{MLM.}  Roughly 15\% of input tokens were masked, resulting in approximately 1,024 input and 154 target tokens. Loss calculations are confined to target tokens.
    \item \textbf{GLM (\texttt{[gMASK]}).} Only the long-text generation objectives (signified by \texttt{[gMASK]}) are utilized, given the compatibility of the span corruption objective (\texttt{[sMASK]}) with the \texttt{[gMASK]} objectives has been verified. The loss computation pertains to the masked regions, encompassing a minimum of 40\% of tokens.
\end{itemize}

We evaluate the compatibility of MLM (in-place token prediction) and GLM ([gMASK], next-token prediction) objectives. Each occupies 50\% of the training batch time, alternating between them. Shifts in objectives occur at 100B and 200B token consumption milestones, facilitated by constant model parameters and architecture, requiring only adjustments in the attention mask. Validation losses indicate that despite their differing natures, both MLM and GLM objectives optimize simultaneously (Supplementary Figure~\ref{fig:ana_unify}(a)(b)).


Furthermore, we investigate the impact of pre-training objectives on convergence speed. Models pre-trained on one objective adapt to another, training over an additional 50B tokens. Comparisons include:
\textit{MLM-adapted GLM versus GLM trained from scratch} and \textit{GLM-adapted MLM versus MLM trained from scratch}.
Our results show significantly faster convergence in adapted models compared to those trained from scratch (Supplementary Figure~\ref{fig:ana_unify}(c)(d)). The MLM-adapted GLM matches the loss of the GLM from-scratch model with a 2.2$\times$ speedup (110B tokens). Similarly, the GLM-adapted MLM achieves a 2$\times$ speedup (100B tokens).

These findings suggest that modeling protein data distribution is not limited to specific training patterns. This bridges the gap between autoencoding PLMs (e.g., ESM~\cite{lin2023evolutionary}) and autoregressive PLMs (e.g., PROGEN2~\cite{nijkamp2023progen2}), underscoring the effectiveness of the \smodel training pipeline.

\subsection*{The Training Stability of Unified Training}
\label{sec:sta}
Training stability is a critical factor for the successful training of large language models (LLMs) at the 100B-scale~\cite{brown2020language, zeng2022glm, chowdhery2022palm}. Given a fixed computing budget, it is essential to balance efficiency and stability, particularly in relation to floating-point (FP) formats. Lower-precision FP formats, such as 16-bit precision (FP16), enhance computational efficiency but are vulnerable to overflow and underflow errors. These vulnerabilities can 
potentially lead to catastrophic collapses during training. \model, drawing on the implementation strategies of GLM-130B~\cite{zeng2022glm}, addresses many unstable training issues. Nonetheless, \model-100B still experiences catastrophic training collapses 
during the transition from the first to the second stage, a challenge not present in smaller-scale models (10B-scale). Incorporating a fixed ratio of GLM loss into pre-training can trigger these collapses, even with a minimal 1\% GLM loss ratio (Supplementary Figure~\ref{fig:st}). To mitigate this issue, we propose the implementation of a smooth transition strategy.

\vpara{Smooth Transition Strategy.}
Our empirical investigations suggest a two-phase smooth transition strategy to integrate GLM loss into training:

\ipara{Gradual Increase in GLM Loss Ratio.} We start by incrementally increasing the GLM loss ratio from 0, aiming to reach the target value $R$ in $K$ steps through linear growth. The GLM loss ratio $R_k$ at each step $k$ is calculated as $R_k = \frac{k \times R}{K}$. Notably, the learning rate remains exceptionally low during this phase. In practice, we set $K$ = 1000 and the learning rate to 1e-7.

\ipara{Normalization of the Learning Rate.} After completing the transition, the learning rate gradually returns to its standard pre-training level as defined in the pre-training script.
The final \model-100B training run demonstrates that loss divergence occurs only at the transition stage, though it initially faces numerous failures due to hardware issues.


\subsection*{Pre-Training Configurations}
\label{sec::pretraining}
Here we introduce the implementation details of pre-training the \model-100B model. 
Since the \model-100B borrows the idea from the GLM-130B~\cite{zeng2022glm} framework, we only emphasize the specific hyper-parameter of \model-100B. For more discussion and design choices please refer to GLM-130B~\cite{zeng2022glm}.

\model-100B is trained on a cluster of 96 DGX-A100 GPU (8$\times$40G) servers in FP16 precision from January 18 to June 30, 2023. During this time, \model-100B has consumed 1 trillion tokens from the dataset consisting of Uniref90 and ColabFoldDB. 
We adopt 3D parallel strategy with the 4-way tensor parallelism~\cite{shoeybi2019megatron},  8-way pipeline parallelism~\cite{narayanan2021memory}, and 24-way data parallelism based on DeepSpeed~\cite{rasley2020deepspeed}.
The model owns 72 transformer layers, 80 attention heads, and 10,240 embedding dims with 31,744 feed-forward embedding dims using GeGLU~\cite{shazeer2020glu}. 
We adopt the Post-LN initialized with the DeepNorm~\cite{wang2024deepnet}. 
We follow the mixed-precision strategy (Apex O2), i.e., FP16 for forwards and backward and FP32 for optimizer states and master weights, to reduce the GPU memory usage and improve training efficiency. 
We also adopt the Embedding Layer Gradient Shrink (EGS) strategy~\cite{zeng2022glm} with $\alpha=0.1$ to stabilize the \model-100B training. 
We warm up the batch size from 240 to 4224 over the first 2.5\% samples. 
We use AdamW~\cite{loshchilov2018decoupled} as our optimizer with $\beta_1$ and $\beta_2$ set to 0.9 and 0.95, and a weight decay value of 0.1. We warm up the learning rate from $10^{-7}$ to $4\times 10^{-5}$ over the first 3.0\% samples, then decay it by a 10 $\times$ cosine schedule to the minimum learning $4 \times 10^{-6}$. We use a dropout rate of 0.1 and clip gradients using a clipping value of 1.0.
Each sample contains a fixed sequence length of 2,048 (We concatenate all protein sequences with a separator into a single
document, and sample protein sequences from this document in such a way that there is virtually
no padding during pre-training.). To adapt the model to the different lengths of proteins in the downstream tasks, we adopt the mix-length pre-training strategy with four different context windows of 256, 512, 1,024, and 2,048. Taking, 512, for example, we concatenate four samples together to cater for the 2,048-sequence-length. The ratio of different context lengths is  $[\#256:\#512:\#1,024:\#2,048 = 0.1:0.4:0.4:0.1]$.
We implement the two-dimensional RoPE from its author blog~\href{https://kexue.fm/archives/8397}{https://kexue.fm/archives/8397} as our position embedding.
For the tokenization of the protein data, we use the residue-level tokenizer which is adopted in several PLMs~\cite{lin2023evolutionary, elnaggarrost}. 
Except for the basic amino acid types, we add special tokens \texttt{[MASK]}, \texttt{[sMASK]}, and \texttt{[gMASK]} for model prediction. 
We also add special tokens \texttt{<sop>}, \texttt{<eop>}, \texttt{<eos>} for sequence separation.
(Cf. Table~S\ref{tb:configs} for the full configurations).

\section*{Pre-Training Datasets}
\label{sec::datasets}
The pre-training dataset of \model-100B is curated from two extensive data repositories: Uniref90~(\url{https://www.uniprot.org/help/downloads}), the Uniref90 version preceding December 2022 is downloaded and ColabFoldDB~\cite{mirdita2022colAbFold}(~\url{https://colabfold.mmseqs.com}). The initial contributions from Uniref90 and ColabFoldDB encompass approximately 153M and 950M~(210M representatives plus 740M members) entries, respectively.

Uniref, a cluster from UniProt, is broadly acknowledged as a high-quality protein dataset often utilized in pre-training PLMs~\cite{lin2023evolutionary, elnaggarrost}. UniRef90 clusters are generated from the UniRef100 seed sequences with a 90\% sequence identity threshold using the MMseqs2(~\url{https://github.com/soedinglab/MMseqs2} algorithm). 
ColabFoldDB is established through an amalgamation of various metagenomic databases including BFD~(\url{https://bfd.mmseqs.com}), MGnify~\cite{mitchell2020mgnify}, SMAG~(eukaryotes)~\cite{delmont2022functional}, MetaEuk~(eukaryotes)~\cite{levy2020metaeuk}, TOPAZ~(eukaryotes)~\cite{alexander2023eukaryotic}, MGV~(DNA viruses)~\cite{nayfach2021metagenomic}, GPD~(bacteriophages)~\cite{camarillo2021massive}, and an updated version of the MetaClust~\cite{steinegger2018clustering} dataset. 
Built upon the foundation of UniProtKB, 
ColabFoldDB is substantially augmented with a large corpus of metagenomic sequences derived from diverse environmental niches.
Metagenomic data introduces a new level of diversity to the database, encompassing numerous environmental niches ranging from the human gut to marine ecosystems. This offers unparalleled opportunities for the discovery of novel proteins.
To comprehensively map the entirety of protein sources in the biological world, the pre-training dataset has been expanded by incorporating protein sequences sourced from ColabFoldDB in addition to those from the Uniref90 dataset.

The left panel (Supplementary Figure~\ref{fig:pretraining_dataset}) illustrates the composition of the dataset used for pre-training the model. The right panel depicts the distribution of taxonomic categories of Uniref90, visualized as concentric circles representing the levels of superkingdom, kingdom, and phylum from innermost to outermost. The innermost circle represents four superkingdoms: Bacteria (67\%), Archaea (3\%), Eukarya (27\%), and Viruses (1\%), with 2\% of sequences labeled as unclassified. The middle circle encompasses 17 classified kingdoms, including an unclassified bacteria category, denoted as ``bacteria*''. The outermost circle denotes the phylum level, marking only those labels with counts over 200,000. In total, Uniref90 includes 273 known phyla.
This comprehensive representation across multiple taxonomic levels demonstrates the rich biodiversity encapsulated within the Uniref90 dataset and affirms its value for wide-ranging biological investigations. 
Protein sequences that are published prior to January 1, 2023, are incorporated into the training set. 
Given its robustness and reliability, our training process also substantially prioritizes this dataset.

\vpara{Training Set.} 
The complete dataset in ColabFoldDB initially contained approximately ~950M sequences. After initial deduplication and short-length filtering, which removed about 150M duplicate sequences, and further refinement by cross-referencing and deduplicating with Uniref90, we narrowed down the dataset to 780M unique sequences, ensuring diversity and representativeness for effective training.
We conduct a composition analysis of each remaining sequence, excluding any that exhibit an individual amino acid composition exceeding 80\% as this may indicate an anomaly or bias in the data. These steps leave us a more representative subset of around ~200M sequences. 
Finally, the pre-trained dataset comprises approximately 939M protein sequences with 200B tokens. Specifically, the UniRef90 dataset contains around 156M protein sequences with 53B residue tokens. The ColabFoldDB cluster includes about 208M protein sequences with 38 tokens, and the ColabFoldDB member contains 575M sequences with 103B tokens. During training, to capitalize on the high-quality data, we assign a greater weight to the Uniref90 data, resulting in a ColabFoldDB sampling ratio of approximately 60\%. This approach triples or quadruples the contribution of Uniref90 data, boosting our model's fine-tuning capability with high-quality data.  


\vpara{Validation Set.} Sequences from UniProt released between January 1 and March 30, 2023, are utilized as the validation datasets. The 18M sequence increment is applied as a query to scrutinize the target database~(i.e., Uniref50 and the training dataset), and sequences over 90\% or 0.5\% similarity are eliminated from the query set~( \texttt{mmseqs easy-search --db-load-mode 2 --min-seq-id 0.9 --alignment-mode 3 --max-seqs 300 -s 4 -c 0.8 }). The remaining after filtering is used as the validation set.

\vpara{Pre-training Data Distribution.}\label{app:data_dist}
The bar charts (Supplementary Figure~\ref{fig:tasks_performance}) represent the distribution of sequence lengths within the Uniref90 and ColabFoldDB datasets. In both datasets, sequences in the '100-400' length category predominate, followed by the '50-100' category. The '0-50' and '400+' categories contain significantly fewer sequences. Note the comparison between the distribution of Uniref90 and ColabFoldDB, indicating the variety of sequence lengths used for model training.

\section*{Data availability}
All data used in this study are publicly available and the usages are illustrated in our methods. 
The pre-training dataset of xTrimoPGLM-100B is curated from two extensive data repositories: Uniref90~(\url{https://www.uniprot.org/help/downloads}), the Uniref90 version preceding December 2022 is downloaded and ColabFoldDB~(\url{https://colabfold.mmseqs.com}).
18 downstream task datasets are all available online (\url{https://huggingface.co/proteinglm}).
All structure prediction datasets are from AlphaFold Database~(\url{https://alphafold.ebi.ac.uk/download}
PDB database~(\url{https://www.rcsb.org/downloads}) that released date is less than May 2020).

\section*{Code availability}
Trained weight for the \smodel model, and downstream datasets are available at \url{https://huggingface.co/proteinglm}.

Model training used DeepSpeed v0.6.1~\url{https://github.com/microsoft/DeepSpeed}. Data analysis used Python v3.8 \url{https://www.python.org/}, NumPy v1.16.4 \url{https://github.com/numpy/numpy}, SciPy v1.2.1 \url{https:// www.scipy.org/}, seaborn v0.11.1 \url{https://github.com/mwaskom/seaborn}, Matplotlib v3.3.4 \url{https://github.com/matplotlib/matplotlib}, pandas v1.1.5 \url{https://github.com/pandas-dev/pandas}, TM-align v20190822 \url{https://zhanglab.dcmb.med.umich.edu/TM-align/} was used for computing TM-scores. Structure visualizations were created in Pymol v2.3.0 \url{https://github.com/schrodinger/pymol-open-source}. 
Protein 3D structures were predicted using AlphaFold2 with the official implementations \url{https://github. com/deepmind/alphafold}.

%% file: extend_fig.tex
\begin{figure*}
    \includegraphics[width=1.0\linewidth]{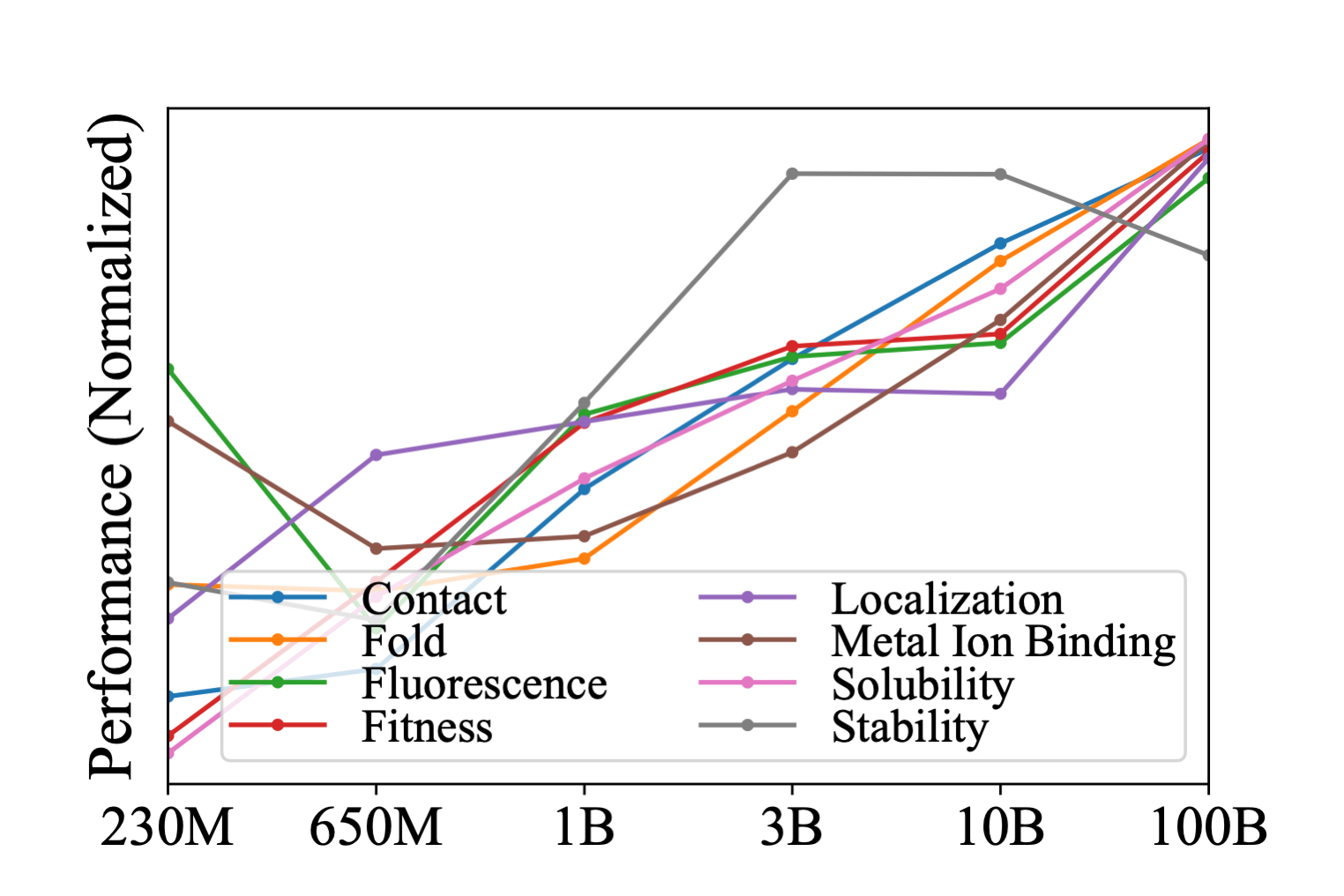}
    \caption[Scaling Analysis on \smodel Family Models]{
    Comparison of \smodel family models with varying sizes and training FLOPs, including 230M (FLOPs: $3 \times 10^{20}$), 650M (FLOPs: $3.8 \times 10^{21}$), 1B (FLOPs: $1.2 \times 10^{21}$), 3B (FLOPs: $1.8 \times 10^{22}$), 10B (FLOPs: $1.8 \times 10^{22}$), and 100B (FLOPs: $6.2 \times 10^{23}$). Models are evaluated across eight downstream tasks using the Linear Probing fine-tuning approach.
    Results indicate that most tasks exhibit a positive correlation between performance and both training FLOPs and model size. Unnormalized performance metrics are provided in Supplementary Table~\ref{tab:unnormalized}.
    }
    \label{fig:glm_scales}
\end{figure*}

\begin{figure*}[!ht]
\centering
\includegraphics[width=1.0\linewidth]{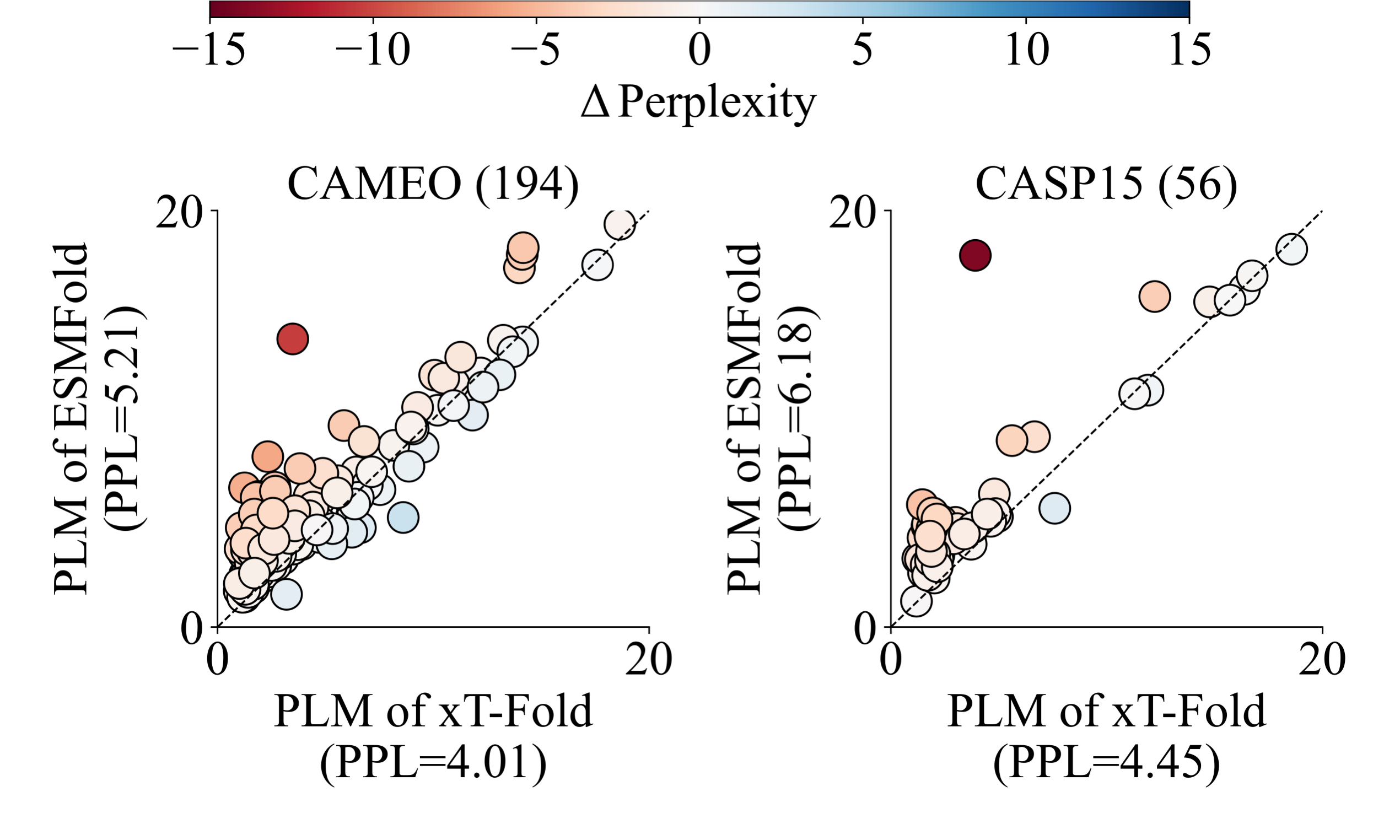} 
\caption[Perplexity Delta Comparison on CAMEO and CASP15]{Scatter plots show delta perplexity for CAMEO and CASP15. The perplexities~(PPL) are from the PLM modules of xT-Fold and ESMFold. 
Points represent proteins, with color gradients indicating perplexity delta by the x-axis PPL minus the y-axis PPL.
}
\label{fig:ppl_comp}
\end{figure*}

\begin{figure*}[!t]
\centering
\includegraphics[width=1.0\linewidth]{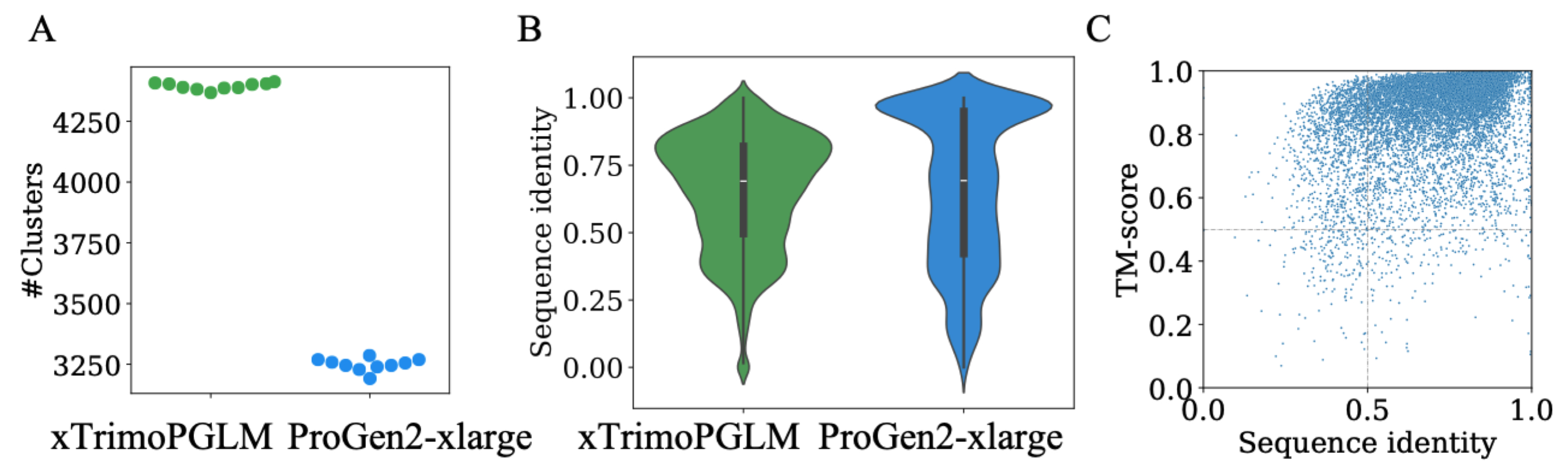} 
\caption[Sequence Identity to Natural Sequence Space and Diversity Analysis]{Sequence identity to natural sequence space and diversity analysis. (A: 5,000 sequences are randomly selected from two sequence set and are clustered using MMseqs2 (--min-seq-id 0.5 -c 0.8). The Y-axis show the number of clusters of 10 repetitions. B: Maximum identity of generated sequences (N=14,626 and 8,466) to UniClust30, UniProt, and BFD databases. C: Scatter plot of maximum sequence identity to natural sequence space and maximum structure similarity to AlphaFold database (UniProt50 subset). 
The bars in the violin plot indicate the median and interquartile range (IQR) for each group with whiskers extending 1.5$\times$ IQR past the upper and lower quartiles.
}
\label{fig:gen_cluster}
\end{figure*}

\begin{figure*}[!t]
\centering
\includegraphics[width=1.0\linewidth]{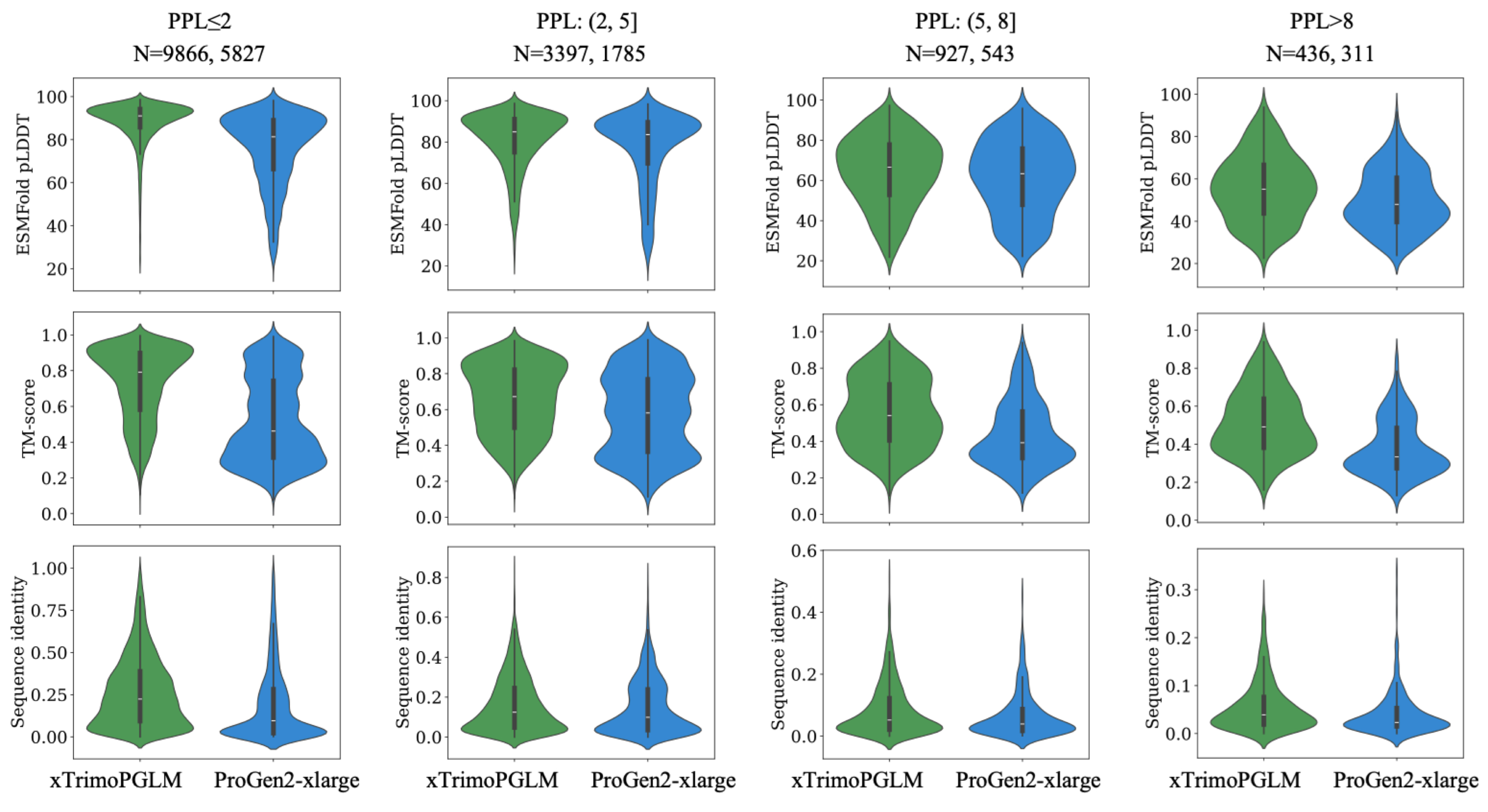} 
\caption[Comparison of sequences generated by xTrimoPGLM and PROGEN2-xlarge]{Comparison of sequences generated by xTrimoPGLM and PROGEN2-xlarge at different PPL ranges. ESMFold predicted confidence (pLDDT scores), the resemblance to proteins cataloged in the Protein Data Bank (TM-score and sequence identity) of generated sequences by xTrimoPGLM (green) and PROGEN2-xlarge (blue). The sequences are divided into four groups according to the perplexity values (<2, 2-5, 5-8 and >8). The number of sequences of both models in each group are shown in the upper of the figure. The bars in the violin plot indicate the median and interquartile range (IQR) for each group with whiskers extending 1.5$\times$ IQR past the upper and lower quartiles.}
\label{fig:generated_vs_ppl}
\end{figure*}

\begin{figure*}[!t]
\centering
\includegraphics[width=1.0\linewidth]{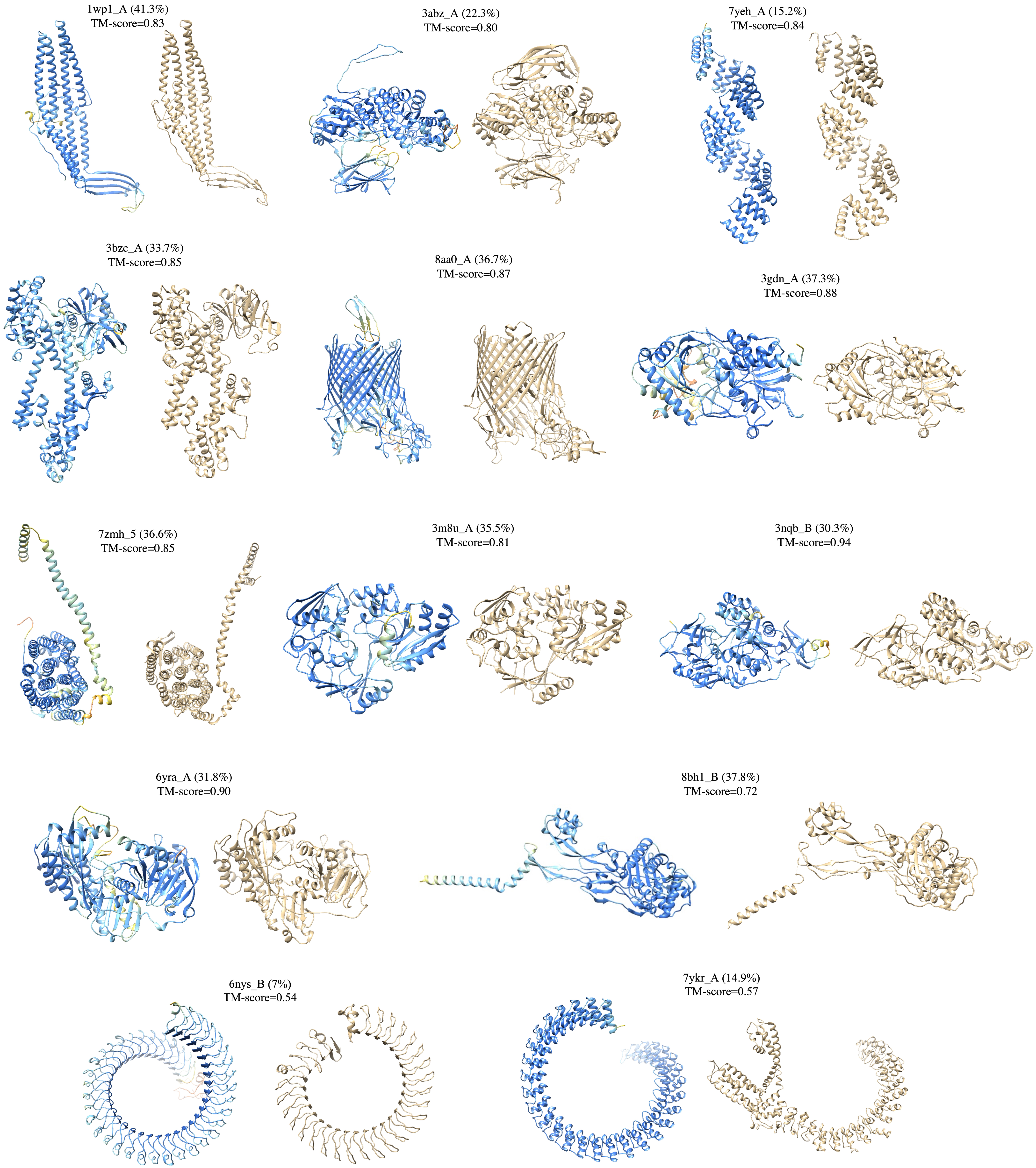} 
\caption[Generated Protein Sequences Cases]{More protein sequences visualizations generated by \model-100B. (Left: Generated by \model-100B. Right: Natural Proteins.)}
\label{fig:app_gen}
\end{figure*}

\begin{figure*}
    \includegraphics[width=1.0\linewidth]{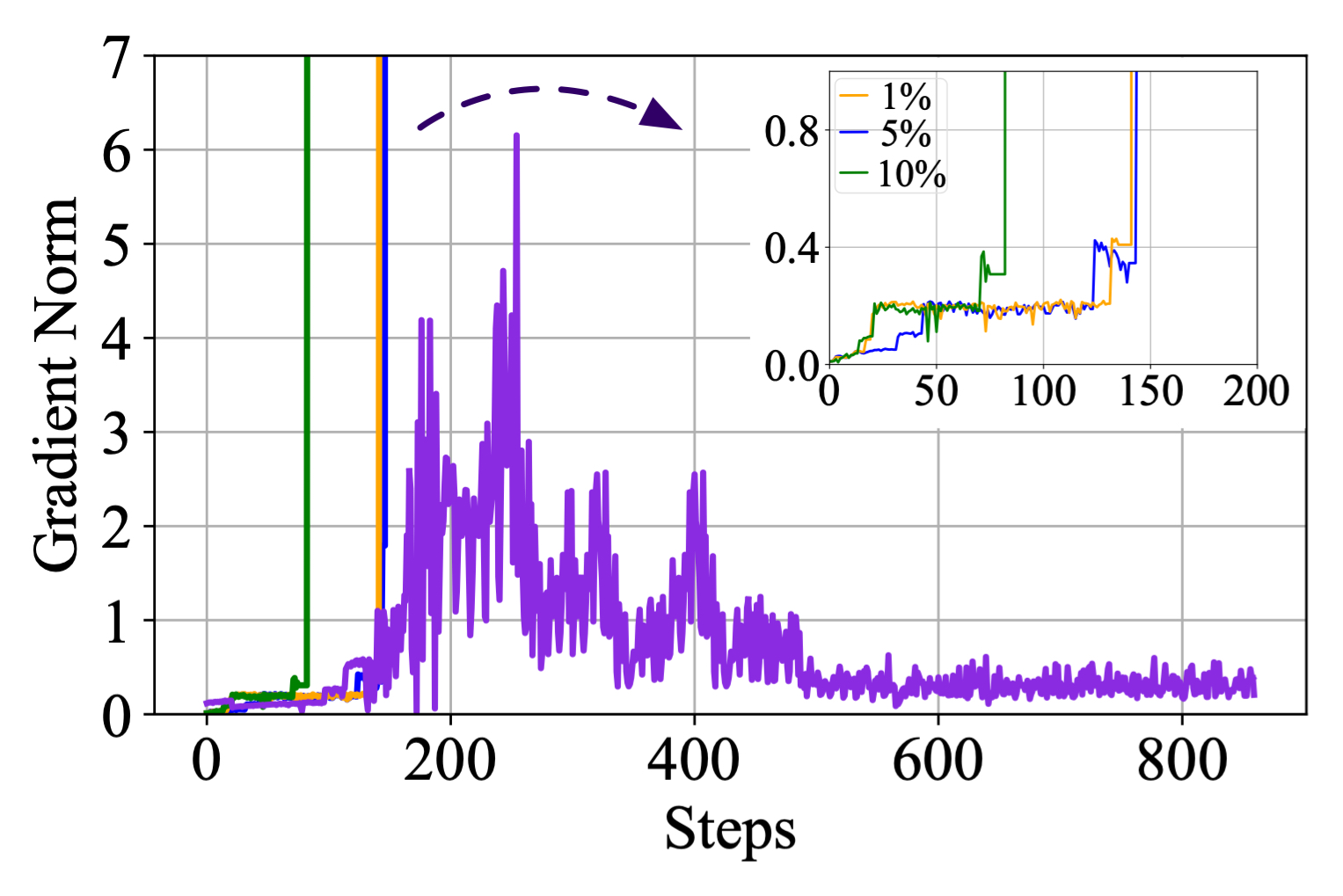}
    \caption{\label{fig:st} Trials on Different Strategies for Transition from Stage-1 to Stage-2. 
    }
\end{figure*}

\begin{figure*}
    \includegraphics[width=1.0\linewidth]{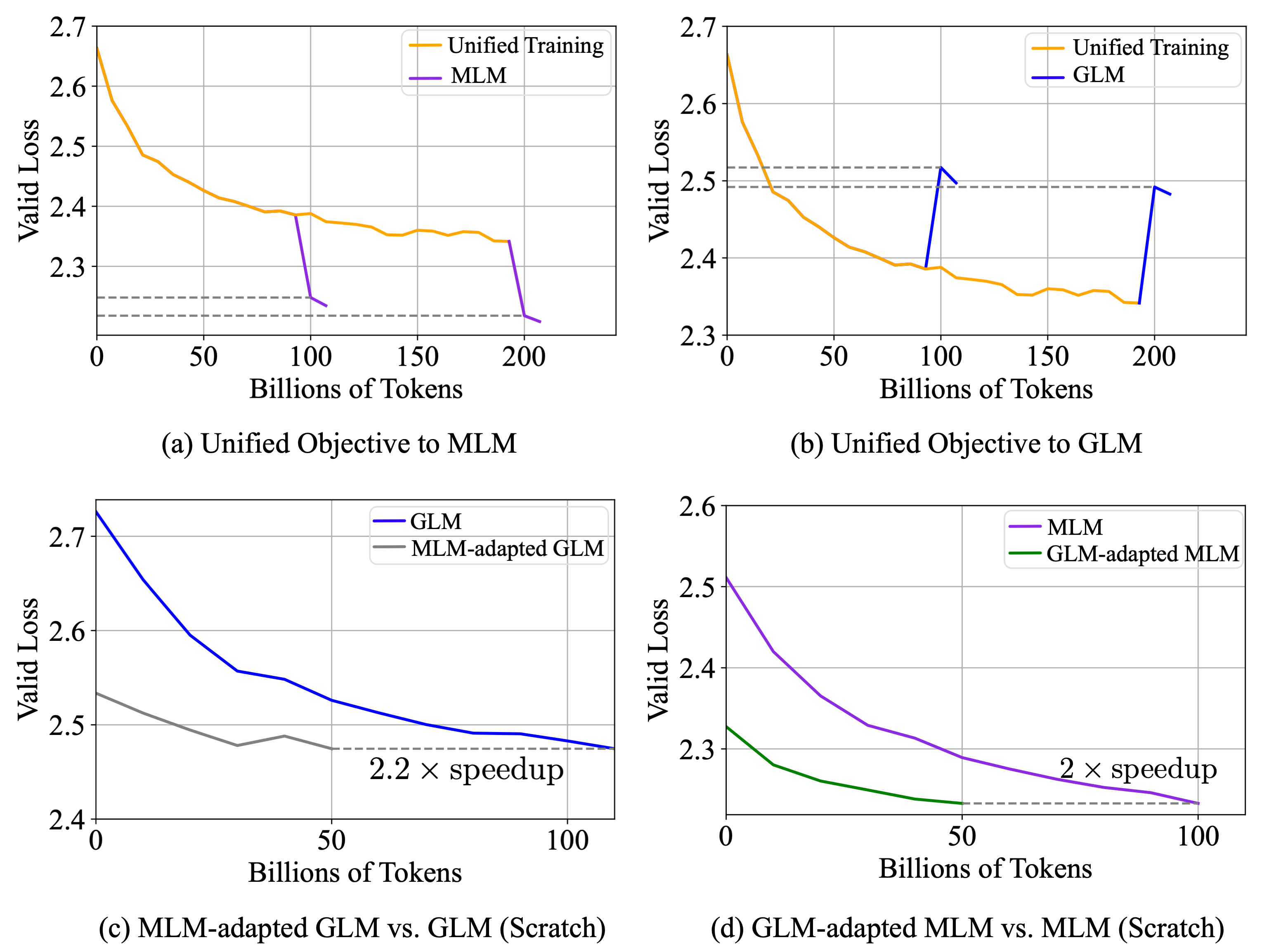}
    \caption[The Empirical Analysis of Unified Training.]{\label{fig:ana_unify} The empirical analysis of unified training. 
 (a)(b) The MLM and GLM objectives are optimized simultaneously. (c)(d) Adapting the model from the pre-trained one significantly accelerates convergence compared to that trained from scratch.
    }
\end{figure*}


\begin{figure*}[htp]
\centering
\includegraphics[width=1.0\linewidth]{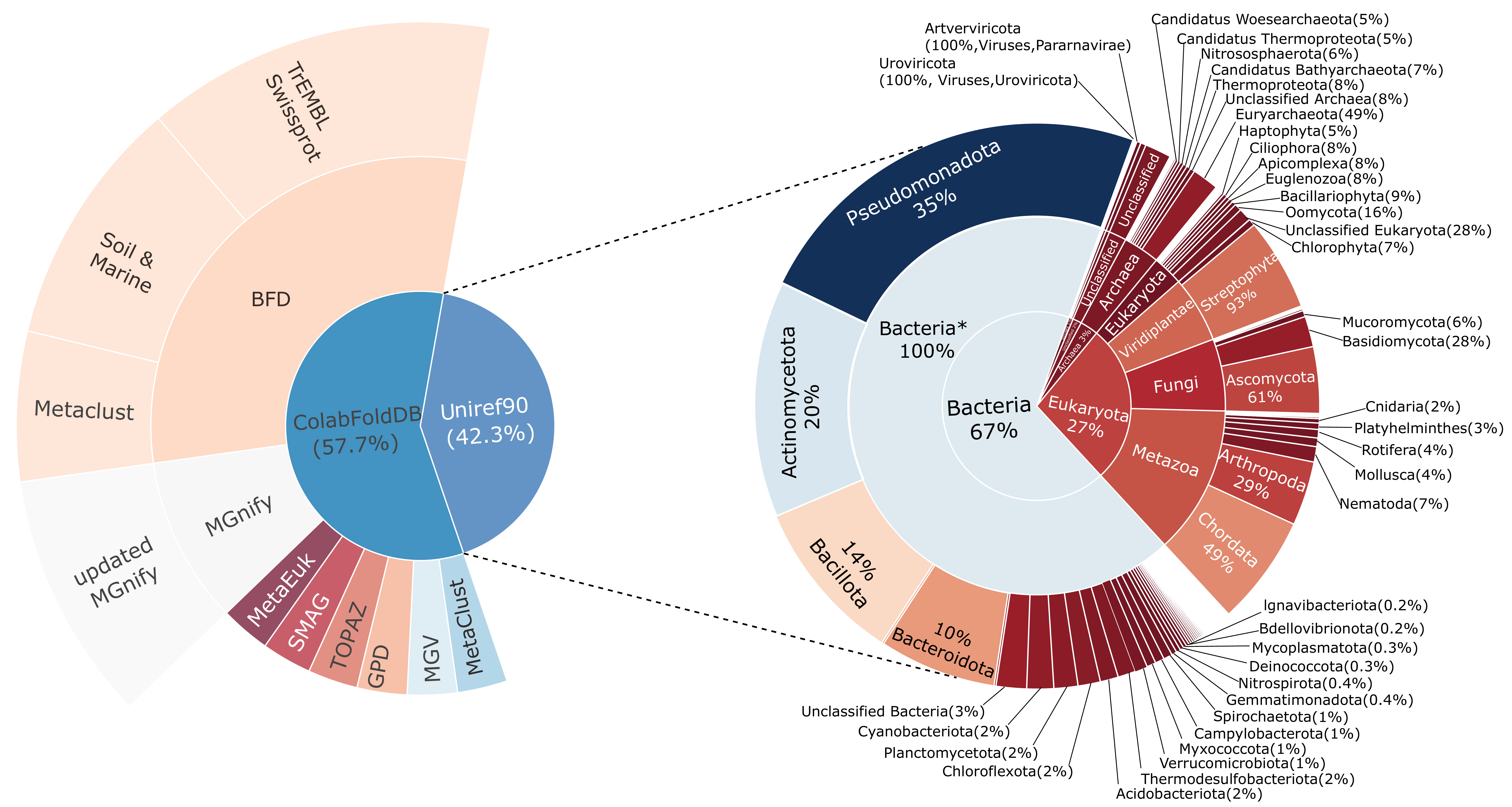} 
\caption[The Pre-training Dataset]{
The left panel illustrates the dataset composition used for pre-training the model. 
The right panel depicts the distribution of taxonomic categories of Uniref90, visualized as concentric circles representing the levels of superkingdom, kingdom, and phylum from innermost to outermost. The innermost circle represents four superkingdoms: Bacteria (67\%), Archaea (3\%), Eukarya (27\%), and Viruses (1\%), with 2\% of sequences labeled as unclassified. The middle circle encompasses 17 classified kingdoms, including an unclassified bacteria category, denoted as ``bacteria*''. The outermost circle denotes the phylum level, marking only those labels with counts over 200,000. In total, Uniref90 includes 273 known phyla.
}
\label{fig:pretraining_dataset}
\end{figure*}

\begin{figure*}[htp]
\centering
\includegraphics[width=1.0\linewidth]{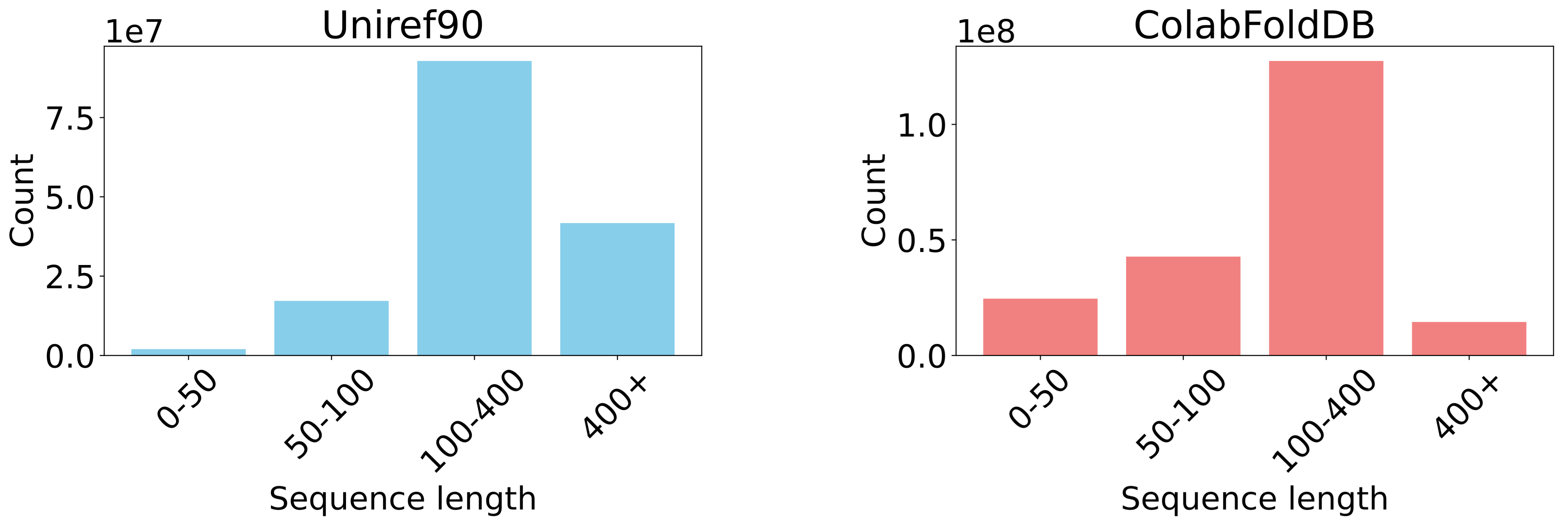} 
\caption{Pre-training Data Distribution}
\label{fig:tasks_performance}
\end{figure*}

%% file: app.tex


\section*{Supplementary materials}
\tableofcontents
\listoffigures
\listoftables

\section{Protein Understanding Benchmarks and Evaluations}
\label{sec::downstream_tasks}
To systematically evaluate \model-100B, we have benchmarked 18 downstream protein-related tasks across multiple domains (Supplementary Table~\ref{tab::tasks}), which are divided into four main categories: protein structure, protein developability, protein interactions, and protein functions. ``Struc.'' represents protein structure, ``Dev.'' represents protein developability, ``Inter.'' represents protein interactions, ``Func.'' represents protein functions, and ``Perf.'' denotes performance (\%). The $\clubsuit$ denotes the results that we produce or reproduce, while the $\blacklozenge$ represents direct citations from original papers with the same split train/valid/test sets. For any dataset without established benchmarks, we employ the results of our own ESM2-15B with LoRA fine-tuning. 
The table elucidates these tasks along with the latest SOTA methodologies employed, their respective performances, and the achievements attained by our proposed \model-100B model. 
We emphasize that this comparison is primarily from a task-based perspective, where \smodel is combined with fine-tuning techniques to achieve the results. The results reveal that \model-100B significantly outperforms current SOTA approaches in most protein-related tasks, hence catalyzing advancements
in this field. 
Next, we individually delve into these subtasks, elaborating on the corresponding task definitions, dataset processing, evaluation metrics, and other relevant details.


\vpara{Contact Map.} 
Contact map prediction~(Cont. P.) aims to determine whether two residues, \textit{i} and \textit{j}, are in contact or not, based on their distance with a certain threshold (<8Å). This task is an important part of the early Alphafold version~\cite{senior2020improved} for structural prediction. The trRosetta dataset~\cite{yang2020improved} is employed and the same split (12,041 training samples, 1,505 validation samples, and 1,505 test samples) as Ankh~\cite{elnaggar2023ankh} is used for this task. The evaluation metric used is the Top L/5 accuracy, considering residue pairs with a separation length greater than 6 and a sequence length cutoff of 512. 



\vpara{Fold Classification.} 
Fold class prediction~(Fold. P.) is a scientific classification task that assigns protein sequences to one of 1,195 known folds. The primary application of this task lies in the identification of novel remote homologs among proteins of interest, such as emerging antibiotic-resistant genes and industrial enzymes~\cite{chen2016profold}. The study of protein fold holds great significance in fields like proteomics and structural biology, as it facilitates the analysis of folding patterns, leading to the discovery of remote homologies and advancements in disease research~\cite{chen2018comprehensive}. The dataset employed for this task is based on SCOP 1.75~\cite{lo2000scop}, a release from 2009, and has been widely adopted by DeepSF~\cite{hou2018deepsf} and Ankh~\cite{elnaggar2023ankh}. We used the same split (12,312 training samples, 736 validation samples, and 3,244 test samples) as Ankh~\cite{elnaggar2023ankh} for evaluation. Total accuracy is employed as the evaluation metric.

\vpara{Secondary Structure.}
The study of a protein's secondary structure~(Sec. Struc. P.) forms a fundamental cornerstone in understanding its biological function. This secondary structure, comprising helices, strands, and various turns, bestows the protein with a specific three-dimensional configuration, which is critical for the formation of its tertiary structure. In the context of this work, a given protein sequence is classified into three distinct categories, each representing a different structural element: Helix (H), Strand (E), and Coil (C). The datasets applied in this study are originally published by NetSurfP-2.0~\cite{klausen2019netsurfp} and have also been utilized by Ankh~\cite{elnaggar2023ankh}. The training set is the same as Ankh~\cite{elnaggar2023ankh}. The datasets employed for testing in our investigation are specifically assembled from the Critical Assessment of Protein Structure Prediction (CASP) editions 12 and 14, which contain 18 and 21 samples. The result we reported is an average of these two datasets. Total accuracy is employed as the evaluation metric.


\vpara{Solubility.} 
This task (Sol. P.) involves a binary classification of a heterogenous set of proteins, assessing them as either soluble or insoluble. The solubility metric is a crucial design parameter in ensuring protein efficacy, with particular relevance in the pharmaceutical domain. We've adhered to the same dataset division as is employed in the development of DeepSol~\cite{khurana2018deepsol} (62,478 training samples and 6,942 test samples). Within this framework, any protein exhibiting a sequence identity of 30\% or greater to any protein within the test subset is eliminated from both the training and evaluation subsets, ensuring robust and unbiased evaluation. Total accuracy is employed as the evaluation metric.

\vpara{Stability.}
The task~(Stab. P.) is to predict the concentration of protease at which a protein can retain its folded state. Protease, being integral to numerous biological processes, bears significant relevance and a profound comprehension of protein stability during protease interaction can offer immense value, especially in the creation of novel therapeutics. The dataset applied in this task is initially sourced from Rocklin et al~\cite{rocklin2017} and subsequently collected within the Task Assessing Protein Embeddings (TAPE)~\cite{tape2019}. We used the same split (53,614 training samples, 2,512 validation samples and 12,851 test samples) as TAPE~\cite{tape2019} for evaluation. In this regression-based task, we employ the SpeaRman Correlation Coefficient~(SRCC) as the evaluation metric to measure the prediction consistency.


\vpara{Temperature Stability.}
The accurate prediction of protein thermal stability~(Temp. Stab.) has far-reaching implications in both academic and industrial spheres. This task primarily aims to predict a protein's capacity to preserve its structural stability under a temperature condition of 65 degrees Celsius. We employed the same database and dataset division strategy used in the development of TemStaPro~\cite{pudvziuvelyte2024temstapro}. The performance of our prediction is evaluated and reported using the Matthews Correlation Coefficient (MCC) score.

\vpara{Optimal Temperature.}
Grasping the catalytic activity of enzymes is pivotal for industrial enzyme design, particularly in predicting the optimal temperature~(Opt. Temp.) for a given enzyme's catalytic effect. The dataset utilized for this task is primarily procured by DeepET~\cite{Li2022DeepET}, a recent advancement in the field that uses deep learning techniques to understand enzyme thermal adaptation. We used the same split (1,706 training samples and 190 test samples) as DeepET~\cite{Li2022DeepET} for evaluation. To quantify the relationship between these variables, we use the SRCC.

\vpara{Optimal PH.}
Enzyme functions normally under a specific range of pH of the surrounding environment. However, an optimal pH for the reaction will largely boost the catalytic ability. We collected a dataset from EpHod~\cite{EpHod2023}, which established a deep learning method to predict the optimal enzyme catalytic pH based on protein sequence only. We used the same split (7,124 training samples, 760 validation samples and 1,971 test samples) as EpHod~\cite{EpHod2023} for evaluation. In the regression-based task, we use the SRCC as the evaluation metric.

\vpara{Clone CLF.}
Protein structure determination includes a series of experimental stages to yield stable proteins for X-ray crystallography. Specifically, the proteins are first selected and expressed, then purified for crystal structure determination. Each step corresponds to a "stage tag" to denote whether the protein is stable under a certain stage. We collected a dataset from PredPPCrys~\cite{PredPPCrys2014}, which manually annotated thousands of proteins with different experimental procedures. We used the same split (23,375 training samples and 4,791 test samples) as PredPPCrys~\cite{PredPPCrys2014} for evaluation. The binary classification task is to predict whether a protein sequence tends to be a cloning failure (Clone CLF.). We use the AUC as the evaluation metric.

\vpara{Material Production.}
The task is to predict whether a protein sequence fails at the protein material stage (MF). The dataset is also collected from PredPPCrys~\cite{PredPPCrys2014} (23,339 training samples and 4,791 test samples) and the AUC metric is employed as the measurement.



\vpara{Metal Ion Binding.}
Metal ion binding~(Metal B.) sites within proteins play a crucial role across a spectrum of processes, spanning from physiological to pathological, toxicological, pharmaceutical, and diagnostic. Consequently, the development of precise and efficient methods to identify and characterize these metal ion binding sites in proteins has become an imperative and intricate task for bioinformatics and structural biology. This task involves a binary classification challenge aimed at predicting the existence of metal-ion binding site(s) on a given protein sequence. We employ data~\cite{cheng2023co} curated from the Protein Data Bank (PDB). We used the same split (6,000 training samples and 1,332 test samples) for evaluation. Total accuracy is employed as the evaluation metric.

\vpara{Enzyme Catalytic Efficiency.} 
This task~(Enzyme Eff.) is focused on predicting kcat values, which are enzymatic turnover numbers denoting the maximum chemical conversion rate of a reaction, for metabolic enzymes originating from any organism. These predictions are based on substrate structures and protein sequences. The underlying importance of this task lies in its potential to yield high-throughput and accurate kcat predictions applicable to any organism or enzyme. Such capabilities are crucial for advancing our understanding of cellular metabolism and physiology. The data, sourced from a variety of repositories including BRENDA, SABIO-RK, KEGG, UniProt, and MetaCyc, are curated by Li et al ~\cite{Li2022DLkcat}. We used the same split (13,470 training samples, 1,684 validation samples and 1,684 test samples) for evaluation. Pearson correlation coefficient (PCC) is employed as the evaluation metric.

\vpara{Peptide-HLA/MHC Affinity.}
The human leukocyte antigen (HLA) gene encodes major histocompatibility complex (MHC) proteins, which can bind to peptide fragments and be presented to the cell surface for subsequent T cell receptors (TCRs) recognition. Accurately predicting the interaction between peptide sequence and HLA molecule will boost the understanding of immune responses, antigen presentation, and designing therapeutic interventions such as peptide-based vaccines or immunotherapies. The classification task aims to predict whether a given paired peptide and HLA sequence can bind or not. The modeling data is from Wu et al~\cite{Wu2023CcBHLA}. The raw dataset contains millions of samples, we used the same split and downsample 1\% for training and 5\% for validation and testing (57,357 training samples, 7,008 validation samples and 8,406 test samples). The AUC metric is employed as the measurement.

\vpara{TCR-pMHC Affinity.}
The interaction between T cell receptors (TCRs) and peptide-major histocompatibility complexes (pMHCs) plays a crucial role in the recognition and activation of T cells in the immune system. TCRs are cell surface receptors found on T cells, and pMHCs are complexes formed by peptides derived from antigens bound to major histocompatibility complexes (MHCs) on the surface of antigen-presenting cells. The classification task is to predict whether a given paired TCR sequence and peptide can bind or not. The evaluated data is major from VDJdb, processed and curated from Pham et al~\cite{Pham2023epiTCR}. We used the vdjdb\_no10x split (19,526 training samples and 4,485 test samples) for evaluation. The AUC metric is employed as the measurement.

\vpara{Antibiotic Resistance.}
 Antibiotic resistance (Antib. Res.) refers to the ability of bacteria and other microorganisms to resist the effects of an antibiotic to which they are once sensitive. In this task~(Antib. Res.), an input protein sequence is categorized according to which of 19 antibiotics it is resistant to. Thus, the scope of antibiotic drug development and research could be explored as an understanding in this topic is accumulated. The Dataset used in this task is curated by CARD~\cite{chhibbar2019generating}. We used the same split (2,072 training samples and 1,344 test samples) for evaluation. Total accuracy is employed as the evaluation metric.
 

\vpara{Fluorescence.}
The Fluorescence Prediction (Fluor. P.) task~\cite{sarkisyan2016local} focuses on predicting the fluorescence intensity of green fluorescent protein mutants, a crucial function in biology that allows researchers to infer the presence of proteins within cell lines and living organisms. This regression task utilizes training and evaluation datasets that feature mutants with three or fewer mutations, contrasting the testing dataset, which comprises mutants with four or more mutations. The partitioning of the datasets mirrors the splitting method implemented in the TAPE~\cite{tape2019}(21,446 training samples, 5,362 validation samples and 27,217 test samples). The quality of these predictions is assessed using the Spearman score as the primary evaluation metric. 


\vpara{Fitness (GB1).}
The task of Fitness Prediction (Fitness P.) is dedicated to anticipating the fitness landscape of the GB1 domain, a process that plays a pivotal role in understanding and enhancing the binding affinity and stability of this domain. As a prevalent protein interaction domain, GB1 finds wide usage in diverse applications such as protein purification, biosensors, and drug delivery~\cite{Luo2021ECNet}. This task is configured as a regression problem, where the goal is to predict the fitness of GB1 binding following mutations at four specific positions. The data for this task is sourced from the FLIP database~\cite{dallago2021flip} and the sampled split is used (6,289 training samples, 699 validation samples and 1,745 test samples). Predictive efficacy is assessed using the Spearman score as the principal evaluation metric.

\vpara{Zero-shot Fitness Prediction.} We utilize the DMS\_ProteinGym\_substitutions dataset from the ProteinGym benchmark \cite{notin2024proteingym} to evaluate zero-shot protein fitness predictions. Due to the ESM models' length constraints, we filtered out sequences exceeding 1024 residues, resulting in 201 assays with over 8 million mutations.
For protein understanding models like the ESM family, we use the metric from ESM-1v \cite{meier2021language} to score mutations based on the log odds ratio at the mutated positions. In sequences with multiple mutations, we assume an additive model. Mutation scores are calculated using the wild type and mutated type, with each mutated position masked in the sequence input.
For generative models like PROGEN2 \cite{nijkamp2023progen2}, we use the log-likelihood of the entire mutated sequence as the predicted mutation score.
Performance is assessed by determining the Spearman correlation coefficient between predicted and ground truth scores. Results are averaged within each assay and then across all assays. For comprehensive comparison, we also include previous supervised fitness predictions on the Fitness GB1 dataset, as shown in Supplementary Table~\ref{tb:zerofit}.
We observe that the scaling effect does not seem to hold in zero-shot fitness prediction, which aligns with observations from other works like PROGEN2. We hypothesize that although large pre-trained models may have a stronger capacity to differentiate between original and mutated sequences, this ability might not directly translate to accurate mutation scores without supervision. Thus, our xT100B model, after SFT, still holds better performance compared with other smaller models on the supervised fitness prediction scenarios.



\vpara{Localization.}
The task of Protein Subcellular Localization Prediction (Loc. P.) bears substantial relevance in bioinformatics, owing to its contributions to proteomics research and its potential to augment our comprehension of protein function and disease mechanisms~\cite{david2021identifying}. In this task, the input to the model is an amino acid sequence of a protein, which is transformed into an output comprising a probability distribution over 10 unique subcellular localization categories. The dataset applied for this task is derived from Uniprot, meticulously curated by Armenteros et al~\cite{almagro2017deeploc} (6,622 training samples and 1,842 test samples). Total accuracy is employed as the evaluation metric.


We evaluated all benchmarked downstream tasks with \model-100B and ESM2 models (Supplementary Table~\ref{tasksComparison}). Metric values are shown in both probing and LoRA (in parentheses) fine-tuning modes, where the \underline{underline} denotes the best performance of probing and \textbf{bold} indicates the best performance of LoRA fine-tuning.

\ipara{Protein Structure.} We collect three datasets, including residual (Secondary Structure), pairing (Contact Map) and ensemble structure level (Fold Classification). 
It is evident that the large pre-trained models (\model-100B and ESM2-15B) bring substantial improvements, as does the application of LoRA.
Concretely, the accuracy of the \model-100B is improved from 76.86 to 93.32 when LoRA is applied. This implies the potential of incorporating LoRA into protein structure prediction models. 
More importantly, the contact map prediction task is intricately interconnected with the task of predicting the three-dimensional structure of proteins, as precise residue contact map prediction can significantly expedite the process. 
Existing structure prediction models may not exhaustively harness the non-linear transfer capabilities intrinsic to the pre-trained model. For instance, a popular model, ESMFold~\cite{lin2023evolutionary}, freezes ESM2 and appends a folding trunk (a transformer encoder with 48 layers) as a representation learner. Conversely, the LoRA technique, by enabling fine-tuning, pioneers a promising trajectory for exploiting the pre-training of large language models to augment the precision of 3D protein structure prediction.

\ipara{Protein Function.}
Several tasks have been established to experimentally assess the consequences of a synthesized protein sequence, with each observation tied to a specific biological function. Accordingly, we evaluate four such tasks within this category.
For instance, the antibiotic resistance task predicts whether a protein sequence is sensitive to a specific bacteria.
The results manifest the consistently superior performance of larger models in comparison to smaller counterparts, such as \model-100B and ESM2-15B vs ESM2-150M. The tendency is evidenced by a notably higher Spearman correlation margin on the fitness task and 10-class classification accuracy on localization prediction. Therefore, we believe that larger PLMs could be deployed in the frontier of biomedical research and pharmaceutical applications. 

\ipara{Protein Interaction.}
Proteins tend to interact with different types of molecules to trigger subsequent bioactivity, including the endogenous catalyzing substrate, metal ions on the cell surface, and exogenous antigen peptides. 
Here we focus on four tasks related to protein interactions. 
Specifically, for enzyme catalytic efficiency and metal ion binding prediction, only the protein sequence is utilized. For immunity-based peptide-MHC and TCR-pMHC interaction prediction, we simply concatenate two sequences with special token \texttt{<eos>} as model input. The results show that LoRA fine-tuning consistently outperforms the probing method, extending its advantage to sequence pair cases where the task pattern has not been seen during the pre-training stage. However, we find that the margin between \model-100B and ESM models tends to be small in peptide-MHC and TCR-pMHC interaction tasks. 
This may be due to the relative simplicity of the task, as the baseline model already achieves high performance (AUC > 0.9).

\ipara{Protein Developability.}
The biophysical condition surrounding protein molecules determines whether they can work normally. Here, we select three related tasks—solubility, stability, and sensitivity—as representatives for evaluation.
The results indicate that \model-100B significantly outperforms ESM models on solubility and stability tasks, even though the two tasks are relatively difficult (ESM-150M performance is around 70). 
However, the improvement in temperature-related tasks remains marginal.
We find a similar performance trend for the two datasets: \model-100B is slightly better than ESM. 
Since both ESM and \model-100B achieve high performance (with MCC > 0.93) in the Temperature Stability task, we could hypothesize that this task may present some challenges for prediction.
On the other hand, the Optimal Temperature task has the smallest training sample size (approximately 1.7k) among all benchmark tasks. Therefore, it could potentially constrain the achievable performance of models.



\section{xT-Fold Acceleration Methods}
\label{xtfold_config}
We developed and trained an xT-Fold model based on xTrimoPGLM-100B. Initially, the model undergoes a 4-bit quantization process, effectively implementing the W8A16 (weights for 8 bits, Activation with 16 bits) strategy before further compressing into a 4-bit model. 
Due to reduced model communication time among the model weights and memory consumption, this approach resulted in at least a 4x increase in training speed with the same devices. However, during the inference phase, it encountered memory overflow issues with longer sequences (exceeding 700 in length). To address this, we incorporated the FlashAttention~\cite{dao2022flashattention} technique into xT-Fold. This technique enhances self-attention through operator fusion and softmax reduction, allowing our model to perform inference on sequences up to 2000 in length in a \textasciitilde200 seconds, at least \textasciitilde5x GPU inference time reduction compared with a standalone AlphaFold2 that not include the time of MSAs feature process. 

\section{xT-Fold Training Settings}
\label{xtfold_data}
The dataset preprocessing follows similar settings with ESMFold and AlphaFold2. We train xT-Fold on  \textasciitilde380K experimentally determined structures of \textasciitilde25k clusters from the PDB~(release date is less than May 2020), further augmented with a distilled dataset of \textasciitilde4M greater than 90\% pLDDT structures predicted from AlphaFoldDB. Predicted structures are sampled 75\% of the time, and real structures 25\% of the time during training. 
The model is trained with the same loss as AlphaFold2 except for omitting masked MSA loss.
We trained on 10M samples with crop size 256 on the first stage, and fine-tuned cropsize 384
on data parallel 64 cards A100 80G with effective batch size 128.
It was trained for the first stage 78K steps on protein crops of size 256 and then fine-tuned at the second stage with the structural violation loss with 0.01 coefficient for 70K steps, on crop sizes of 384. 
In the second stage, we use a cosine decay to make the learning rate from 5e-5 to 1e-5 after a warm-up of 3000 steps.

Below, we provide additional methodological insights into the architecture and training procedures of xT-Fold.

\noindent \textbf{Model Architecture:} The key modification involves employing MSARowAttentionWithPairBias exclusively for the input query sequence, rather than attending to each sequence individually in MSAs, and we further exclude the MSAColumnAttention block. We employ a 48-layer version of this block with up to three recycling stages during training and utilize $\text{recycle}_{\text{num}}=3$ during inference. This framework is a refined adaptation based on AF2's Evoformer framework. Despite minor adjustments, we retain the Evoformer terminology, but it more closely resembles a simplified version within the current AF3 framework, referred to as Pairformer~\cite{abramson2024accurate}. This adjustment significantly enhances training and inference speeds.

\noindent \textbf{Initialization:} During initialization, we treat the Evoformer as a residual block by initializing its output parameters to zero. This initialization establishes a quasi-identity mapping from PLM to the structure module, leveraging Evoformer's residual modules. Throughout the training, Evoformer modules are dynamically selected based on gradients. We froze the entire PLM during training without fine-tuning.
Hyperparameters: Hyperparameter settings closely follow those of AF2 and ESMFold, including: Exponential Moving Average (EMA): Decay 0.999; Adam Optimizer Parameters: Base learning rate: 1e-3, $\beta_1$ = 0.9, $\beta_2$ = 0.999, eps = 1e-8.

\vpara{Data Distillation:} For data distillation in training, we initially filtered approximately 60 million samples from AFDB based on AF2's pLDDT  0.9 criteria, resulting in approximately 4 million samples after clustering with a sequence max identity of 0.5. Subsequently, we filtered out parts similar to the test set by 90\%, and integrated 38K samples from PDB in a 75:25 (\#Distill:\#PDB) ratio blend. Throughout this process, we employed cluster sampling based on sequence length and cluster size, with a sampling probability calculated as: 
\beq{
\text{prob} = \left( \frac{1}{\text{cluster\_size}} \right) \times \frac{1}{512} \times \max\left(\min(\text{seq\_len}, 512), 256\right)
}

\noindent \textbf{Training Procedure:} The training process consists of two stages: 
Stage 1: Trained on 10 million samples with Crop size: 256 and Batch size: 128;
Stage 2: Increased crop size to 384. Incorporated pLDDT and a minimally weighted violent loss (coefficient set to 0.01) Trained on an additional 1 million samples. Used cosine decay with the learning rate decreasing from 5e-5 to 1e-5.

\section{AlphaFold2 with FlashAttention}
\label{subsec:af2_flashatten}
Since we hope to use FlashAttention to speed up the running of AlphaFold2, we use OpenFold's AlphaFold2 implementation code to load DeepMind's AlphaFold2 weights (by --jax\_param\_path option) for inference. In our previous tests, we have confirmed that structure prediction by using OpenFold code + Jax weight and directly using DeepMind's official version for inference have exactly the same results. FlashAttention significantly reduces the time and memory consumption of original Self-Attention by reducing the number of reads and writes between high bandwidth memory (HBM) and GPU on-chip SRAM, without changing the results of numerical computation.

For each query sequence, we use DeepMind's official MSA retrieval pipeline to obtain the MSA sequences and PDB templates and limited the maximum PDB template release date to 2020-05-01. We used five checkpoints of AF2 to run structure prediction. For each checkpoint, we set different random number seeds and run five times, obtaining a total of 25 predicted models (without AMBER relaxation). We calculated the TM-score and GDT\_TS score of all models and found that there was little difference between all models (CAMEO TM-score 0.87 \textasciitilde 0.88, CASP15 TM-score 0.74 \textasciitilde 0.77). It should be noted that AlphaFold2's model\_4 and model\_5 have PDB templates as input, and our xT-Fold doesn’t. For a more fair comparison, we use the first run result of model\_3 (model\_3\_0) as the representative result of AlphaFold2.

\section{The SFT \& ReST Pipelines}
\label{sec:sft}
\subsection{Dataset}
For the comprehensive assessment of the alignment ability of \model-100B, we curate 5 datasets from the 18 benchmarks for supervised fine-tuning and reinforcement self-training (Supplementary Section~\ref{sec::downstream_tasks}). 

\vpara{Strong Fluorescence Intensity Proteins.}
Targeting generating protein sequences characterized by elevated fluorescence intensity, we select sequences exhibiting fluorescence intensity scores surpassing 0.8 in the Fluorescence task dataset, yielding a final dataset comprising 21,446 sequences for training and 5,362 sequences for validation. 

\vpara{High Fitness Mutations.}
Targeting generating protein sequences with mutations of high fitness, we selected sequences with fitness scores exceeding 0.5 in the Fitness task dataset, yielding a final dataset comprising 4,163 sequences for training and 469 sequences for validation. 

\vpara{Low Thermal Stability Proteins.}
Targeting generating protein sequences with reduced stability at 65 degrees Celsius, we selected sequences labeled as '0' in the Temperature Stability task dataset, yielding a final dataset comprising 141,598 sequences for training and 32,790 sequences for validation.

\vpara{Immunoglobulins.}
Targeting generating protein sequences that correspond to the specific fold class of immunoglobulins, we selected sequences labeled as '36' in the Fold Classification task dataset, yielding a final dataset comprising 981 sequences for training and 221 sequences for validation.

\vpara{Nucleus Localization Proteins.}
Targeting generating protein sequences that localize in the nucleus, we select sequences with labeled as '7' in the Localization task dataset, yielding a final dataset comprising 2,312 sequences for training and 610 sequences for validation.

\subsection{Supervised Fine-tuning.}

\vpara{Supervised Fine-tuning.}
We trained the xTrimoPGLM-100B model with the GPT objective using LoRA fine-tuning with $\text{LoRA}_r = 16$ and $\text{LoRA}_\alpha = 32$.
In addition, We use Adam as our optimizer with $\beta_1$ and $\beta_2$ set to 0.9 and 0.95, and a weight decay value of 0.01. We warm up the learning rate from $5\times 10^{-7}$ to $10^{-5}$ over the first 10.0\% samples, then decay it linearly to the minimum learning rate $10^{-6}$.

We trained the Progen2-xlarge model with the GPT objective in full scale. We use Adam as our optimizer with $\beta_1$ and $\beta_2$ both set to 0.9, $\epsilon$ set to $10^{-7}$, and a weight decay value of 0.01 (except for the Fold Classification task, which corresponds to a weight decay value of $10^{-4}$). We warm up the learning rate from $10^{-7}$ to $10^{-5}$ over the first 500 samples, then decay it by a cosine schedule to the minimum learning rate $5\times 10^{-6}$. The gradient accumulation step is 10 for the Fold task and 4 for others.

We trained the ProtGPT2 model with the official fine-tuning script on Huggingface. We use Adam as our optimizer with $\beta_1$ and $\beta_2$ set to 0.9 and 0.999, and a constant learning rate $10^{-5}$.

\subsection{Reinforcement Self-Training}
To explore the capacity for iterative improvement of \model-100B, we implement a 1-step reinforcement self-training regime based on the initial supervised fine-tuning, leveraging feedback from task predictors as the reward model to further enhance the model's performance. The task predictor on each task is fine-tuned by qLoRA~\cite{dettmers2024qlora}. 
More specifically, the reward models assess the generated sequences by predicting the numeric features for regression tasks (Fluorescence and Fitness), or the likelihood of the desired label for classification tasks (Temperature Stability, Fold and Localization). 
Then we apply a filtering process with strategies detailed below, and using the selected high-quality sequences as training data, fine-tune the checkpoint saved after supervised fine-tuning until the model converges.

\vpara{Filtering Generated Sequences.}
For the Fluorescence task, we select sequences with predicted fluorescence intensity scores greater than 2.5 (1,386 sequences in total). For the Fitness task, we select the top 300 sequences in descending order of their predicted fitness scores. For the Temperature Stability task, we select the top 200 sequences in descending order of their predicted likelihood of being classified as unstable at 65 degrees Celsius. For the Fold task, we select the top 500 sequences in descending order of their predicted likelihood of being classified as corresponding to the fold class of DNA-binding domains. For the Localization Task, we select sequences with a likelihood of being classified as localizing in the nucleus surpassing 0.5 (1,084 sequences in total).


\section{Statistical Analysis of Generated Sequences}
\label{subsec: more_gen}
In this section, we examine both the sequence and structural attributes of generated sequences, shedding light on their statistical properties.

\vpara{Statistical properties of the sampled artificial sequences.}
More specifically, a diverse set of sequences is sampled using a cross product of temperature ($T$ $\in$ {0.8, 1.0, 1.2, 1.4, 1.6}) and nucleus sampling probability ($P$ $\in$ {0.5, 0.7, 0.9, 1.0}) parameters.
For each combination of $T$ and $P$, we sample 600 sequences for the comprehensive sequence analysis. 

\begin{figure*}[!ht]
\centering
\includegraphics[width=1.0\linewidth]{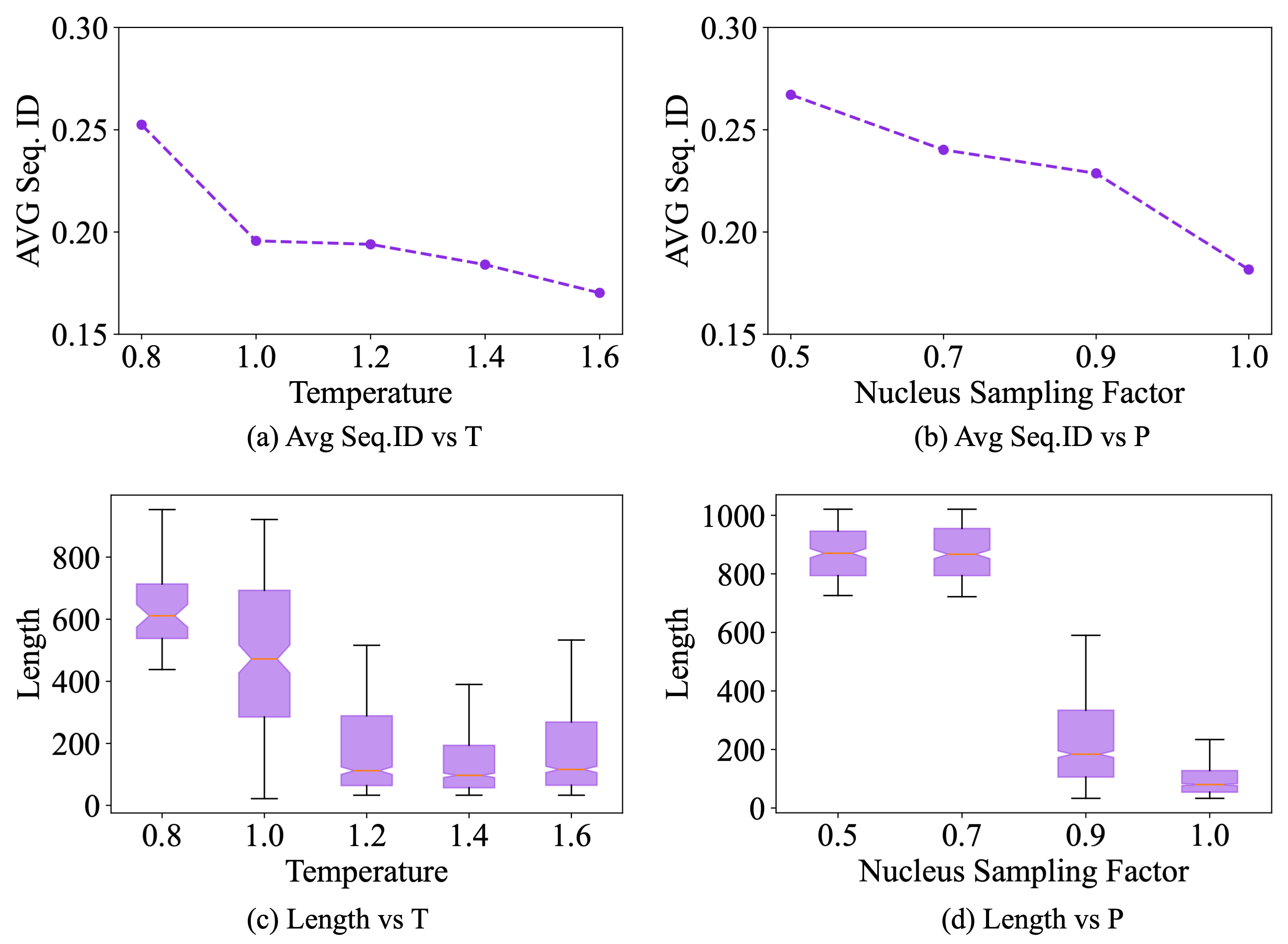} 
\caption[Distributions of generated sequences]{\smodel generates a diverse set of sequences by varying temperature ($T \in {0.8, 1.0, 1.2, 1.4, 1.6}$) and nucleus sampling probability ($P \in {0.5, 0.7, 0.9, 1.0}$) parameters. For each combination of $T$ and $P$, 600 sequences are sampled for comprehensive analysis. This figure displays the distributions of sequence identity (a, b) and length for the generated sequences across the different temperature and nucleus sampling probability settings.
The box plots (c, d) mark the median, the orange horizontal line; upper and lower quartiles, purple horizontal line; upper and lower bound, black horizontal line; 
and 1.5$\times$ interquartile range (whiskers). 
}
 \label{fig:pro_gen}
\end{figure*}

We present the pairwise sequence identity analysis of generated sequences obtained through various combinations of temperature and nucleus sampling factors (Supplementary Figure~\ref{fig:pro_gen}(a)(b)). 
We observe that higher nucleus sampling factors and temperatures, which flatten the token probability distribution during the generation process, lead to a broader range of sequence diversity. However, it should be noted that the likelihood of selecting the \texttt{<eos>} token also increases under these conditions. Consequently, higher factors may result in shorter sequences (Supplementary Figure~\ref{fig:pro_gen}). 
Furthermore, our empirical study suggests that the pre-trained model tends to generate repetitive tokens when the temperature drops below 0.8 and the nucleus sampling factor falls under 0.7, which results in abnormal long sequences. 
Therefore, we recommend a careful calibration of the hyperparameters, specifically the balance between temperature and nucleus sampling factors, to generate protein sequences that conform to the desired specifications.

\vpara{Intrinsically unstructured/disordered proteins/domains.}
Intrinsically unstructured or disordered proteins/domains (IUPs)~\cite{dosztanyi2005pairwise} exist in a largely disordered structural state and carry out basic protein functions. Hence, it is essential to identify IUPs by the commonly used disorder prediction methods, IUPred3~\cite{erdHos2021iupred3}, to reflect the biological evolutionary behaviors. Without extra functional annotations, we generate a dataset of protein sequences to evaluate our proposed method in the protein disorder task. For comparison, we also simulate a natural dataset by uniformly sampling from the original training dataset. Our generated dataset and the natural dataset consist of 6,523 and 10,000 sequences, respectively.

In order to compare the two datasets comprehensively, all three prediction types are provided in Table~S\ref{tb:gene_dis_structure}, i.e., short disorder, long disorder, and globular structured domains~\cite{dosztanyi2018prediction}. 
Short disorder (SHORT) emphasizes predicting short-disordered regions, while long disorder (LONG) chiefly targets global structural disorder encompassing a minimum of 30 consecutive protein residues.
The prediction corresponding to globular domains (GLOBULAR) is a structured prediction for structural studies and structural genomics projects. 
We also present the ordered content (the proportion of ordered regions over the entire protein, termed ORDERED) from globular disorder predictions, to analyze the structural and biochemical attributes of sequences generated by \model. This approach diverges from the definition of ordered content (ratio of ordered to disordered regions) employed in ProtGPT2~\cite{ferruz2022protgpt2}.

Consequently, the two datasets show similar disorder prediction results as reported in Table~S\ref{tb:gene_dis_structure}. 
Our generated sequences have close prediction results to the natural dataset in all four metrics, with the largest gap of 3.89\% in LONG between them. 
The experimental results affirm that sequences generated by \model-100B exhibit comparable tendencies for minimal, maximal, and structured predicted disorder, akin to natural sequences.

\begin{figure*}[!ht]
\centering
\includegraphics[width=1.0\linewidth]{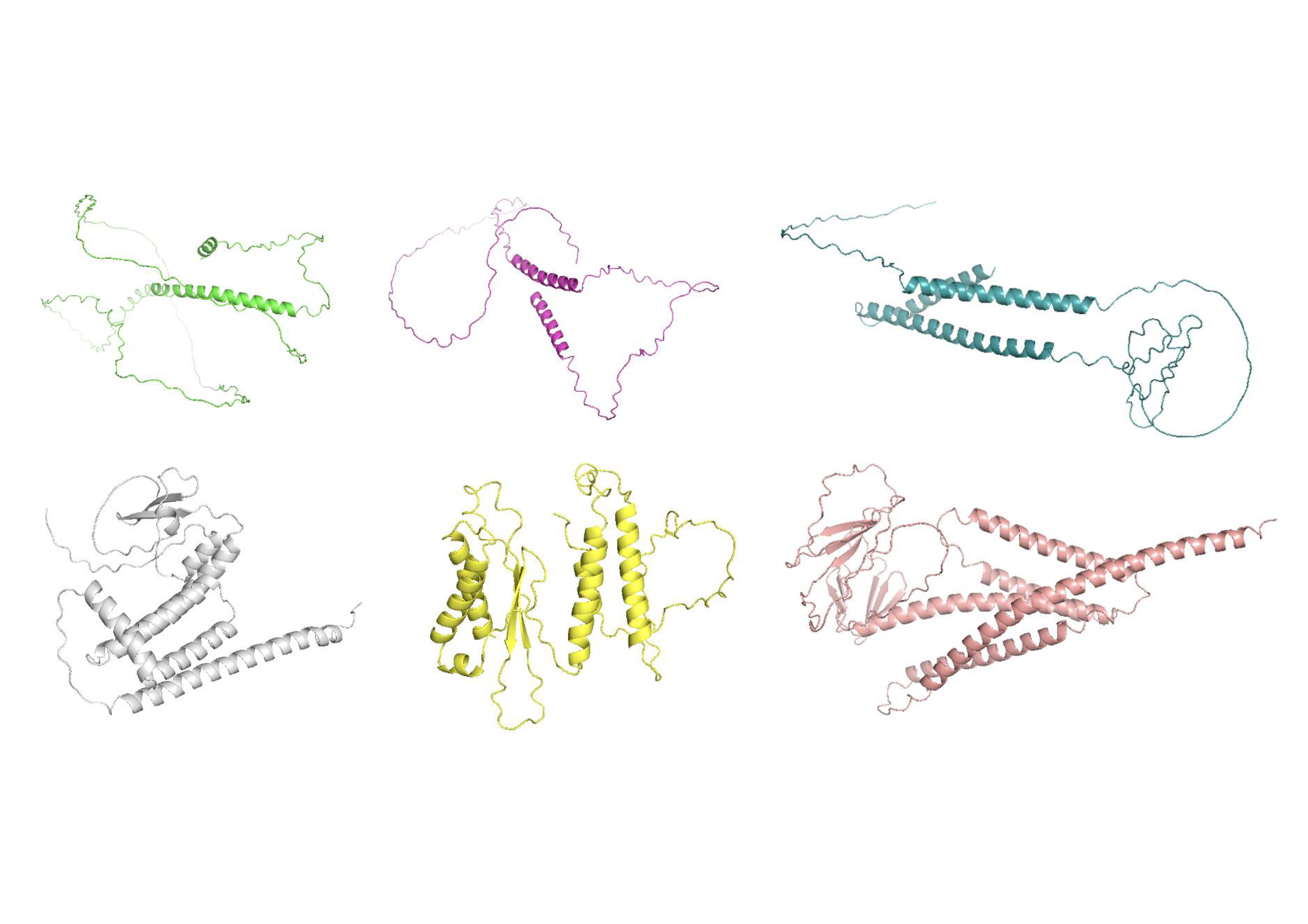} 
\caption[Structure Examples of Generated Protein Sequences]{Structure examples of generated protein sequences with different parameter configurations. 
}
\label{fig:fail_case_long_disorder}
\end{figure*}

\vpara{Trial \& Error of Generated Structures with N-gram penalty.}
\label{sec::fail_case_long_disorder}
We first produced batches of samples with an n-gram penalty (N-gram=3) to reduce the probability of generating repetitive sequences (Supplementary Figure~\ref{fig:fail_case_long_disorder}). The first row depicts sequences with parameter ($T$=1.0, $P$=1.0, N-gram-penalty=3), while the second row removes the n-gram constraints to reduce long loop disorder regions. We find many examples exhibiting low-complexity sequences (e.g., local repeats), where the predicted structures contain long loop disorder regions. We hypothesize that the n-gram penalty potentially impedes the model's capacity to generate grammatically correct sequences with ease.
Once we remove the n-gram penalty, the generated structures tend to be more natural.


\section{\model-Ab: OAS Fine-tuning for Antibody}
\label{abfold}
We further adopt the \smodel framework to explore a special family of proteins: antibodies. 
Antibodies are ubiquitous yet vital proteins that are generated by the adaptive immune system to bind and neutralize foreign pathogens such as SARS-CoV-2~\cite{zhu2022international, li2022immune}. It functions via binding to antigens on the pathogen, recognizing it, and finally inhibiting it. 
Generally, antibodies, composed of two identical heavy chains and two identical light chains, form a large Y-shaped structure.
The specificity of antibody binding is determined by CDRs at the tips of the Y-shaped proteins (VH and VL).
The estimated number of potential human antibodies ranges from $10^{13}$ to $10^{16}$, signifying the incredible diversity of the immune response. Their specificity, combined with this abundance, renders antibodies invaluable for the development of targeted therapies.
By utilizing the \smodel framework, we've made advancements in predicting antibody naturalness and structure prediction with our antibody-specific model, \model-Ab. 

We do not directly fine-tune on \model-100B, mainly due to limitations in computational budgets and considering the inherent lack of diversity in OAS antibody data, most of which are of similar length and have similar framework areas.
Hence, we first pre-train \model-1B model on the general protein dataset SM Section~\ref{sec::datasets} 
This process undertakes 500B tokens. 
Since antibodies represent a distinct subset of general proteins, then we finetuned the model using the OAS dataset\footnote{Observed Antibody Space (OAS)~\cite{kovaltsuk2018observed} data. Following the paper, we filter OAS data with IMGT schema \cite{lefranc2009imgt} and therefore get 678m sequences without disorder and incompletion.}, comprising 1 billion antibody sequences. 
Considering that the CDRs are the most important parts of an antibody, we randomly mask one or two whole CDRs for 40\% of samples with \texttt{[sMASK]}.
A further 40\% of the samples undergo a random span masking process, while the remaining 20\% are subjected to the MLM objective. 
We exclude the \texttt{[gMASK]} loss from consideration, as it is not required for downstream antibody-related tasks involving long-text generation.
When fine-tuning the \model-Ab-1B model on OAS data, we decrease the maximum learning rate to 2e-5 and make the model consume 500B tokens with 2,048 samples per batch and the 1,024 input length per sample. 
It takes about 168 hours to use 128 Nvidia A100 80G GPU cards with mixed precision training. We carry out evaluations on two critical undertakings within the realm of drug discovery including assessing the zero-shot naturalness of antibodies and predicting the structural conformation of antibodies.





\subsection{Zero-shot Naturalness}
In protein design and expression, a crucial step involves filtering out proteins with low expression while retaining those with high naturalness.
Perplexity (PPL) given by a protein language model can be used to predict the naturalness of proteins~\cite{nijkamp2023progen2, bachas2022antibody}. For the GLM objective, PPL is calculated by:
\begin{equation}
\text{PPL}(\mathbf{x}) = \exp\left(-\sum_{i=1}^{l}{\log{P_{\text{model}}(x_i | x_{\hat{i}},x_i=\texttt{[sMASK]})}}\right),
\end{equation}
where $P_{\text{model}}(x_i | x_{\hat{i}},x_i=\texttt{[sMASK]})$ is the probability of the $i$-th amino acid, denoted by $x_i$, as predicted by the model. Here, the context $x_{\hat{i}}$ is given, with a \texttt{[sMASK]} token in the $i$-th position. Note that $x_{\hat{i}}$ represents all tokens excluding the $i$-th token.
For the MLM objective, pseudo-perplexity \cite{salazar2019masked} is usually utilized as a substitute for perplexity since perplexity is only computed via the auto-regressive model. Pseudo-perplexity (PPPL) of a protein is defined as
\begin{equation}
\text{PPPL}(\mathbf{x}) = \exp\left(-\sum_{i=1}^{l}{\log{P_{\text{model}}(x_i | x_{\hat{i}},x_i=\texttt{[MASK]})}}\right),
\end{equation}
where $P_{\text{model}}(x_i | x_{\hat{i}},x_i=\texttt{[MASK]})$ represents the probability of the $i$-th amino acid $x_i$ predicted by the model given the context $x_{\hat{i}}$ with a \texttt{[MASK]} in $i$-th position.

\vpara{Datasets.} 
To assess the performance of various models, we construct two datasets derived from protein expression experiments conducted in a wet laboratory. Any samples that yield less than 10 mg/L of the purified proteins in the supernatant liquid are categorized as unexpressed, whereas those yielding more than 10 mg/L are deemed as successfully synthesized.
The first dataset (Dataset 1) comprises 601 antibody sequences, derived from experiments conducted on CHO and HEK293 cells. These sequences include 114 proteins from humans, 90 from mice, 1 from rhesus, and 396 from undefined species (not directly sourced from specific species). Of these, 516 are successfully expressed.
The second dataset (Dataset 2) – sourced from HEK293 cells – contains 98 human antibody sequences targeting a specific antigen, of which 90 are successfully expressed.

\vpara{Metrics.} 
Each sample comprises both a heavy chain and a light chain. For models that do not incorporate chain types, we calculate the perplexity of each chain individually, then multiply these values to obtain the overall perplexity for the pair.
For models incorporating chain types, we concatenate both chains in the following format: 
$\texttt{[human]}\texttt{[heavy]}\text{sequence1}\texttt{<eos>}\texttt{[human]}$ $\texttt{[light]}\text{sequence2}\texttt{<eos>}$,
where \texttt{[human]} is a special token to indicate the species of sequences, \texttt{[heavy]} and \texttt{[light]} are two tokens to represent the types of sequences, \texttt{<eos>} means the end of sequences. 
We use the area under the receiver operating characteristic (ROC) curve (AUC) as a measure to evaluate the models' ability to predict protein naturalness.
Notably, Iglm~\cite{shuai2021generative} and PROGEN2 \cite{nijkamp2023progen2} are auto-regressive models, while AbLang \cite{olsen2022ablang}, ESM2 \cite{lin2023evolutionary}, and AntiBERTy \cite{ruffolo2021deciphering} are auto-encoder models. 
Thus we evaluate Iglm and PROGEN2 using PPL, while the remaining models are tested using PPPL.
As \model-Ab-1B can function as either an auto-regressive or an auto-encoder model, we employ both PPL and PPPL to calculate its AUC score.

\vpara{Results.} 
The results are shown in Supplementary Table~\ref{tb:zero_shot}. Among these, \model-Ab-1B surpasses other baselines in two datasets. 
Moreover, we further fine-tune \model-Ab-1B with the GLM objective with 30 billion tokens to gain \model-Ab-1B-GLM. Analogously, we fine-tune it with the MLM objective with 100 billion tokens to get \model-Ab-1B-MLM. 
Since the consumed tokens (80\% tokens) of the GLM objective is 4 times more than that (20\% tokens) of the MLM objective in the pre-training stage, \model-Ab-1B-MLM is fine-tuned with more tokens than \model-Ab-1B-GLM for a relatively fair comparison. 
Consequently, \model-Ab-1B-GLM and \model-Ab-1B-MLM keep similar results on Dataset 1 with little difference of AUC on pair test, while they benefit from additional training on Dataset 2, as the AUC scores are improved by 0.02 consistently.

\vpara{Ablation Study.} To justify the contribution of different components, i.e, \texttt{[sMASK]} within random spans or \texttt{[sMASK]} with CDR regions, of the GLM objective, 
we train \model-Ab-1B-GLM-CDR only with the CDR span task and \model-Ab-1B-GLM-Random with the random span task, based on the pre-trained \model-Ab-1B. 
xTrimoPGLM-Ab-1B-GLM (50\% CDR span task and 50\% random span task) outperforms these two models on Dataset 1 and Dataset 2. These distinctions highlight the importance of the combination of CDR span task and random span task.

\subsection{\model-AbFold: Antibody structure prediction}\label{sec:AbFold}
In this section, our aim is to predict antibody structures based on their sequences.
The study of protein structure assists in the design and modification of proteins, as well as in target identification and structural analysis for protein-based drug design.
A popular method to predict protein structures is leveraging Multiple Sequence Alignment (MSA) and structure templates to encode sequences and then using encoded matrices to generate structures.  
However, MSA requires significant computational resources and time.
Given that \smodel is trained using the MLM task, it is naturally suited to serve as an encoder for understanding tasks
Therefore, 
we develop \model-AbFold, which is based on \model-Ab-1B, with the aim of predicting three-dimensional antibody structures directly from amino acid sequences.
Our experiments encompass both single-chain structure prediction and complex structure prediction, i.e., the VH-VL complex.

\vpara{Datasets \& Metrics.}
The structure prediction dataset for single chains is derived from the RCSB Protein Data Bank (PDB) \cite{Berman2008ThePD} prior to April 13, 2022, which consists of both amino acid sequences and structures. We collect all antibody data in PDB, which contains 19k antibody chains (VL or VH). Structural data with missing resolution values or resolutions greater than 9 Å are excluded to maintain quality. Additionally, sequences with an amino acid repetition rate exceeding 90\% are filtered out. 
Finally, we obtain about 7.5k unique sequences (VL or VH chains). The training set consists of 7,234 sequences, and 350 sequences are left as the test set.
The dataset for VH-VL complexes includes approximately 4.7k antibodies from PDB, which are released before January 2022. We selected 68 samples as the test set, which are released between January 2022 and November 2022.

Root mean square deviation (RMSD) and TM-score \cite{zhang2004scoring} are used as evaluation metrics for both tasks. Another important metric DockQ \cite{basu2016dockq} is involved in the structure prediction of complexes.

\vpara{Model Architecture.}
Our principal hypothesis is that with an adequately proficient encoder, structure prediction models can accommodate complex structures using shallow Evoformer layers and structure modules.
Therefore, compared with the current prevailing folding structures, such as ESMFold, AlphaFold2, we introduce the following modifications to \model-AbFold: 1) We eliminate MSA and template search modules, as they offer minimal benefit for antibody folding in our pre-training and fine-tuning paradigm; 2) Unlike Alphafold2, which employs 48 blocks of Evoformer, and ESMfold, which utilizes 48 layers of folding trunk, we significantly \textbf{reduce the number of downstream folding blocks from 48 to 1}. The architecture of xTrimoPGLM-AbFold is depicted in Supplementary Figure~\ref{fig:structure_prediciton_model}.


\vpara{Training Settings.}
For single-chain structure prediction, we convert protein sequences of length $L$ into the format of $\texttt{[human]}\texttt{[chain type]}\text{sequence}\texttt{<eos>}$, and feed it into the \model-Ab-1B model to obtain the hidden representation $\mathbf{M}$ of the last layer. The information corresponding to \texttt{[human]}, \texttt{[chain type]} and \texttt{<eos>} are removed from $\mathbf{M}$, where $\mathbf{M}$ $\in \mathbb{R}^{L\times D}$ and $D$ is the size of the hidden dimension of the \model-Ab-1B model. Then, we extend $\mathbf{M}$ along its $L$ dimension in a pairwise manner to obtain a tensor $\mathbf{Z}\in\mathbb{R}^{L\times L\times 2D}$ (Supplementary Figure~\ref{fig:structure_prediciton_model}).
After that, $\mathbf{M}$ and $\mathbf{Z}$ are fed into a single-block Evoformer module for information fusion and then into the structure module for prediction of the angle and position of each residue. 
For the VH-VL complex, it should be noted that the input is converted into the format of $\text{vh\_sequence}\texttt{[linker]}\text{vl\_sequence}$, where the \texttt{[linker]} is composed of four groups of residue sequences, each of which is composed of four G residues and one S residue, just like GGGGSGGGGSGGGGSGGGGS. 

For structure prediction of single chains, the loss function of structure prediction mainly follows the work of AlphaFold2~\cite{Tunyasuvunakool2021HighlyAP} , which contains Frame Aligned Point Error~(FAPE) and a number of auxiliary losses but excludes MSA loss. The loss can be formalized as follows:
\beq{\mathcal{L} = 0.5\mathcal{L}_{\text{FAPE}} + 0.5\mathcal{L}_{\text{aux}} + 0.3\mathcal{L}_{\text{dist}} + 0.01\mathcal{L}_{\text{conf}} + 
0.5\mathcal{L}_{\text{rmsd\_ca}}
}
\noindent where $\mathcal{L}_{\text{aux}}$ is the auxiliary loss from the structure module, $\mathcal{L}_{\text{dist}}$ is an averaged cross-entropy loss for distogram prediction, $\mathcal{L}_{\text{conf}}$ is the model confidence loss,  $\mathcal{L}_{\text{angle\_norm}}$ is the side chain and backbone torsion angle loss ~\cite{Tunyasuvunakool2021HighlyAP} and $\mathcal{L}_{\text{rmsd\_ca}}$ is the rmsd for carbo alpha. 
In addition to the loss described by the formula above, the VH-VL complex replaces the rmsd-ca loss with a chain center-of-mass loss ~\cite{Evans2021ProteinCP} and a structural violation loss~\cite{Tunyasuvunakool2021HighlyAP}, with weights of 1.0 and 0.03, respectively. The concrete loss is shown as follows:
\beq{\mathcal{L}_{\text{vh-vl}} = 0.5\mathcal{L}_{\text{FAPE}} + 0.5\mathcal{L}_{\text{aux}} + 0.3\mathcal{L}_{\text{dist}} + 0.01\mathcal{L}_{\text{conf}} + \mathcal{L}_{\text{centre\_mass}} + 0.03\mathcal{L}_{\text{violation}}
}
\vpara{Baselines.}
For single-chain structure prediction tasks, we conduct a comparison of existing influential folding models, including Alphafold2 and four PLM-based models: OmegaFold~\cite{wu2022high}, ESMFold~\cite{lin2023evolutionary}, IgFold~\cite{Ruffolo2022FastAA}, and xTrimoAbFold~\cite{Wang2022xTrimoAbFoldDN}. We use public checkpoints AlphaFold2~\footnote{https://github.com/deepmind/alphafold},
ESMFold~\footnote{https://github.com/facebookresearch/esm},
IgFold~\footnote{https://github.com/Graylab/IgFold},
OmegaFold~\footnote{https://github.com/HeliXonProtein/OmegaFold} to infer the test set.


For the prediction of VH-VL complex structures, we compared ZDock~\cite{Chen2003ZDOCKAI} , a rigid protein docking algorithm based on fast Fourier transform correlation techniques, ClusPro~\cite{Kozakov2017TheCW}, a docking method using bioinformatics and computational chemistry techniques, EquiDock~\cite{Ganea2021IndependentSM}, a genetic evolution-based algorithm, HDOCK~\cite{Yan2020TheHS}, an algorithm that combines global and local search, and AlphaFold-Multimer~\cite{Evans2021ProteinCP}, which predicts complex structures based on protein sequence, MSA, and template information.


\vpara{Results.}
Each experiment is conducted 5 times with different random seeds and reports the averaged results. 
As demonstrated in Supplementary Table~\ref{table:vh_vl_structure}, 
\model-AbFold significantly outperforms all other models, notably xTrimoAbFold—an existing state-of-the-art model—in every metric related to antibody structure prediction.
The impressive performance of \model-AbFold implies that a pre-trained antibody model, when fine-tuned with merely a single additional Evoformer~\cite{Tunyasuvunakool2021HighlyAP} block, can emerge as a leading model for antibody structure prediction, even without the presence of MSA and templates.

Supplementary Table~\ref{tb:vh_vl_complex_structure} shows the performance of VH-VL complex in different models. AlphaFold-Multimer, which relies heavily on MSA and template information, outperforms most of protein docking algorithms. However, \model-AbFold, which does not use any MSA or template information, performs comparable with AlphaFold-Multimer, indicating that \model-Ab-1B has learned sufficient and rich information on antibodies. 
Crucially, \model-AbFold achieves a speedup of \textbf{6,300}$\times$ over the original AlphaFold-Multimer and \textbf{103}$\times$ over the faster MSA-searching AlphaFold-Multimer~\cite{luo2023xtrimodock}, owing to the original AlphaFold-Multimer consumes a long time to search MSA (0.8 hours per sample).
When we increase the number of Evoformer blocks to 16, \model-AbFold attains the best performance on all metrics while still maintaining a \textbf{2,400}$\times$ speedup than the original AlphaFold-Multimer and \textbf{40}$\times$ speedup than the accelerated AlphaFold-Multimer. It is noteworthy that only a marginal improvement is attained when the number of Evoformer blocks is increased from 1 to 16, which indicates that \smodel-Ab has already captured sufficient information for downstream tasks to predict atom positions with precision.




\subsection{Antibody-specific Generation using \model-Ab.}
\label{subsec:anti_des}
To demonstrate the generation capacity of \model-Ab, we select a heavy chain antibody sequence (specifically 368.04.B.0106) that interacts with SARS-CoV-2-WT. We implement four distinctive masking strategies to redesign the Complementarity Determining Region 3 (CDR3) of the selected sequence, as the CDR3 region is a critical element in the structure of an antibody or T cell receptor. This sequence is characterized by significant variability and plays an integral role in the specificity of antigen recognition. 
The four strategies are defined as follows,

\begin{itemize}
    \item \textbf{CDR3 Short Masking (CSM).} This strategy involves masking a partial segment of the CDR3 region. We select the length of the masked region based on a uniform distribution within the interval $[3,6]$. Subsequently, a segment of the CDR3 region is randomly replaced with the \texttt{[sMASK]} token. Upon feeding this modified antibody sequence into \model-Ab-1B, the masked segment of the CDR3 region undergoes a redesign. The comparison between the conformations of the CDR3-redesigned antibodies and the original sequence is depicted in Supplementary Figure~\ref{fig::sm}.
    \item \textbf{CDR3 Whole Masking (CWM).} This strategy involves masking the entirety of the CDR3 region with the \texttt{[sMASK]} token, thus necessitating a de novo design approach. Given the increased complexity of this setting, compared to the CSM, the CWM requires more sophisticated computational models. This method provides a comprehensive and integrative methodology to delve deeper into the complexities of antibody functionality, as shown in Supplementary Figure~\ref{fig::wm}.
    \item \textbf{CDR3 Random Mutation (CRM).} This strategy adopts a random mutagenesis approach focusing on specific sites within the CDR3 region. It involves randomly selecting 3-6 positions within the CDR3 domain and subsequently introducing random mutations at these sites. This method can be seen as a stochastic baseline that operates at a comparable edit distance. The result is shown in Supplementary Figure~\ref{fig::wmr}.
    \item \textbf{CDR3 Random Retrieval (CRR).}
    This strategy comprises the random substitution of the CDR3 region using sequences from other antibodies present in the SARS-CoV-2 wild-type library. The predicted structures are illustrated in Supplementary Figure~\ref{fig::rr}.
\end{itemize}

Under the aforementioned settings, we generate a set of 6,000 antibodies via \model-Ab-1B. Six antibodies are randomly selected as depicted in Supplementary Figure~\ref{fig:cdr_cases}. \model-AbFold is utilized as the structure prediction model. 
In response to the observation that using CDR3 short masking tends to generate antibodies closely resembling the ground truth with a small edit distance, we implemented a filter to exclude any antibodies with an edit distance of 2 or less. A series of generated sequences and their corresponding edit distances from the ground truth is presented in Supplementary Table~\ref{table:CDR3_Comparison}. Importantly, it is noteworthy that both the CSM and CWM policies are capable of generating sequences of varying lengths without resorting to mutations or deletions. In contrast, the sequences generated by the two parallel baselines, CRM and CRR, display considerable disorder, regardless of whether there are few mutations or a complete replacement of the entire CDR3 fragment. Our analysis further identifies a relationship between the edit distance and the structure of the generated antibody's CDR3 region. 
Specifically, as the edit distance grows, the organization of the CDR3 region tends to degrade, suggesting that even large generative models currently face limitations.



\section{Model FLOPs Comparisons}
\label{sec::model_flops}
We conduct a comparative analysis of computational resources utilized by different pre-trained protein language models (Supplementary Table~\ref{tab:flos_comparison}). The parameters detailed in this table are meticulously calculated by implementing the models as per the configurations outlined in their respective source papers and accompanying resources, such as code and model checkpoints. 
When discrepancies arise between a paper's theoretical account and its practical application, we favor the metrics provided in the paper.
From the right-hand side, the total training tokens are computed by multiplying the training steps, global batch size, and sequence length, given that all models listed are sequence language models. 
The model's parameters are estimated directly by following the authors' released implementations and hyperparameters, with the sum of the training parameters calculated while disregarding tied weights and buffers.
The total training computed is estimated by first approximating the FLOPs for one forward propagation (1F) of a single training sample. This is then multiplied by three to account for one forward and one backward propagation without activation recomputation (1F1B). The resulting number is then multiplied by the number of samples used during the entire pre-training process, which is equivalent to the total training tokens divided by the sequence length during pre-training.
Only matrix multiplication (matmul) operations are considered in the compute statistics, with other operations such as embedding, element-wise addition, softmax, and layer normalization excluded from the FLOP count. The matmuls considered within the attention block include key, query, and value transformations, attention matrix computation, attention over values, and post-attention linear transformation. Hidden size transformations in the feed-forward block, a projection from hidden dimensions into vocabulary dimensions, and a linear transformation in the language model head (if one exists), are also included in the matmul FLOPs. As an example, if hidden states of size (B, L, D) are multiplied by a weight matrix of size (D, 4D), the resulting FLOPs is BLD4D2 (the factor of 2 accounts for multiplication and addition operations). The total training compute for ProtGPT2 is estimated assuming each A100 GPU performs 120 TFLOPs per second. Consequently, 128 A100 GPUs would achieve approximately 5.3e+21 FLOPs over four days of training.

\begin{figure*}[!t]
\centering
\includegraphics[width=1.0\linewidth]{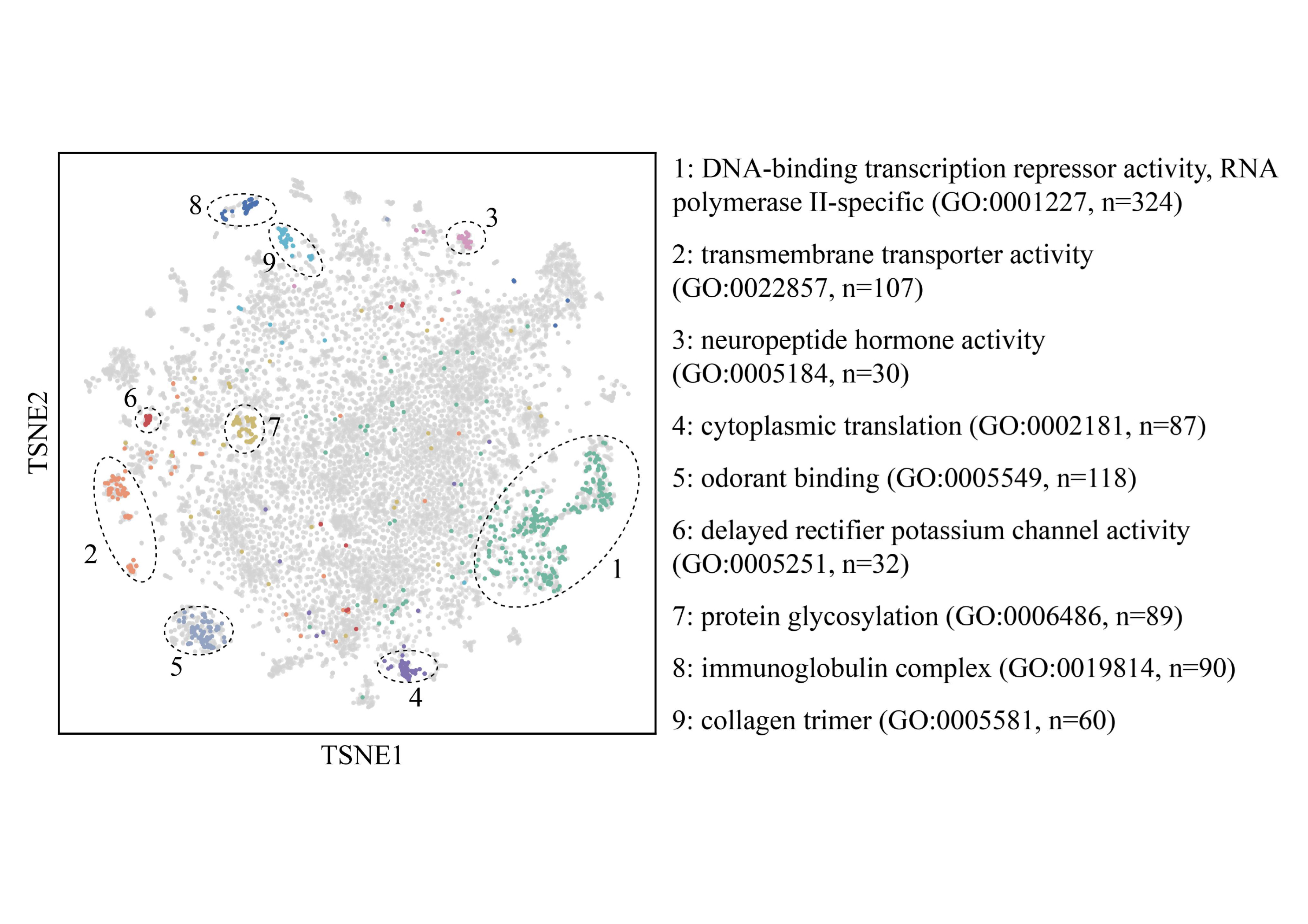} 
\caption[\label{fig:cluster_emb} Human Protein Sequence Mapping.]{ Human Protein Sequence Mapping. This panel illustrates the t-SNE visualization of \model-100B embeddings for human protein sequences. The visualization demonstrates \model's ability to capture biologically significant latent embeddings across a variety of functional protein sequences. Using a dataset of 20,255 human protein sequences from UniProt, each protein is represented as a distinct dot. The figure highlights nine clusters, each correlating with specific Gene Ontology annotation terms, differentiated by unique colors.}
\end{figure*}



\begin{figure*}[t]
\centering
\includegraphics[width=1.0\linewidth]{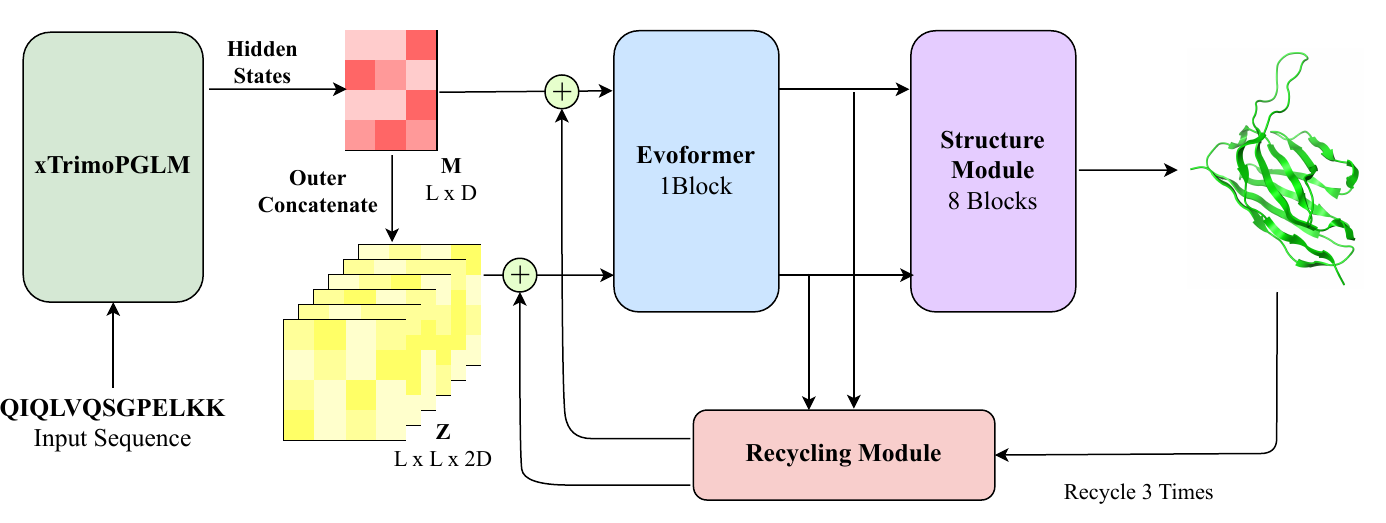} 
\caption[\model-Abfold Architecture]{Architecture of \model-AbFold for structure prediction.}
\label{fig:structure_prediciton_model}
\end{figure*}


\begin{figure}[t]
\centering

\subfigure[CDR3 Short Masking]{\label{fig::sm}
    \includegraphics[width=0.41\textwidth]{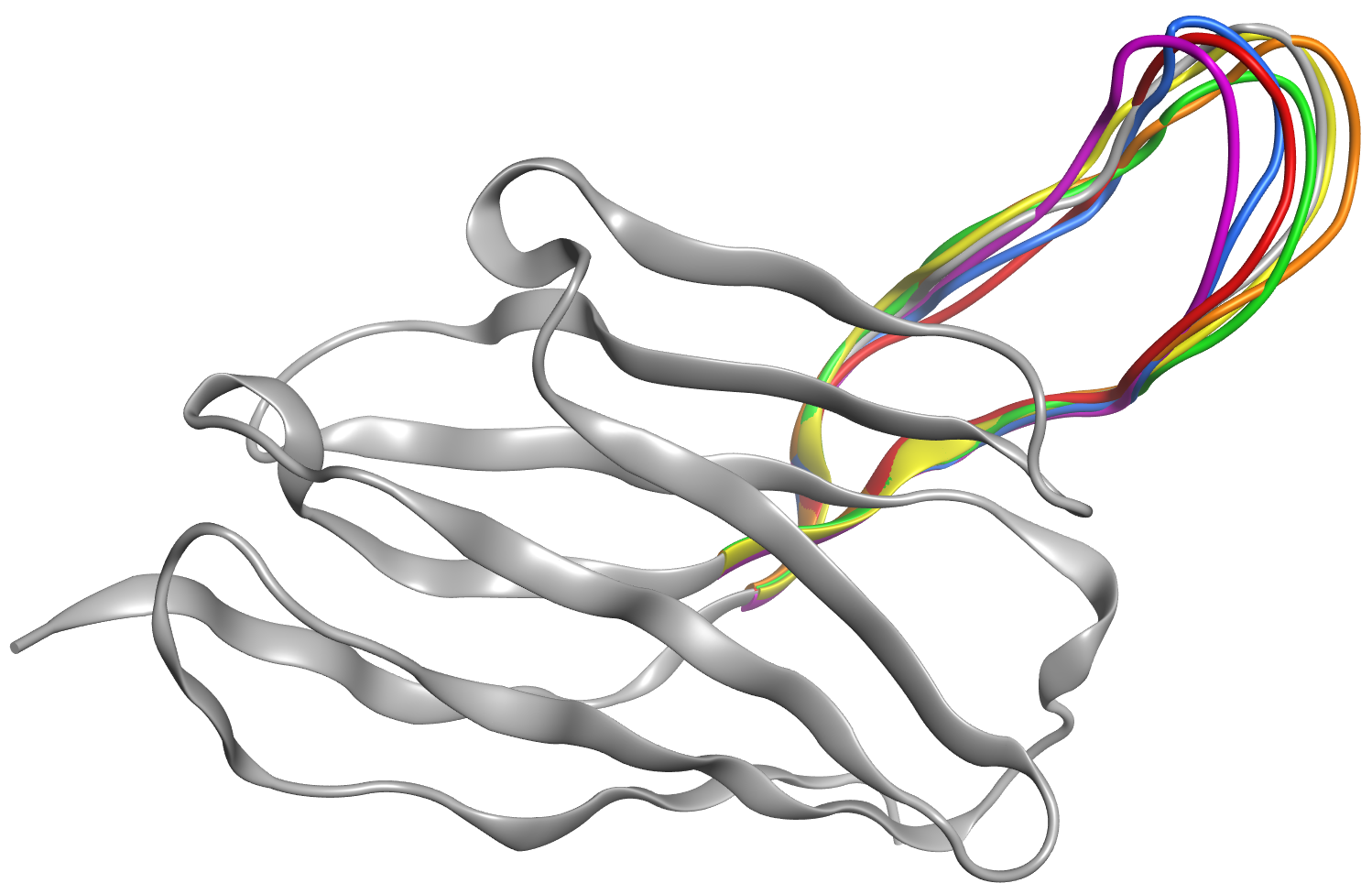}
}
\hspace{-0.1in}
\subfigure[CDR3 Whole Masking]{\label{fig::wm}
    \includegraphics[width=0.41\textwidth]{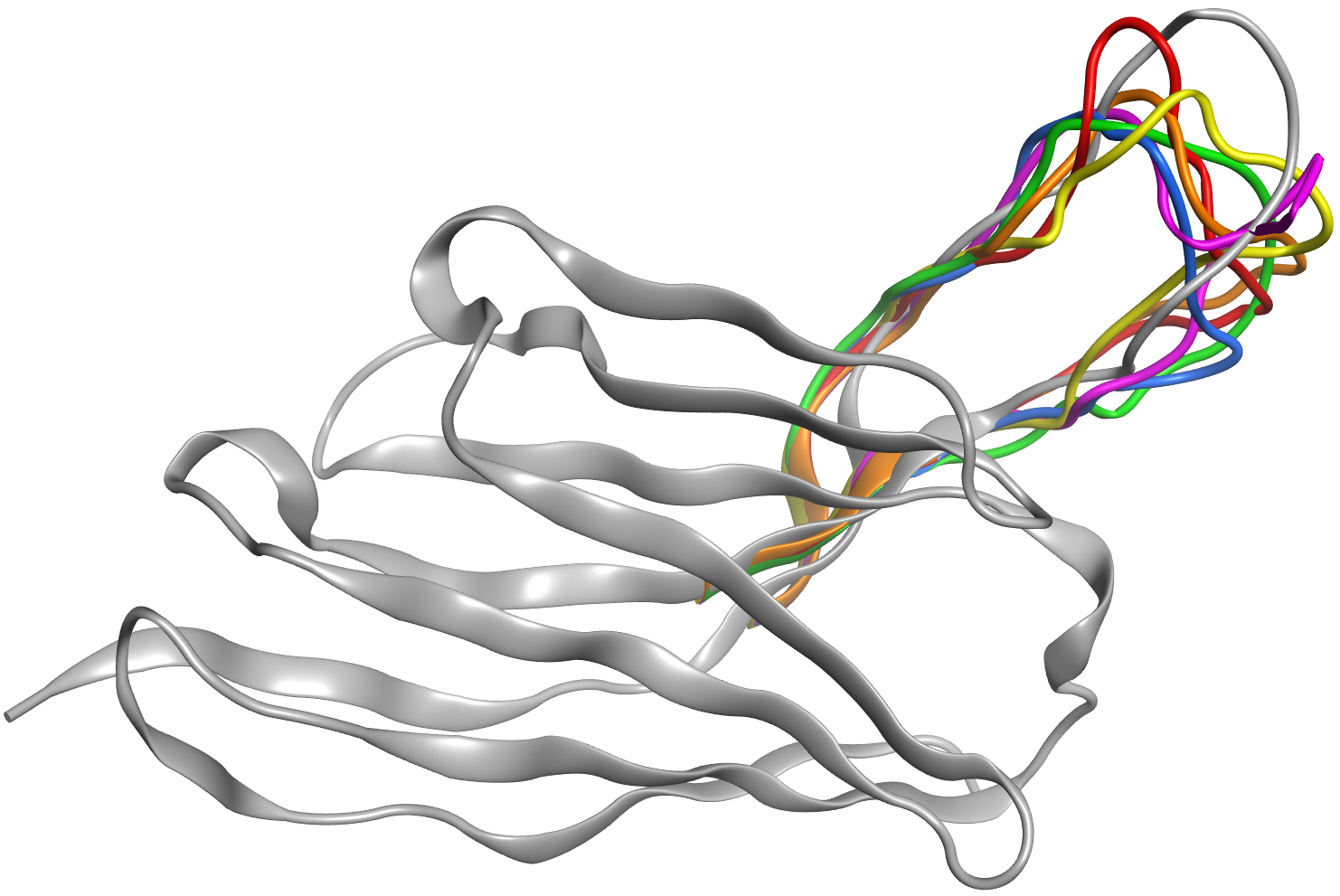}
}
\vspace{-0.1in}
\subfigure[CDR3 Random Mutation]{\label{fig::wmr}
    \includegraphics[width=0.41\textwidth]{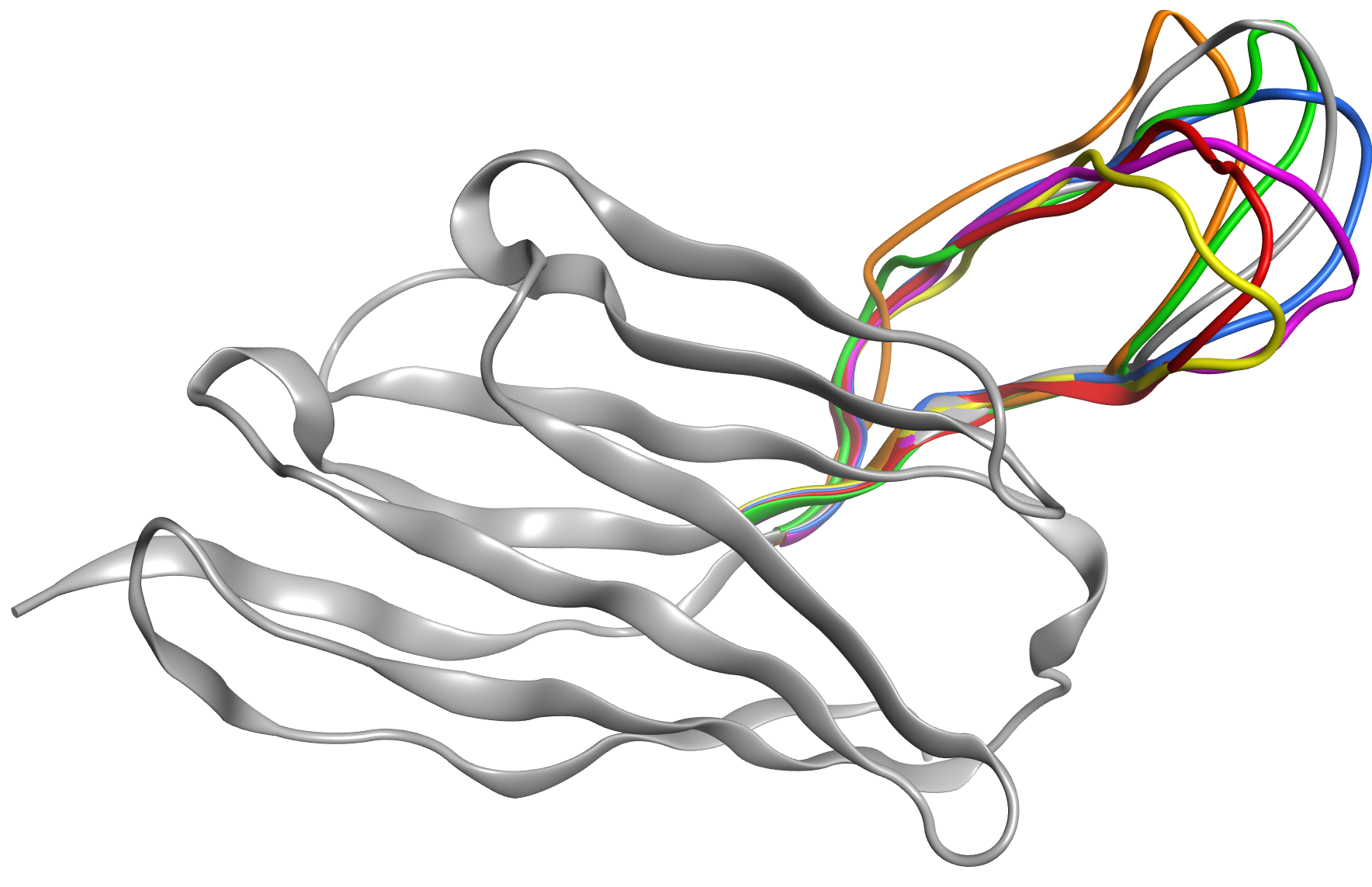}
}
\subfigure[CDR3 Random Retrieval]{\label{fig::rr}
    \includegraphics[width=0.41\textwidth]{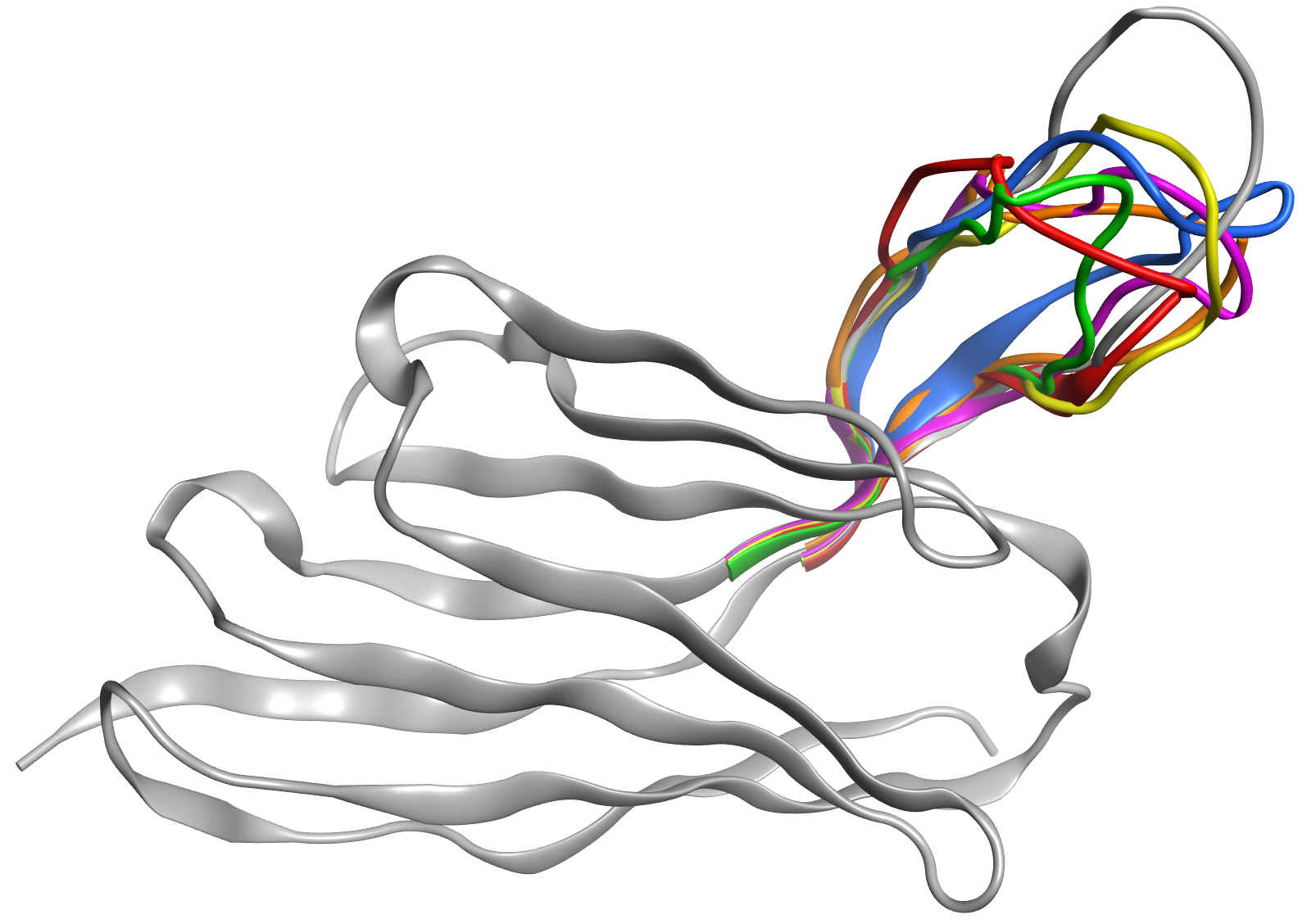}
}
\caption[Conformation of Antibody Generation]{\label{fig:cdr_cases} Conformations of antibody generation. The conformations of various methodologies implemented for \model-AbFold. 
}
\end{figure}

\begin{table*}[ht]
\centering
\caption[The TM-score and GDT\_TS score on CAMEO and CASP15.]{The TM-score and GDT\_TS score on CAMEO and CASP15.}
\label{tb:gdt}
\begin{tabular}{@{}ccccc@{}}
\toprule
\multirow{2}{*}{Methods} 
& \multicolumn{2}{c}{CAMEO (194)} & \multicolumn{2}{c}{CASP15 (56)} \\ 
\cmidrule{2-5}
& TM-score        & GDT\_TS       & TM-score        & GDT\_TS       \\
\midrule
xT-Fold              & 0.86            & 0.83          & 0.70            & 0.64          \\
ESMFold              & 0.85            & 0.81          & 0.65            & 0.59          \\
OmegaFold            & 0.80            & 0.76          & 0.60            & 0.54          \\
AF2 (Single Seq.)    & 0.39            & 0.28          & 0.33            & 0.23          \\
AF2\_model\_3 (MSAs) & 0.87            & 0.85          & 0.76            & 0.70          \\
RoseTTAFold2 (MSAs)  & 0.86            & 0.83          & 0.73            & 0.68          \\ 
\bottomrule
\end{tabular}
\end{table*}

\begin{table}
	\newcolumntype{?}{!{\vrule width 1pt}}
	\newcolumntype{C}{>{\centering\arraybackslash}p{5em}}
 \newcolumntype{G}{>{\centering\arraybackslash}p{8em}}
	\caption{
		\label{tb:comp} Comparisons between different architectures of PLMs.} 
	\footnotesize
	\centering 
	\renewcommand\arraystretch{1.0}
	\begin{tabular}{@{~}l?@{~}*{1}{CCCCG}@{~}}
		\toprule
		\textbf{Downstream Task} & \textbf{Autoenc.} & \textbf{Autoreg.} & \textbf{Enc.-Dec.} & \textbf{GLM} & Example\\
            \midrule
            Protein Understanding & \checkmark & $\times$ & \checkmark & \checkmark & Contact Prediction\\ 
            Protein Generation & $\times$ & \checkmark &    --- & \checkmark & Antibody Re-design\\
            \midrule
            \tabincell{c}{\textbf{Representatives}} & \tabincell{c}{ESM-1b\cite{rives2021biological}, \\ ESM2\cite{lin2023evolutionary}, \\ Pro.BERT\cite{brandes2022proteinbert}} & \tabincell{c}{PROGEN\cite{madani2023large}, \\ PROGEN2\cite{nijkamp2023progen2}, \\ ProtGPT2\cite{ferruz2022protgpt2}} & \tabincell{c}{ProtTrans\cite{elnaggarrost}, \\ Ankh\cite{elnaggar2023ankh}} & \tabincell{c}{\model} & / \\ 
	
		\bottomrule
	\end{tabular}
\end{table}

\begin{table*}[t]
\centering
\caption[Overview of 18 Downstream Benchmark Tasks]{Summary information for 18 benchmarked downstream tasks.}
\vspace{0.2cm}
\resizebox{1.0\columnwidth}{!}{
\begin{tabular}{ccccccccc}
\toprule
Type & Task & Metric & Train & Valid & Test & Prev.Method & Perf. & xT-100B Perf.\\
\midrule
\multirow{3}{*}{Struc.} & Cont. P.  & Top L/5 ACC & 12K & 1.5K & 1.5K & ESM2-15B~\cite{lin2023evolutionary} & 92.19$^{\clubsuit}$ & \textbf{93.32}\\
 &Fold. P. & 12K-cls ACC & 12.3K & 736 & 3.2K & Ankh\_{L}~\cite{elnaggar2023ankh} & 61.10$^{\blacklozenge}$ & \textbf{75.61}\\
 & Sec. Struc. P. & 3-cls ACC & 11K & - & 39 & Ankh\_{L}~\cite{elnaggar2023ankh} & 80.70$^{\blacklozenge}$ & 75.33\\
\cmidrule{2-9}
\multirow{5}{*}{Dev.} & Sol. P. & 2-cls ACC & 62.4K & - & 6.9K & ESM2-15B~\cite{lin2023evolutionary} & 76.49$^{\clubsuit}$ & \textbf{79.45} \\
 &Stab. P.  & SRCC & 53.6K & 2.5K & 12.8K & ESM2-15B~\cite{lin2023evolutionary} & 80.75$^{\clubsuit}$ & \textbf{84.21}\\
 & Temp. Stab. & MCC & 283K & 63K & 73.2K & TemStaPro~\cite{pudvziuvelyte2024temstapro} & 83.80$^{\blacklozenge}$ & \textbf{94.22}\\
 & Opt. Temp. & SRCC & 1.7K & - & 190 & DeepET~\cite{Li2022DeepET} & 62.40 $^{\blacklozenge}$& \textbf{73.96}\\
 & Opt. PH & AUC & 7.1K & 760 & 1.9K & ESM2-15B~\cite{lin2023evolutionary} & 62.48 $^{\blacklozenge}$& \textbf{64.99}\\
 & Clo. CLF & AUC & 23.3K & - & 4.7K & ESM2-3B~\cite{lin2023evolutionary} & 77.09 $^{\blacklozenge}$& \textbf{84.82}\\
 & Mat. Pro. & AUC &23.3K & - & 4.7K & ESM2-15B~\cite{lin2023evolutionary} & 79.17 $^{\blacklozenge}$& \textbf{86.48}\\
\cmidrule{2-9}
\multirow{4}{*}{Inter.} & Metal B. & 2-cls ACC & 6K & - & 1.3K & ESM2-15B~\cite{lin2023evolutionary} & 79.35$^{\clubsuit}$ & \textbf{82.78} \\
 & Pept.-HLA Aff. & AUC & 57.4K & 7K & 8.4K & CcBHLA~\cite{Wu2023CcBHLA} & 95.00$^{\blacklozenge}$ & \textbf{96.68}\\
 & TCR-pMHC Aff. & AUC & 19.5K & - & 4.5K & epiTCR~\cite{Pham2023epiTCR} & 92.50$^{\blacklozenge}$ & \textbf{95.10}\\
\cmidrule{2-9}
\multirow{4}{*}{Func.} &Antib. Res.  & 19-cls ACC & 2K & - & 1.3K & ESM2-15B~\cite{lin2023evolutionary} & 98.28$^{\clubsuit}$ & \textbf{98.38}\\
 & Fluor. P. & SRCC & 21.4K & 5.4K & 27.2K &  Ankh\_{L}~\cite{elnaggar2023ankh} & 62.00 $^{\blacklozenge}$ & \textbf{66.00}\\
 & Fitness P. & SRCC & 6.3K & 699 & 1.7K & Ankh\_{L}~\cite{elnaggar2023ankh} & 84.00 $^{\blacklozenge}$ & \textbf{96.10}\\
 &Loc. P.  & 10-cls ACC & 6.6K & - & 1.8K & Ankh\_{L}~\cite{elnaggar2023ankh} & 83.20$^{\blacklozenge}$ & 81.60\\
 & Enzyme Eff. & PCC & 13.5K & 1.7K & 1.7K & DLKcat~\cite{Li2022DLkcat} & 71.00$^{\blacklozenge}$ & \textbf{74.79}\\
\bottomrule
\end{tabular}}

\label{tab::tasks}
\end{table*}

\begin{table*}[ht]
\centering
\caption[Parameter Efficient Fine-tuning on Downstream Tasks]{Performance of different models across all benchmarked downstream protein-related tasks. 
xT100B depicts \model-100B model, E15B and E150M for ESM-15B and ESM-150M model respectively. Metric values are shown in both probing and LoRA (in parentheses) fine-tuning modes, where the \underline{underline} denotes the best performance of probing and \textbf{bold} indicates the best performance of LoRA fine-tuning.
}
\vspace{0.3cm}
\resizebox{1.0\columnwidth}{!}{
\small
\begin{tabular}{cccccc}
\toprule
\multirow{2}{*}{Type} & \multirow{2}{*}{Task} & \multirow{2}{*}{Metric} & \multicolumn{3}{c}{Model} \\
\cmidrule(l){4-6} 
& & & xT100B~(LoRA) & E15B~(LoRA) & E150M~(LoRA) \\
\midrule

\multirow{3}{*}{P. Struc.} & Cont. Pred.  & Top L/5 ACC & \underline{76.86} (\textbf{93.32}) & 73.52 (92.19) & 63.60 (84.72) \\
 &Fold Pred. & 12-cls ACC & \underline{71.57} (\textbf{75.61}) & 67.39 (69.20) & 54.87 (59.25)  \\
 & Sec. Struc. Pred. & 3-cls ACC & \underline{74.63} (75.33)  & 74.40 (\textbf{75.85}) & 73.31 (74.15) \\
\midrule

\multirow{4}{*}{P. Func.} &Antib. Res.  & 19-cls ACC & \underline{98.29} (\textbf{98.38}) & 98.13 (98.28) & 97.54 (96.94)  \\
 & Fluor. & SRCC &  \underline{65.16} (\textbf{66.00}) & 63.84 (63.71) & 52.68 (54.54)  \\
 & Fitness & SRCC &  \underline{81.69} (\textbf{96.10}) & 77.12 (94.75) &  69.60 (94.65) \\
 &Localization  & 10-cls ACC &  79.99 (81.60) & \underline{80.78} (\textbf{82.35}) &  77.85 (78.88) \\
\midrule

\multirow{4}{*}{P. Inter.} & Enzyme eff. & PCC & \underline{71.44} (\textbf{74.79}) & 68.95 (74.58) & 65.77 (71.72) \\
& Metal Bind. & 2-cls ACC & \underline{81.70} (\textbf{82.78})  & 79.35 (80.85) & 73.94 (81.53) \\
 & Pept.-HLA/MHC Aff. & AUC & 87.22 (96.68) & 90.48 (\textbf{97.28}) & \underline{91.39} (97.12) \\
 & TCR-pMHC Aff. & AUC & 89.76 (\textbf{95.10}) & \underline{91.10} (94.05) & 87.81 (90.40) \\
\midrule

\multirow{6}{*}{P. Dev.} & Solubility & 2-cls ACC & \underline{76.04} (\textbf{79.45}) & 74.76 (74.63) & 71.50 (72.47) \\
 &Stability  & SRCC & \underline{75.52} (\textbf{84.21})  & 71.69 (80.75) & 69.08 (77.69) \\
 & Temp. Sta. & MCC & \underline{93.07} (\textbf{94.22}) & 93.01 (93.24) & 86.28 (85.93) \\
 & Opt. Temp. & SRCC & \underline{73.64} (\textbf{73.96}) & 73.08 (73.29) & 68.57 (69.47) \\
 & Opt. PH & SRCC & \underline{61.39} (\textbf{64.99}) & 59.35 (62.48) & 59.60 (56.20) \\
 & Clone CLF. & AUC & \underline{84.55} (\textbf{84.82}) & 76.06 (76.64) & 73.55 (76.07) \\
 & Mat. Pro. & AUC & \underline{86.52} (\textbf{86.48}) & 79.35 (79.17) & 76.18 (77.91) \\

\bottomrule
\end{tabular}}
\label{tasksComparison}
\end{table*}

\begin{table*}[ht]
\centering
\caption[Downstream Task Comparison with ProtT5.]{Performance comparisons between ProtT5~\cite{elnaggarrost} with ESM2-15B, ESM2-3B, and xTrimoPGLM-100B on all eight downstream protein-related tasks via the Linear Probing approach. We utilize the best-performing protein language model, ProtT5-XL-U50, with 3 billion parameters. xT100B depicts \model-100B model, E15B and E3B for ESM2-15B and ESM2-3B model respectively. The \textbf{bold} denotes the best performance and  \underline{underline} indicates the second-best performance.
}
\vspace{0.3cm}
\resizebox{1.0\columnwidth}{!}{
\small
\begin{tabular}{ccccccc}
\toprule
\multirow{2}{*}{Type} & \multirow{2}{*}{Task} & \multirow{2}{*}{Metric} & \multicolumn{4}{c}{Model} \\
\cmidrule(l){4-7} 
& & & xT100B & E15B & E3B & ProtT5-XL3B \\
\midrule

\multirow{2}{*}{P. Struc.} 
& Cont. Pred.  & Top L/5 ACC & \textbf{76.86} & 73.52 &  \underline{76.01} & 74.52 \\
 &Fold Pred. & 12-cls ACC & \textbf{71.57}  & 67.39  & 69.20 & \underline{67.54}  \\
\midrule

\multirow{3}{*}{P. Func.} 
& Fluor. & SRCC &  \textbf{65.16}  & \underline{63.84}  & 56.68 & 52.67 \\
 & Fitness & SRCC &  \textbf{81.69} & \underline{77.12} &  73.16 &74.39  \\
 &Localization  & 10-cls ACC &  79.99 & \underline{80.78} &  80.07 & \textbf{80.89}  \\
\midrule

\multirow{1}{*}{P. Inter.} 
& Metal Bind. & 2-cls ACC & \textbf{81.70}  & 79.35 & \underline{79.72} & 77.78\\
\midrule

\multirow{2}{*}{P. Dev.} 
& Solubility & 2-cls ACC & \textbf{76.04} & 74.76 & 72.74 & \underline{74.91}\\
 &Stability  & SRCC & \textbf{75.52} & 71.69 & 69.93 & \underline{73.26} \\

\bottomrule
\end{tabular}}
\label{tasksComparisont5}
\end{table*}

\begin{table*}[ht]
\centering
\caption[Unnormalized Performance of xTrimo-series Models]{Comparison of xTrimo-series models across eight downstream protein-related tasks using the Linear Probing approach. We include models with 230M, 650M, 1B, 3B, 10B, and 100B parameters.\textbf{Bold} indicates the best performance, while \underline{underline} signifies the second-best.}
\vspace{0.3cm}
\resizebox{1.0\columnwidth}{!}{
\small
\begin{tabular}{ccccccccc}
\toprule
\multirow{2}{*}{Type} & \multirow{2}{*}{Task} & \multirow{2}{*}{Metric} & \multicolumn{6}{c}{Model} \\
\cmidrule(l){4-9} 
& & & 230M & 650M & 1B & 3B & 10B & 100B \\
\midrule

\multirow{2}{*}{P. Struc.} 
& Cont. Pred.  & Top L/5 ACC 
& 59.46&60.57&69.49&72.23&\underline{75.10}&\textbf{76.86}\\
 &Fold Pred. & 12-cls ACC 
 & 30.65&29.98&34.53&54.63&\underline{66.10}&\textbf{71.57} \\
\midrule

\multirow{3}{*}{P. Func.} 
& Fluor. & SRCC 
& 54.59&45.59&55.62 & 56.65 & \underline{58.85} & \textbf{65.00} \\
 & Fitness & SRCC 
 & 73.60 & 75.86 & 76.90  & 77.26 & \underline{77.98} & \textbf{81.69}  \\
 &Localization  & 10-cls ACC 
 & 73.00 & 74.12 & 74.80 & \underline{75.13} & 74.65 & \textbf{79.99}   \\
\midrule

\multirow{1}{*}{P. Inter.} 
& Metal Bind. & 2-cls ACC 
& 74.2 & 71.61 & 73.00 & 75.37 & \underline{78.22} & \textbf{81.70}\\
\midrule

\multirow{2}{*}{P. Dev.} 
& Solubility & 2-cls ACC 
& 71.21 & 72.63 & 73.43 & 74.00 & \underline{74.56} & \textbf{76.04}\\
 &Stability  & SRCC 
 & 74.20 & 73.70 & 74.61 & \textbf{80.10} & \underline{78.36} & 75.52 \\

\bottomrule
\end{tabular}}
\label{tab:unnormalized}
\end{table*}

\begin{table*}[ht]
\centering
\caption[Performance comparisons among zero-shot and supervised fitness prediction.]{Comparison of performance (Spearman Coefficient) between zero-shot and supervised fitness prediction models. 'E1V' refers to ESM-1V, while 'E650M', 'E1B', 'E3B', and 'E15B' denote ESM2 models with 650M, 1B, 3B, and 15B parameters, respectively. 'xT1B', 'xT3B', 'xT10B', and 'xT100B' represent xTrimo-series models. P.GEN2 denotes the PROGEN2 model. Results are based on single models, without ensemble methods. For ESM-1V, the average performance across its 5 model variants is reported.}
\label{tb:zerofit}
\begin{tabular}{@{}cccc@{}}
\toprule
\multirow{2}{*}{Methods} 
& \multicolumn{1}{c}{Zero-Shot (ProteinGym)} & \multicolumn{2}{c}{Supervised (GB1)} \\ 
\cmidrule{2-4}
& -       & LP       & LoRA        \\
\midrule
ESM-1V &0.4045 & - & -\\
ESM2-650M &0.4172 & 0.7533 & 0.9515\\
ESM2-3B &\textbf{0.4220} & 0.7669 & 0.9471\\ 
ESM2-15B  &0.4169 & 0.7712 & 0.9475 \\ 
\midrule
P.GEN2-small &0.3223&-&-\\
P.GEN2-base &0.3764&-&-\\
P.GEN2-large &0.3761&-&-\\
P.GEN2-xlarge &0.3781&-&-\\
\midrule
xT1B  & 0.4067 & 0.7690 & 0.9577 \\
xT3B   & 0.4192 & 0.7726 & 0.9542 \\
xT10B &0.4176  & 0.7798 & \textbf{0.9624} \\
xT100B &0.3987 & \textbf{0.8169} &0.9610\\
\bottomrule
\end{tabular}
\end{table*}

\begin{table*}[!ht]
\center
\caption[Compute Comparisons of Pre-training Language Models]{Comparison of training computes between different pre-trained protein language models.} 
\vspace{0.3cm}
\label{tb:training_flops_comparison}
\small
\begin{tabular}{cccc}
\hline
 Model & Total train compute (FLOPs) & Params & Training tokens \\
\hline
ESM2-150M & 1.1E+21 & 150M & 1,000B \\
ESM2-650M & 4.4E+21  & 650M & 1,000B\\
ESM2-3B & 1.8E+22  & 2.8B & 1,000B\\
ESM2-15B & 5.1E+22  & 15B & 864B\\

ProtBert & 7.6E+21  & 420M & 2516B\\
ProtT5-xl & 1.7E+22  & 2.8B & 1,929B\\
ProtT5-xxl & 3.7E+22  & 11B & 1,039B\\

Ankh-base & 2.6E+21  & 740M & 952B\\
Ankh-large & 6.5E+21  & 1.9B & 952B\\

ProtGPT2 & -  & 740M & 4.8B (per epoch)\\

PROGEN & 7.6E+21  & 1.2B & 1,049B\\

PROGEN2-small & 1.8E+20  & 150M & 170B\\
PROGEN2-medium & 8.9E+20  & 760M & 170B\\
PROGEN2-base & 1.1E+21  & 760M & 200B\\
PROGEN2-large & 3.4E+21  & 2.8B & 200B\\
PROGEN2-xlarge & 1.4E+22  & 6.4B & 350B\\

xTrimoPGLM-100B & 6.2E+23  & 101B & 1,000B\\
\hline
\label{tab:flos_comparison}
\end{tabular}
\end{table*}

\begin{table*}[ht]
\center
\small
\caption[Zero-shot performance of xTrimoPGLM-AB]{\label{tb:zero_shot}Performance of different antibody pre-training models in zero-shot naturalness datasets. 
}
\vspace{0.2cm}
\begin{tabular}{c|ccc|ccc}
\hline
\multirow{2}*{Model} & \multicolumn{3}{c|}{\textsc{Dataset 1}} & \multicolumn{3}{c}{\textsc{Dataset 2}} \\
 & H Chain & L Chain & Pair & H Chain & L Chain & Pair \\
\hline
Iglm \cite{shuai2021generative} & 0.698 & 0.651 & 0.683 & 0.703 & 0.594 & 0.665 \\
AbLang \cite{olsen2022ablang} & 0.655 & 0.497 & 0.613 & 0.713 & 0.671 & 0.679 \\
ESM2-15B \cite{lin2023evolutionary} & 0.682 & 0.552 & 0.686 & 0.716 & 0.510 & 0.626 \\
AntiBERTy \cite{ruffolo2021deciphering} & 0.763 & 0.549 & 0.699 & 0.723 & 0.678 & 0.679 \\
Progen2-oas \cite{nijkamp2023progen2} & 0.703 & \textbf{0.734} & 0.748 & 0.701 & 0.565 & 0.644 \\
\hline
xTrimoPGLM-Ab-1B PPL & 0.745 & 0.696 & \textbf{0.756} & 0.702 & 0.688 & 0.704 \\
xTrimoPGLM-Ab-1B PPPL & 0.754 & 0.683 & 0.750 & 0.741 & 0.668 & 0.700 \\
xTrimoPGLM-Ab-1B-GLM PPL & \textbf{0.763} & 0.676 & 0.742 & 0.703 & 0.685 & \textbf{0.724} \\
xTrimoPGLM-Ab-1B-MLM PPPL & 0.733 & 0.682 & 0.746 & \textbf{0.766} & \textbf{0.704} & 0.722 \\
\hline
\multicolumn{1}{c}{Ablation Study} \\
\hline
xTrimoPGLM-Ab-1B-GLM-CDR PPL & 0.652 & 0.700 & 0.689 & 0.699 & 0.647 & 0.671 \\
xTrimoPGLM-Ab-1B-GLM-Random PPL & 0.736 & 0.666 & 0.725 & 0.715 & 0.640 & 0.708 \\
\hline
\end{tabular}
\end{table*}

\begin{table*}[!t]
\begin{center}
\caption[Structure Prediction of VH and VL Individuals]{Structure prediction of VH and VL in antibodies. RMSD H1-3 means RMSD on CDR1-3 of heavy chains and RMSD L1-3 means RMSD on CDR1-3 of light chains.}
\vspace{0.3cm}
\label{table:vh_vl_structure}
\small
\setlength{\tabcolsep}{4pt}
\begin{tabular}{ccc|ccc|cccc}
\toprule
 \multirow{2}*{Model}&   
 \multirow{2}*{\textsc{RMSD$\downarrow$}}  &  \multirow{2}*{\textsc{TM-score$\uparrow$}}  &  \multicolumn{3}{c|}{\textsc{Heavy Chain} RMSD$\downarrow$}                            & \multicolumn{3}{c}{\textsc{Light Chain} RMSD$\downarrow$} \\ \cline{4-9}
                                                & & & \textsc{H1}   & \textsc{H2}    & \textsc{H3}    & \textsc{L1}  & \textsc{L2}  & \textsc{L3 } \\  \midrule
AlphaFold2      & 1.225  & 0.951                  & 1.254                 & 1.091         & 2.826          & 0.89                                                                               & 0.723 & 1.329\\
 OmegaFold                   & 1.337           & 0.946                 & 1.418                  & 1.183         & 3.246          & 0.860                                                                              &0.598 & 1.360\\
 ESMFold                       & 1.421           & 0.943                  & 1.464                  & 1.320          & 3.409          & 1.048                                                                              & 0.679 & 1.520\\
 IgFold               &1.261           & 0.945                 & 1.324                  & 1.126          & 2.998          & 0.948                                                                              & 0.589 & 1.318 \\ 
 xTrimoAbFold                    & 1.089            & 0.958                  & 1.176                  & 0.912        & 2.472 & 0.811 &\textbf{0.566}                                                                               & 1.038 \\
\midrule
xTrimoPGLM-AbFold                                           & \thead{\textbf{0.9823} \\ $\pm$0.007}           & \thead{\textbf{0.961} \\ $\pm$0.001}        & \thead{\textbf{1.089} \\ $\pm$0.012}       & \thead{\textbf{0.866} \\ $\pm$0.011}         & \thead{\textbf{2.230} \\ $\pm$0.04}          & \thead{\textbf{0.779} \\ $\pm$0.017} & \thead{ {0.573} \\ $\pm$0.008} 
&\thead{\textbf{0.937} \\ $\pm$0.014} \\  

\bottomrule
\end{tabular}
\end{center}
\end{table*}

\begin{table*}[!t]
\center
\caption[Structure Prediction of VH-VL Complexes]{Structure prediction of VH-VL complexes. The inference time is calculated on the whole test set with a single A100 GPU. xTrimoPGLM-AbFold (evo 1) and xTrimoPGLM-AbFold (evo 16) are xTrimoPGLM-AbFold with 1 Evoformer block and 16 Evoformer blocks respectively.}
\vspace{0.3cm}
\label{tb:vh_vl_complex_structure} 
\small
\begin{tabular}{ccccc}
\toprule
 & {\textsc{RMSD$\downarrow$}} & {\textsc{TM-score$\uparrow$}}  & {\textsc{DockQ $\uparrow$}} & {\textsc{Inference time $\downarrow$}} \\
\midrule
ZDock & 10.982 & 0.596 & 0.108 & 34h \\
ClusPro & 5.899 & 0.792  & 0.404 & 1.3h \\
EquiDock & 18.293 & 0.559  & 0.032 & 2m \\
HDOCK & 2.032 & 0.926 & 0.705 & 3.2h \\
\thead{AlphaFold-Multimer\\\ } & \thead{1.325\\\ } & \thead{0.962\\\ } & \thead{0.765\\\ } & \thead{56.6h (original) \\55m (faster MSA)}\\
\midrule
xTrimoPGLM-AbFold (evo 1) & 1.304   & 0.962 & 0.762 & \textbf{32s}\\
xTrimoPGLM-AbFold (evo 16) & \textbf{1.234}   & \textbf{0.966} & \textbf{0.770} & 82s\\
\bottomrule
\end{tabular}
\end{table*}

\begin{table*}[!t]
\centering
\tiny
\caption{Full Configurations for \model-100B Training.}
\label{tb:configs}
\begin{tabular}{|p{0.5\linewidth}|p{0.5\linewidth}|}
\hline
\textbf{KEY} & \textbf{Value} \\
\hline
glu\_activation & GeGLU  \\
hidden dim. & 10,240 \\
ffn size & 31,744 \\
\# layers & 72 \\
\# attention heads & 80 \\
sequence\_length & 2,048 \\
global batch size & 4,224 \\
max learning rate & 4e-05 \\
min learning rate & 4e-06 \\
\hline
adam\_beta1&0.9\\
adam\_beta2&0.95\\
adam\_eps&1e-08\\
aggregated\_samples\_per\_sequence& 1,2,4,8\\
attention\_dropout&0.1\\ 
attention\_softmax\_in\_fp32&True\\
average\_block\_length&6\\
bias\_dropout\_fusion&True\\
checkpoint\_activations&True\\
checkpoint\_in\_cpu&False\\
checkpoint\_num\_layers&9\\
clip\_grad&1.0\\
tensor\_parallel\_size&4\\
pipeline\_parallel\_size&8\\
data\_parallel\_size&24\\
deepnorm&True\\
distributed\_backend&nccl\\
eval\_interval&300\\
fp16&True\\
mlm\_prob&0.1\\
span\_prob&0.2\\
gpt\_prob&0.7\\
hidden\_dropout&0.1\\
init\_method\_std&0.0052\\
initial\_loss\_scale&65536\\
layernorm\_epsilon&1e-05\\
rotary\_embedding&2D\\
learnable\_rotary\_embedding&False\\
length\_per\_sample&2048\\
log\_interval&1\\
lr\_decay\_iter&None\\
lr\_decay\_samples&439,453,125\\
lr\_decay\_style&cosine\\
lr\_warmup\_samples& 14,648,437\\
make\_vocab\_size\_divisible\_by&128\\
masked\_softmax\_fusion&True\\ 
micro\_batch\_size&1\\
min\_gmask\_ratio&0.4\\
min\_loss\_scale&1.0\\
optimizer&adamw\\
partition\_activations&True\\
rampup\_batch\_size&240,24,12207031\\
save\_interval&300\\
seed&1234\\
short\_seq\_prob&0.02\\
shrink\_embedding\_gradient\_alpha&0.1\\ 
single\_span\_prob&0.02\\
split&949,50,1\\
tokenizer\_type& ProteinTokenizer\\
weight\_decay &0.1\\
zero\_stage & 1\\
\hline
\textsc{Finetune}&\\
\hline
lora\_($R$, $\alpha$)&(8,16),(16,32)\\
\hline
\end{tabular}
\end{table*}

\definecolor{brightpurple}{rgb}{0.74, 0.2, 0.64}

\begin{table}[h!]
\centering
\caption[Edit Distance of Antibody Generation]{A collection of sequences produced via four distinct masking approaches. 
}
\vspace{0.3cm}
\begin{tabular}{c|l|c}
\hline
\textbf{Marker} & \textbf{CDR3 Short Masking} & \textbf{Edit Distance} \\
\hline
\textcolor{gray}{Ground truth} & AKDKDYGDLPTVDYYYHYGMDV & - \\
\hline
\textcolor{red}{Red} & AKDKDYGDLPTVLRYYYYGMDV & 3 \\
\textcolor{green}{Green} & AKDKDYGDLPQYYYYHYGMDV & 3 \\
\textcolor{blue}{Blue} & AKDKDYGDLPSLSYYYHYGMDV & 3 \\
\textcolor{yellow}{Yellow} & AKDKDYGDLPTVDYFFLLGMDV & 4 \\
\textcolor{brightpurple}{Purple} & AKDKDYGDLSLSPPYYHYGMDV & 5 \\
\textcolor{orange}{Orange} & AKDKDYGDLPTVDYYDYYGLDV & 3 \\
\hline
 & \textbf{CDR3 Whole Masking} & \\
\hline
\textcolor{red}{Red} & AKDSYYGSGSYYNPDQGYYYYYGMDV & 12 \\
\textcolor{green}{Green} & AKDGPGGSGSYSADYYYYYGMDV & 10 \\
\textcolor{blue}{Blue} & AKDKDCGGDCYLLDYHYYYGMDV & 8 \\
\textcolor{yellow}{Yellow} & AKDSTVTPLPAAIRTYYYYYYGMDV & 12 \\
\textcolor{brightpurple}{Purple} & AKDLNRRGISIFGVDNDYYFYGLDV & 13 \\
\textcolor{orange}{Orange} & AKDSYYGSGSYSYVSYYYYYYGMDV & 11 \\
\hline
 & \textbf{CDR3 Random Mutations} & \\
\hline
\textcolor{red}{Red} & AKDKDHVGFMTVDYYYHYGMDV & 4 \\
\textcolor{green}{Green} & AKDILFIDLPTVDYYYHYGMDV & 5 \\
\textcolor{blue}{Blue} & AKDKDYGDLPTVDYYYLQLIPC & 6 \\
\textcolor{yellow}{Yellow} & AKDKDYGDLPTVDYDIGYGMDV & 3 \\
\textcolor{brightpurple}{Purple} & AKDKDYRHRETVDYYYHYGMDV & 4 \\
\textcolor{orange}{Orange} & AKDKDYGDLPTVDYYYALRRRR & 6 \\
\hline
 & \textbf{CDR3 Random Retrieval} & \\
\hline
\textcolor{red}{Red} & ARDRSGKDVLTGYPMFPAGMDV & 14 \\
\textcolor{green}{Green} & ARDLSAGHCTGGVCYTAGGIDY & 16 \\
\textcolor{blue}{Blue} & ARGVITMVRGVIRDYYYYGMDV & 13 \\
\textcolor{yellow}{Yellow} & ARDLGGGYSNVYVNHYYGMDV & 12 \\
\textcolor{brightpurple}{Purple} & ARDEITVTAGAWGNYYYGMDY & 14 \\
\textcolor{orange}{Orange} & AKGYCGGDCYSGLLDWYFDL & 16 \\
\hline

\end{tabular}

\label{table:CDR3_Comparison}
\end{table}


\begin{table*}[!t]
\center
\caption[Distorder Proteins and Domains]{Disorder proteins/domains predictions $(\%)$.}
\vspace{0.3cm}
\label{tb:gene_dis_structure} 
\small
\begin{tabular}{ccccc}
\hline
 & {\textsc{Short}} & {\textsc{Long}}  & {\textsc{Globular}} & {\textsc{Ordered}} \\
\hline
Natural Data (10K) & 59.84  & 64.27 & 64.96 & 34.56\\
Generated & 63.38 & 68.16 & 68.57 & 34.20 \\
Random Generation & 56.10 & 55.11 & 54.84 & 92.58\\
\hline
\end{tabular}
\end{table*}